\documentclass{article}

\PassOptionsToPackage{numbers, compress}{natbib}

\usepackage[preprint]{neurips_2026}

\usepackage[utf8]{inputenc}
\usepackage[T1]{fontenc}
\usepackage{microtype}
\usepackage{url}

\usepackage{amsmath}
\usepackage{amssymb}
\usepackage{amsfonts}
\usepackage{amsthm}
\usepackage{mathtools}
\usepackage{nicefrac}

\PassOptionsToPackage{table,dvipsnames,xcdraw}{xcolor}
\usepackage{xcolor}

\definecolor{oxfordblue}{rgb}{0.0, 0.13, 0.28}
\definecolor{linkblue}{rgb}{0.0, 0.20, 0.40}
\definecolor{limegreen}{rgb}{0.2, 0.8, 0.2}
\definecolor{bondiblue}{rgb}{0.0, 0.58, 0.71}
\definecolor{britishracinggreen}{rgb}{0.0, 0.26, 0.15}
\definecolor{faintgray}{RGB}{245,245,245}
\definecolor{faintborder}{RGB}{230,230,230}
\definecolor{lightblack}{gray}{0.4}
\definecolor{pinegreen}{rgb}{0.0, 0.47, 0.44}
\definecolor{codegreen}{rgb}{0,0.6,0}
\definecolor{codegray}{rgb}{0.5,0.5,0.5}
\definecolor{codepurple}{rgb}{0.58,0,0.82}
\definecolor{backcolour}{rgb}{0.97,0.97,0.97}
\colorlet{darkgreen}{green!45!black}
\colorlet{darkred}{red!70!black}
\colorlet{limegreen}{LimeGreen}

\definecolor{paperVanilla}{HTML}{1F4E79}    \definecolor{paperPriceAux}{HTML}{C74C2A}   \definecolor{paperPriceRt}{HTML}{2C8C5D}    \definecolor{paperUnsafe}{HTML}{7A3A8A}     

\definecolor{bleudefrance}{rgb}{0.19, 0.55, 0.91}
\definecolor{darkcyan}{rgb}{0.0, 0.55, 0.55}
\definecolor{darkgreen}{rgb}{0.0, 0.2, 0.13}
\definecolor{darkspringgreen}{rgb}{0.09, 0.45, 0.27}
\definecolor{darkmidnightblue}{rgb}{0.0, 0.2, 0.4}
\definecolor{MidnightBlue}{RGB}{25,25,112}
\definecolor{MidnightBlueComplementingGreen}{RGB}{25,112,25}
\definecolor{MidnightBlueComplementingPurple}{RGB}{112,25,112}
\definecolor{MidnightBlueComplementingRed}{RGB}{112,25,69}
\definecolor{WowColor}{rgb}{.75,0,.75}
\definecolor{MildlyAlarming}{rgb}{0.85,0.25,0.1}
\definecolor{SubtleColor}{rgb}{0,0,.50}
\definecolor{antiquefuchsia}{rgb}{0.57, 0.36, 0.51}
\definecolor{fashionfuchsia}{rgb}{0.96, 0.0, 0.63}
\definecolor{jade}{rgb}{0.0, 0.66, 0.42}
\definecolor{caribbeangreen}{rgb}{0.0, 0.8, 0.6}
\definecolor{aquamarine}{rgb}{0.5, 0.8, 0.85}
\definecolor{lightseagreen}{rgb}{0.13, 0.7, 0.67}
\definecolor{attentioncolor}{RGB}{152,90,81}
\definecolor{burgred}{RGB}{40,3,22}
\definecolor{AnnieGreen}{RGB}{17,123,92}
\definecolor{Turquoise}{RGB}{64,224,208}
\definecolor{darkjade}{RGB}{0,122,84}
\definecolor{RedAlizarin}{rgb}{0.82, 0.1, 0.26}
\definecolor{Window1}{RGB}{92,150,31}
\definecolor{Window1dark}{RGB}{41,67,13}
\definecolor{Window2}{RGB}{255,168,28}
\definecolor{Window2dark}{RGB}{114,75,12}
\definecolor{Window3}{RGB}{255,96,33}
\definecolor{Window3dark}{RGB}{97,36,12}
\definecolor{InputColor}{RGB}{20,255,177}
\definecolor{InputColorlight}{RGB}{222,237,229}

\definecolor{darkcerulean}{rgb}{0.03, 0.27, 0.49}
\definecolor{smokyblack}{rgb}{0.06, 0.05, 0.03}
\definecolor{warmblack}{rgb}{0.0, 0.26, 0.26}
\definecolor{cobalt}{rgb}{0.0, 0.28, 0.67}
\definecolor{darkcobalt}{rgb}{0.1, 0.38, 0.77}

\definecolor{darkerred}{RGB}{139, 0, 0}
\definecolor{darkergreen}{RGB}{0, 100, 0}
\definecolor{warmred}{RGB}{205, 92, 92}
\definecolor{warmgreen}{RGB}{34, 139, 34}
\definecolor{warmblue}{RGB}{70, 130, 180}

\definecolor{arrowredfig1}{HTML}{B85450}
\definecolor{expertfig1}{HTML}{6A9153}
\definecolor{mixturefig1}{HTML}{6C8EBF}

\definecolor{warmorange}{RGB}{210, 125, 75}
\colorlet{warmorangedark}{warmorange!80!black}   \colorlet{warmorangelight}{warmorange!15!white}  

\newcommand{\colorwblk}[1]{\textcolor{warmblack}{#1}}

\newcommand{\colorwblkb}[1]{\textcolor{warmblack}{\textbf{#1}}}

\usepackage{graphicx}
\usepackage{rotating}
\usepackage{lscape}
\usepackage{adjustbox}
\usepackage{subcaption}   \usepackage{float}

\usepackage{tikz}
\usetikzlibrary{matrix,chains,positioning,arrows,calc, shadings}

\usepackage{booktabs}
\usepackage{array}
\usepackage{tabularx}
\usepackage{multirow}
\usepackage{colortbl}     \usepackage{siunitx}
\usepackage{nicematrix}
\usepackage{algorithm}
\usepackage{algpseudocode}

\usepackage{listings}

\lstnewenvironment{rsverbatim}[1][]{\lstset{
    basicstyle=\small\ttfamily,
    columns=flexible,
    breaklines=true,
    frame=tb,
    #1
  }}{}

\usepackage[toc,page]{appendix}

\usepackage{titletoc}
\newcommand\DoToC{\startcontents
  \printcontents{}{1}{\textbf{Appendix Contents}\vskip3pt\hrule\vskip5pt}
  \vskip3pt\hrule\vskip5pt
}

\usepackage{enumitem}
\usepackage{wrapfig}
\usepackage{lipsum}
\usepackage{fancyvrb}
\usepackage{silence}
\usepackage[UKenglish]{isodate}
\usepackage{etoolbox}
\usepackage{comment}

\usepackage{hyperref}
\usepackage{cleveref}

\usepackage[framemethod=TikZ]{mdframed}

\newcounter{theo}[section] 
\renewcommand{\thetheo}{\arabic{section}.\arabic{theo}}

\newcounter{lem}[section] 
\renewcommand{\thelem}{\arabic{lem}}

\newcounter{prf}[section]
\renewcommand{\theprf}{\arabic{section}.\arabic{prf}}

\theoremstyle{remark}

\theoremstyle{definition}

\usepackage{bbm}             

\hypersetup{colorlinks=true,
            linkcolor=cobalt,
            urlcolor=darkcerulean,
            citecolor=cobalt}

\usepackage{tikz-cd}
    \pgfdeclarelayer{edgelayer}
    \pgfdeclarelayer{nodelayer}
    \pgfsetlayers{edgelayer,nodelayer,main}
\tikzstyle{new style 0}=[fill={rgb,255: red,255; green,94; blue,247}, draw=black, shape=circle]
    \tikzstyle{pointy}=[fill=white, draw=black, shape=circle]
\tikzstyle{pointy}=[->]
\usepackage{fontawesome}

\makeatletter
\newcommand{\pushright}[1]{\ifmeasuring@#1\else\omit\hfill$\displaystyle#1$\fi\ignorespaces}
\newcommand{\pushleft}[1]{\ifmeasuring@#1\else\omit$\displaystyle#1$\hfill\fi\ignorespaces}
\makeatother

\newcommand{\1}{\mathbbm{1}}

\renewcommand{\phi}{\varphi}

\newcounter{termcounter}
\renewcommand{\thetermcounter}{\Roman{termcounter}}
\crefname{term}{term}{terms}
\creflabelformat{term}{#2\textup{(#1)}#3}

\makeatletter
\def\term{\@ifnextchar[\term@optarg\term@noarg}\def\term@optarg[#1]#2{\textup{#1}\def\@currentlabel{#1}\def\cref@currentlabel{[][2147483647][]#1}\cref@label[term]{#2}}
\def\term@noarg#1{\refstepcounter{termcounter}\textup{(\thetermcounter)}\cref@label[term]{#1}}
\makeatother

\crefname{lemma}{lemma}{lemmata}
\Crefname{lemma}{Lemma}{Lemmata}
\crefname{assumption}{assumption}{assumptions}
\Crefname{assumption}{Assumption}{Assumptions}
\crefname{example}{Example}{Examples}
\crefname{proposition}{Proposition}{Proposition}

\usepackage{bm}

\renewcommand{\1}{\bm{1}}

\DeclareMathAlphabet{\mathsfit}{\encodingdefault}{\sfdefault}{m}{sl}
\SetMathAlphabet{\mathsfit}{bold}{\encodingdefault}{\sfdefault}{bx}{n}

 \graphicspath{{figs/}}

\usepackage[utf8]{inputenc} \usepackage[T1]{fontenc}    \usepackage{url}            \usepackage{booktabs}       \usepackage{amsfonts}       \usepackage{nicefrac}       \usepackage{microtype}      \usepackage{xcolor}         \usepackage{ulem}

\usepackage{titlesec}
\titlespacing*{\subsection}{0pt}{0.4em}{0.2em}

\lstdefinestyle{pythonstyle}{
    backgroundcolor=\color{backcolour},
    commentstyle=\color{codegreen},
    keywordstyle=\color{blue},
    numberstyle=\tiny\color{codegray},
    stringstyle=\color{codepurple},
    basicstyle=\ttfamily\footnotesize,
    breakatwhitespace=false,
    breaklines=true,
    captionpos=b,
    keepspaces=true,
    numbers=left,
    numbersep=5pt,
    showspaces=false,
    showstringspaces=false,
    showtabs=false,
    tabsize=2,
    language=Python,
    extendedchars=true,
    inputencoding=utf8,
    literate={—}{{---}}1
        {–}{{--}}1
        {“}{{``}}1
        {”}{{''}}1
        {‘}{{`}}1
        {’}{{'}}1
        {…}{{\ldots}}1
        {×}{{$\times$}}1
        {≤}{{$\leq$}}1
        {≥}{{$\geq$}}1
        {≠}{{$\neq$}}1
        {±}{{$\pm$}}1
        {α}{{$\alpha$}}1
        {β}{{$\beta$}}1
        {σ}{{$\sigma$}}1
}
\hypersetup{
    colorlinks  = true,
    linkcolor   = linkblue,
    anchorcolor = linkblue,
    citecolor   = linkblue,
    filecolor   = linkblue,
    urlcolor    = linkblue
}

\crefname{ass}{Assumption}{assumptions}
\crefname{algorithm}{algorithm}{algorithms}
\Crefname{appendix}{Appendix}{Appendices}

\newcommand{\vega}{\mathcal{V}}

\newcommand{\fv}{\textsc{fast-vollib}~}
\newcommand{\bs}{\mathrm{BS}}
\newcommand{\bsm}{\mathrm{BSM}}
\newcommand{\iv}{\sigma_{\mathrm{imp}}}
\newcommand{\price}{P}
\newcommand{\mkt}{P^{\star}}

\title{PIVOT: Bridging Black-Scholes Implied-Volatility and Price Objectives via Differentiable Jäckel Operator}

\author{Raeid Saqur\textsuperscript{1,3}\thanks{\textit{Corresponding author}: \texttt{raeid.saqur@maths.ox.ac.uk}} \\ [0.3em]
  Yannick Limmer\textsuperscript{4} \quad
  Anastasis Kratsios\textsuperscript{2,3} \quad
  Blanka Horvath\textsuperscript{1} \quad 
  Hans Buehler\textsuperscript{1} \\
  \vspace{1pt}\\
  \textsuperscript{1}Mathematical Institute, University of Oxford \\
  \textsuperscript{2}McMaster University \\
  \textsuperscript{3}Vector Institute for AI \\
  \textsuperscript{4}DRW \\ [0.3em]
}

\begin{document}

\maketitle
\begin{abstract}
Modern option-learning systems operate in two coordinates: \emph{price space}, where markets quote and no-arbitrage constraints are most naturally enforced, and \emph{implied volatility (IV) space}, where volatility surfaces are smoothed, regularized, and evaluated.
The bottleneck is interface, not approximation: J\"{a}ckel's seminal
\textit{Let's Be Rational} (LBR) solver already inverts the
Black--Scholes price to machine precision efficiently.
What is missing is a differentiable layer that preserves LBR in the forward
pass and avoids backpropagating through its branch logic.
Such a layer must also confront the unavoidable singularity of the inverse
map in the low-vega regime, where
$\partial\sigma/\partial P = 1/\vega$ diverges as $\vega\to 0$.
We close this gap with \colorwblkb{PIVOT}, \textbf{P}rice--\textbf{I}mplied-\textbf{V}olatility
\textbf{O}bjective \textbf{T}ranslator. 
PIVOT keeps the LBR forward pass intact and supplies the backward pass by \colorwblk{implicit differentiation} through the smooth Black--Scholes/Black-76 price map, with an explicit gating contract: 
invalid domains return \texttt{NaN}, well-conditioned rows receive the exact
$1/\vega$ gradient, and low-vega rows are attenuated rather than silently
regularized.
On a single H100, a fused Triton kernel reaches $1.79{\times}10^{9}$ IV/s at
machine precision ($9.3{\times}10^{-14}$ max relative error vs.\ the
reference C solver); end-to-end label generation sustains 48.9M/s on
synthetic chains and 16.6M/s on SPX OptionMetrics. In
a HyperIV-style one-day reproduction on SPX, PIVOT-augmented objectives
Pareto-dominate the baselines, reducing held-out price MAE by up
to 43.4\% and the strongest three-seed gated objective improving price MAE by
38.8\% and IV MAE by 21.3\% jointly; cross-asset results on RUT, VIX, and NDX show
directional price-MAE gains of 40.1\%, 24.2\%, and 16.7\%, while an ungated
IV-roundtrip control collapses to a degenerate near-zero surface, confirming
the gate as a correctness contract rather than a tuning knob. 
\end{abstract}

\section{Introduction}\label{sec:intro}
Implied volatility is the inverse coordinate chart that maps an observed option
price into the volatility parameter of the Black--Scholes or Black-76 pricing
model~\citep{black1973pricing,black1976pricing}.  In practice, option datasets
are stored, smoothed, filtered, and learned in IV coordinates.  Recent
machine-learning work on volatility surfaces makes this dependence explicit:
deep smoothing methods construct surfaces from sparse quote
clouds~\citep{ackerer2020deep,wiedemann2025operator}, HyperIV uses a
hypernetwork to produce (approximately) arbitrage-free surfaces in real time~\citep{hyperiv_yang2025},
and generative approaches model distributions over (approximately) arbitrage-free volatility
surfaces~\citep{ning2023arbitrage,vuletic2024volgan}.  These models differ
architecturally, but they share a low-level dependency: millions of market
prices $\mkt$ must be converted to accurate pointwise IV labels $\iv$, and many
objectives compare models in both \textit{price} and \textit{volatility} coordinates. This creates a mismatch between the numerical layer and the ML layer.

J\"{a}ckel's ``Let's Be Rational'' (LBR) algorithm is the gold-standard IV
solver~\citep{jackel2013implementing, jackel2015let}, but the reference
implementation is scalar C code exposed in Python through CPU wrappers such as
\texttt{py\_vollib}~\citep{pyvollib}.  That design is excellent for single-option
analytics, but weak for modern (ML) workloads: historical option universes contain
millions of contracts $(S, K, \tau, r, q)$, neural models train on GPU tensors,
and differentiable objectives require gradients through the pointwise inverse
$\mkt \mapsto \iv$.
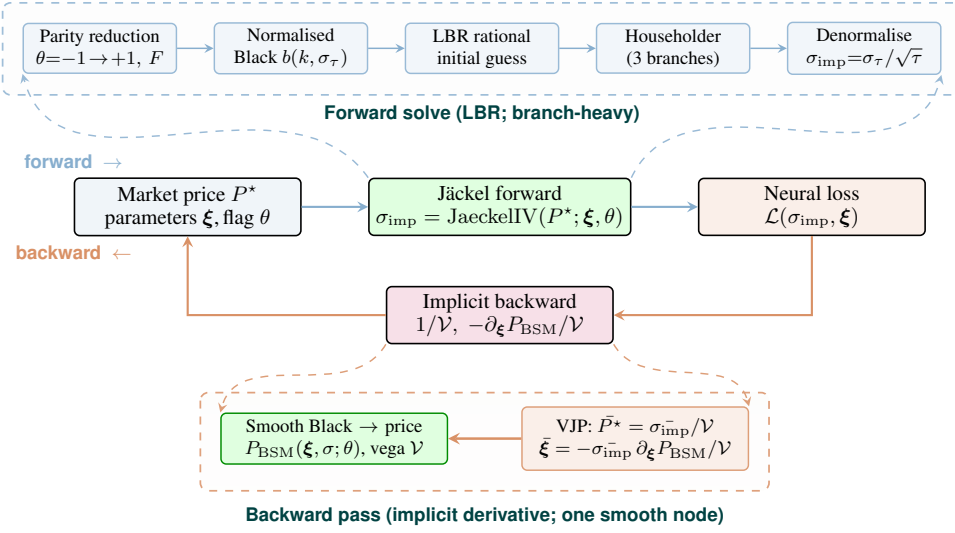
\begin{figure}[t]
    \centering

\tikzset{
    box/.style={
        draw,
        rounded corners=2pt,
        align=center,
        line width=0.55pt,
        inner sep=2.4pt,
        minimum height=0.65cm,
        minimum width=2.30cm,
        font=\footnotesize
    },
    wide/.style={
        draw, rounded corners=2pt, align=center,
        minimum height=0.78cm, minimum width=3.4cm,
        font=\small, line width=0.6pt, inner sep=3pt
    },
    detail/.style={
        draw, rounded corners=2pt, align=center,
        line width=0.55pt, inner sep=2.6pt,
        minimum height=0.68cm, minimum width=2.30cm,
        font=\footnotesize
    },
    core/.style={detail, minimum width=3.15cm},
    fwd/.style={box, fill=warmblue!7, draw=warmblue!55},
    backward/.style={box, fill=warmorange!12, draw=warmorange!75, minimum width=3.4cm, minimum height=0.95cm, },
    bypass/.style={box, fill=green!12, draw=green!55!black, minimum width=3.4cm},
    input/.style={core, fill=warmblue!7, draw=warmblue!45!black},
    fwdcore/.style={core, fill=green!10, draw=green!45!black},
    losscore/.style={core, fill=warmorange!10, draw=warmorange!70!black},
    bwdcore/.style={core, minimum width=3.35cm, fill=red!7, draw=red!45!black},
    smooth/.style={detail, minimum width=3.10cm, fill=green!10, draw=green!55!black},
    vjp/.style={detail, minimum width=3.45cm, fill=warmorange!10, draw=warmorange!75},
    block-title/.style={font=\sffamily\small\bfseries, color=black!75},
    title/.style={font=\sffamily\small\bfseries, color=black!74},
    note/.style={font=\scriptsize\itshape, color=black!58, align=center},
    arr/.style={->, line width=0.55pt, draw=black!75},
    fwdarr/.style={arr, draw=warmblue!65},
    bwdarr/.style={arr, draw=warmorange!85, line width=0.95pt},
    looparr/.style={->, thick},
    loopfwd/.style={->, thick, draw=warmblue!65},
    loopbwd/.style={->, thick, draw=warmorange!85},
    callfwd/.style={->, draw=warmblue!62, line width=0.70pt},
    callbwd/.style={->, draw=warmorange!78, line width=0.70pt},
    dottedlink/.style={draw=black!45, dotted, line width=0.90pt},
    zoom/.style={->, dashed, line width=0.5pt, draw=black!45},
    zoomfwd/.style={->, dashed, line width=0.55pt, draw=warmblue!55},
    zoombwd/.style={->, dashed, line width=0.55pt, draw=warmorange!75},
    nograd/.style={draw=red!58, dashed, line width=0.58pt},
    panel/.style={draw=black!25, dashed, rounded corners=3pt},
    panelbase/.style={
        dashed,
        rounded corners=3pt,
        line width=0.65pt,
        fill opacity=0.20,
    },
    panelfwd/.style={panelbase, draw=warmblue!52,   fill=warmblue!3!gray!5!white},
    panelbwd/.style={panelbase, draw=warmorange!72,
            top color=warmorange!4, bottom color=warmorange!1, fill=warmorange!2},
}
\pgfdeclarelayer{background}
\pgfsetlayers{background,main}

\begin{tikzpicture}[
    scale=0.88,
    transform shape,
    >=stealth,
    node distance=0.55cm,
    every node/.style={font=\small},
]
\node[fwd] (parity) at (0,0) {Parity reduction\\$\theta{=}{-}1\!\to\!{+}1,\,F$};
\node[fwd, right=of parity] (norm)   {Normalised\\Black $b(k,\sigma_\tau)$};
\node[fwd, right=of norm]   (init)   {LBR rational\\initial guess};
\node[fwd, right=of init]   (hh)     {Householder\\(3 branches)};
\node[fwd, right=of hh]     (denorm) {Denormalise\\$\iv{=}\sigma_\tau/\sqrt{\tau}$};

\draw[fwdarr] (parity) -- (norm);
\draw[fwdarr] (norm)   -- (init);
\draw[fwdarr] (init)   -- (hh);
\draw[fwdarr] (hh)     -- (denorm);

\begin{pgfonlayer}{background}
\draw[panelfwd]
    ($(parity.north west)+(-0.30,0.30)$)
    rectangle
    ($(denorm.south east)+(0.30,-0.30)$);
\end{pgfonlayer}

\node[block-title, align=center, font=\sffamily\bfseries\footnotesize, color=warmblack] (fwd-title)
    at ($(parity.south)!0.5!(denorm.south) + (0,-0.60)$)
    {Forward solve (LBR; branch-heavy)};

\node[wide, fill=warmblue!8, below=1.55cm of norm, xshift=-1.55cm]  (inputs)
    {Market price $\mkt$\\parameters $\bm{\xi}$,\,flag $\theta$};
\node[wide, fill=green!12,  right=1.0cm of inputs]               (forward)
    {J\"{a}ckel forward\\$\iv=\mathrm{JaeckelIV}(\mkt;\bm{\xi},\theta)$};
\node[wide, fill=warmorange!12, right=1.0cm of forward]              (lossnode)
    {Neural loss\\$\mathcal{L}(\iv,\bm{\xi})$};
\node[wide, fill=purple!12, below=0.75cm of forward]             (backnode)
    {Implicit backward\\$1/\vega,\;-\partial_{\bm{\xi}}\price_{\bsm}/\vega$};

\draw[loopfwd] (inputs)  -- (forward);
\draw[loopfwd] (forward) -- (lossnode);
\draw[loopbwd] (lossnode.south) |- (backnode.east);
\draw[loopbwd] (backnode.west)  -| (inputs.south);

\node[bypass, below=2.65cm of inputs, xshift=2.2cm] (bsbox) {Smooth Black\,$\,\to\,$\,price\\
    $\price_{\bsm}(\bm{\xi},\sigma;\theta)$, vega $\vega$};

\node[backward, right=1.1cm of bsbox] (vjp) {VJP: $\bar{\mkt}=\bar{\iv}/\vega$\\
    $\bar{\bm{\xi}}=-\bar{\iv}\,\partial_{\bm{\xi}}\price_{\bsm}/\vega$};

\draw[bwdarr] (vjp) -- (bsbox);

\begin{pgfonlayer}{background}
\draw[panelbwd]
    ($(bsbox.north west)+(-0.30,0.30)$)
    rectangle
    ($(vjp.south east)+(0.30,-0.30)$);
\end{pgfonlayer}

\node[block-title, align=center,
        font=\sffamily\bfseries\footnotesize,
        color=warmblack,
        ] (bwd-title)
    at ($(bsbox.south)!0.5!(vjp.south) + (0,-0.72)$)
    {Backward pass (implicit derivative; one smooth node)};

\draw[zoomfwd] (forward.north west) to[out=125, in=-75]  (parity.south west);
\draw[zoomfwd] (forward.north east) to[out=65,  in=-105] (denorm.south east);

\draw[zoombwd] (backnode.south west) to[out=-120, in=75]  (bsbox.north west);
\draw[zoombwd] (backnode.south east) to[out=-60,  in=105] (vjp.north east);

\node[font=\footnotesize\bfseries, color=warmblue!65,
      above=0.02cm of inputs.north west, anchor=south]
    {\textsf{forward}\;\;$\rightarrow$};
\node[font=\footnotesize\bfseries, color=warmorange!85,
      below=0.02cm of inputs.south west, anchor=north]
    {\textsf{backward}\;\;$\leftarrow$};

\end{tikzpicture}
    \caption{Autograd-native J\"{a}ckel IV. \colorwblkb{Top:} The compact
    autograd layer expands into a branch-heavy LBR forward pass that solves
    $\price_{\bsm}(\bm{\xi}, \iv; \theta) = \mkt$ for $\iv$ in normalized
    Black coordinates $(k, \sigma_\tau)$. \colorwblkb{Bottom:} The backward
    pass bypasses every solver branch and applies the implicit derivative
    $\partial\iv/\partial\mkt = 1/\vega$ through one smooth Black price node,
    delivering the implicit gradient.}
    \label{fig:autograd_pipeline}
\end{figure}
 The contribution of this paper is therefore not a learned replacement for
J\"{a}ckel's solver.  That would be a poor trade: the analytic--numerical solver
is already accurate, deterministic, and -- once GPU-vectorized -- fast.  The
contribution is instead to make exact pointwise IV inversion an autograd-native
systems primitive. The numerically delicate inverse remains in the forward
pass; the gradient is supplied analytically by the implicit function theorem.

At a high level, \colorwblkb{PIVOT} separates the two jobs that IV inversion normally
entangles.  The forward pass delegates each quote to the trusted J\"{a}ckel
solver, vectorized over modern tensor backends, and returns both the IV value
and explicit validity and conditioning information.  The backward pass does
not differentiate through the solver's rational approximants, branch masks, or
Householder updates. Instead, it treats the IV as the exact local inverse of a
smooth pricing map and supplies the corresponding analytic sensitivity
directly with appropriate boundary behaviour as implied volatility or variance approaches zero. 
This design keeps the mature numerical inverse intact while making the 
singular \textit{low-vega} regime visible to the training loop: 
well-conditioned quotes receive the exact inverse sensitivity, and ill-conditioned or invalid
quotes can be masked or attenuated before their IV-space gradients reach the optimizer.

\paragraph{Contributions.}
We propose \colorwblkb{PIVOT}
(\colorwblkb{P}rice--\colorwblkb{I}mplied-\colorwblkb{V}olatility
\colorwblkb{O}bjective \colorwblkb{T}ranslator), which makes exact IV inversion usable
as a differentiable bridge between price-space and IV-space objectives.  The
paper makes three linked contributions.

\textbf{(C1) J\"{a}ckel as a batched GPU primitive.}  We provide a
J\"{a}ckel-focused vectorized implementation of pointwise implied-volatility
inversion.  The forward pass retains LBR's normalized-Black construction
$b(k, \sigma_\tau)$ rather than replacing it by a neural surrogate, and
parallelizes the solver over heterogeneous option chains on NumPy, PyTorch,
JAX, and Triton-style execution paths.  The point is not to learn a new scalar
IV formula -- the analytic inverse is already exact in the bulk and the only
loss from a neural surrogate would be in the tails -- but to make the trusted
numerical inverse available at the scale and interface required by neural
pipelines.

\textbf{(C2) Autograd-native implicit backward.}  We turn IV inversion into an
autograd-native primitive.  The PyTorch and JAX operators call the vectorized
J\"{a}ckel solver in the forward pass and use implicit differentiation through
the smooth Black--Scholes / Black-76 price in the backward pass, applying the
analytic inverse-price sensitivity directly.  This gives the exact inverse-function
derivative without backpropagating through rational branch logic, masks, or
Householder iterations, and decouples gradient cost from the number of solver
iterations.

\textbf{(C3) Conditioning-aware low-vega gating mechanism.}  We specify and test
the gating mechanism needed to use IV inside losses.  In existing pipelines,
the standard remedy for the low-vega singularity is to \emph{hand-engineer it
out} of the training set -- pre-filter low-vega rows, drop short-dated deep
wings, generate offline IV labels only on the surviving subset -- a workflow
that is laborious, dataset-specific, and discards rows that price-space
supervision could otherwise use.  Our gating mechanism removes that step:
below-intrinsic and arbitrage-violating quotes are flagged as invalid and
return \texttt{NaN} from the inverse, the well-conditioned set
$\mathcal{G}_\tau = \{|\vega| > \tau\}$ receives the exact $1/\vega$ gradient,
and rows outside $\mathcal{G}_\tau$ are attenuated by either a hard mask
$\1\{|\vega| > \tau\}$ or the smooth gate
$w_\tau(\vega) = \vega^2/(\vega^2 + \tau^2)$ so that no NaN gradients enter
the optimizer.  Low-vega rows therefore stay in the batch and continue to
contribute through price-space residuals, where the price-to-volatility map
is well-conditioned regardless, while their singular IV-space channel is
suppressed automatically.  The training loop is freed from bespoke filtering
and offline label preparation; the gating mechanism handles the conditioning
in-line.

\paragraph{Empirical Results Summary.}  On a single H100 GPU (80\,GB) under
Linux/Ubuntu, the system produces 48.9\,M synthetic IV labels/s and 16.6\,M SPX
OptionMetrics labels/s on 3.03\,M real quotes, with self-consistency
round-trip residual Q99 $\le 1.8\!\times\!10^{-7}$ in price units across the
2018--2023 SPX panel.  In a controlled HyperIV-style one-day reproduction, a
PIVOT-augmented objective combining a price-MSE auxiliary with a $w_\tau$-gated
IV-roundtrip term Pareto-dominates the local vanilla baseline on SPX and
repeats the directional improvement on RUT, VIX, and NDX, with best-run price
MAE reductions of $43.4\%/40.1\%/24.2\%/16.7\%$ at IV MAE preserved within
$10\%$ of vanilla.  An ungated diagnostic, included as a negative control,
collapses to a degenerate near-zero surface ($96.5\%$ of test rows below
$|\vega|\le 10^{-14}$) and confirms that the gating mechanism is a
correctness contract rather than a tuning detail.

\section{Background and Preliminaries}\label{sec:background}

\paragraph{European options and put--call parity.} An option is a financial contract that gives its holder the right, but not the
obligation, to transact an underlying asset at a fixed strike price $K \in
(0, \infty)$, on a specified maturity date $T \in (0, \infty)$.  A call option
gives the right to buy the underlying, whereas a put option gives the right to
sell it.  Underlyings may include equity indices, single stocks, currencies,
commodities, futures, or other derivative contracts.  For simplicity of
exposition, we use ``option'' to mean a European call option unless stated
otherwise; European put prices can be transformed to (mathematically equivalent)
call prices through the well-known \emph{put--call parity}.

\paragraph{Setup and notation.}
For a given option contract, we use $S \in \mathbb{R}_+$ for the current (aka.\
spot) price of the underlying asset, $K \in (0, \infty)$ to denote its strike
price, $\tau$ to denote time-to-expire, with $\tau = T - T_0$, where $T \in
(T_0, \infty)$ is the expiry date in the future.  In practice, options are
traded for fixed expiries $T_1, ..., T_m$; and only a finite range of strikes
$K^i_1, ... K^i_{n_i}$ for each expiry $T_i$ is available.

The forward price of the underlying asset for expiry $T$ is denoted
$F_{T_0, T} \in \mathbb{R}_+$ and, under the standard no-arbitrage assumption,
is given by
\begin{equation}
    F_{T_0, T} = S \, e^{(r - q)\tau},
    \label{eq:app_forward_price}
\end{equation}
where $r \in \mathbb{R}$ is the risk-free interest rate and $q \in \mathbb{R}$
is the continuous dividend yield (or, more generally, the cost-of-carry) over
the period $[T_0, T]$.  We then define the \emph{log-forward-moneyness} as
$k = \log\!\left(\nicefrac{K}{F_{T_0, T}}\right)$, and write $\Phi(\cdot)$ for
the standard normal CDF.  Throughout this paper, we use $\sigma \in
\mathbb{R}_+$ to denote the (annualized) volatility of the underlying, and
$\sigma_\tau := \sigma \sqrt{\tau}$ to denote the corresponding \emph{total
volatility}, which captures the cumulative volatility over the life of the
option.

We work with European-style options on the parameter tuple
$\bm{\xi}=(S,K,\tau,r,q)$ -- \textit{spot, strike, time-to-expiry, risk-free rate,
dividend yield} -- and put--call flag $\theta\in\{+1,-1\}$.

\paragraph{Black--Scholes, Black-76.}
The seminal Black--Scholes~\citep{black1973pricing} provides a closed-form,
analytic formula for pricing European options -- characterized by the single
volatility $\sigma \in \mathbb{R}_+$ parameter under the so-called risk-free
$\mathbb{Q}$-measure and the assumptions of constant volatility, lognormal
returns, and continuous trading.  Using the notation introduced above, the
Black--Scholes price of a European call is
\begin{equation}
    \price_{\bsm}(S, K, \tau, r, q, \sigma)
    \;=\; S\,e^{-q\tau}\,\Phi(d_1) \;-\; K\,e^{-r\tau}\,\Phi(d_2),
    \label{eq:app_black_scholes}
\end{equation}
where $d_{1,2} = \bigl[\log(S/K) + (r - q)\tau \pm \tfrac{1}{2}\sigma_\tau^{2}\bigr] / \sigma_\tau$.
The corresponding put price follows from put--call parity.

In \emph{Black-76}, the same equation is written in forward coordinates: the
spot is replaced by the forward price $F$, the carry term cancels, and both
option legs are discounted by $e^{-r\tau}$~\citep{black1976pricing}.  This
distinction matters in data pipelines because market prices are discounted
cash prices, while many surface models operate on forwards, log
forward-moneyness ($k = \log(\nicefrac{K}{F})$), and implied volatilities.
Equivalently, in forward coordinates with log-forward-moneyness $k = \log(K /
F_{T_0,T})$, the Black-76 form gives
\begin{equation}
    \price_{\mathrm{B76}}(F_{T_0,T}, K, \tau, r, \sigma)
    \;=\; e^{-r\tau}\bigl[\, F_{T_0,T}\,\Phi(d_1) \;-\; K\,\Phi(d_2)\,\bigr],
    \qquad
    d_{1,2} \;=\; \frac{-k \pm \tfrac{1}{2}\sigma_\tau^{2}}{\sigma_\tau}.
    \label{eq:app_black76}
\end{equation}

\paragraph{Normalized Black function.}
For numerical stability and analytical tractability, it is often preferable to
work with a dimensionless form of the Black price rather than the raw cash
price.  Following~\citet{jackel2015let}, and restricting attention to European
calls, we work with the \emph{undiscounted} (forward) Black price $B :=
e^{r\tau}\,\price_{\mathrm{B76}}$ and define the \emph{normalized Black
function}
\begin{equation}
    b(k, \sigma_\tau) \;:=\; \frac{B(F, K, \tau, \sigma)}{\sqrt{F K}} \;=\;
    e^{-k/2}\,\Phi\!\left(\frac{-k}{\sigma_\tau} + \frac{\sigma_\tau}{2}\right)
    \;-\; e^{k/2}\,\Phi\!\left(\frac{-k}{\sigma_\tau} - \frac{\sigma_\tau}{2}\right),
    \label{eq:app_normalized_black}
\end{equation}
where $k = \log(K / F_{T_0,T})$ is the log-forward-moneyness and $\sigma_\tau =
\sigma\sqrt{\tau}$ is the total volatility.  The function $b$ depends only on
the two dimensionless quantities $(k, \sigma_\tau)$: the rates curve enters
only through the discount factor $e^{-r\tau}$ and the forward $F = S\,e^{(r-q)\tau}$,
both of which are absorbed before $b$ is evaluated.

This separation of concerns is precisely the appeal of the normalization.
Discounting and the forward are determined by the rates and dividend curves
--- a separate, linear calibration problem --- while $b$ isolates the
genuinely nonlinear, volatility-dependent part of the price.  The function
admits the tight bounds
\begin{equation}
    0 \;\le\; b(k, \sigma_\tau) \;\le\; b_{\max}(k) \;\le\; 1,
    \qquad b_{\max}(k) \coloneqq e^{-k/2}, \quad k \ge 0,
    \label{eq:app_b_bounds}
\end{equation}
for out-of-the-money calls,\footnote{For in-the-money calls ($k < 0$) the
put--call invariance maps the problem back to the OTM regime, so it suffices
to analyze $k \ge 0$.  The bounds in~\eqref{eq:app_b_bounds} underpin the
region-based asymptotic expansions and branching strategy used in our J\"ackel
solver following the \emph{Let's Be Rational} algorithm~\citep{jackel2015let}.}
which keeps $b$ numerically well-behaved across the full range of strikes and
maturities encountered in practice --- including deep wings and short expiries
where the cash price spans many orders of magnitude.  The input space
dimension is also reduced from the six raw Black-76 parameters $(F, K, \tau,
r, q, \sigma)$ to just two $(k, \sigma_\tau)$, which is particularly
attractive for learned surface models.  For these reasons, the normalized form
is the standard parameterization in modern implied-volatility
routines~\citep{jackel2015let} and in much of the recent ML-for-options
literature.

Following~\citet{jackel2015let}, the inverse problem reduces in forward coordinates to the dimensionless
\emph{normalized Black function} $b(k,\sigma_\tau)$, which is bounded in
$[0,b_{\max}(k)]$ on out-of-the-money strikes and isolates the
volatility-dependent part of the price from the rates and dividend curves.

\paragraph{Implied volatility, vega, and the inverse map.}
Define $\vega:=\partial\price_{\bsm}/\partial\sigma$.  Vega is strictly
positive on the admissible domain, so $\sigma\mapsto\price_{\bsm}(\bm{\xi},\sigma;\theta)$
is a strictly monotone bijection from $\mathbb{R}_+$ onto the no-arbitrage
price interval.  For an observed mid $\mkt$, the implied volatility $\iv$ is
the unique solution of
\begin{equation}
    \price_{\bsm}(\bm{\xi}, \iv; \theta) = \mkt,
    \label{eq:iv_inverse_problem}
\end{equation}
existence and uniqueness following directly from the bijection above
\citep{black1973pricing,black1976pricing}.  IV is the working coordinate of
volatility-surface modelling: it collapses the put--call distinction (a single
scalar field $\sigma(k,\tau)$ describes both wings), lies in a narrow band
$\sim[0.05,1.0]$ regardless of moneyness, and exhibits stable structural
features (smile, skew, term structure) that price space does not.  The
expanded justification with the put--call collapse, conditioning, and
structural-stability arguments is in
Appendix~\ref{app:sec:bg_iv_coordinate}.  By the implicit function theorem,
whenever $\vega\ne 0$ the inverse map's derivatives are
\begin{equation}
    \frac{\partial\iv}{\partial\mkt} \;=\; \frac{1}{\vega},
    \qquad
    \frac{\partial\iv}{\partial\bm{\xi}}
    \;=\; -\,\frac{\partial\price_{\bsm}/\partial\bm{\xi}}{\vega} ,
    \label{eq:implicit_iv}
\end{equation}
the closed-form gradient an autograd-native IV layer should expose without
backpropagating through any solver internals.  Equation~\eqref{eq:implicit_iv}
is the foundation of the autograd-native treatment we develop in
\S\ref{sec:jackel-iv}; it also exposes the only unavoidable failure mode of
an exact differentiable IV inverse: as $\vega\to 0$ the price-to-IV chart
becomes singular and $1/\vega$ diverges.  We treat this regime explicitly in
\S\ref{sec:lowvega_contract}.

\paragraph{Why J\"ackel as the forward solver.}
J\"ackel's ``Let's Be Rational'' (LBR) algorithm is the reference scalar IV
solver: rational seeds in normalized Black coordinates, region-specific
transformed objectives on the bounded domain
$0\le b\le b_{\max}(k)\le 1$, and at most two Householder iterations to
machine precision~\citep{jackel2015let,jackel2013implementing}.  Existing
open-source wrappers such as \texttt{py\_vollib}~\citep{pyvollib} expose
this as scalar CPU code lacking autograd, batched-tensor input, and explicit
validity masking, a workflow burden that even recent state-of-the-art neural
surface models inherit~\citep{wiedemann2025operator}; PIVOT closes that
interface gap.  The full positioning relative to alternative scalar
solvers is in Appendix~\ref{app:sec:bg_why_jackel}.

\section{Differentiable J\"{a}ckel Solver}\label{sec:jackel-iv}

Each input row is a market parameter tuple $\bm{\xi}=(S,K,\tau,r,q)$ with
put--call flag $\theta\in\{+1,-1\}$ and observed mid $\mkt$.  The forward
problem evaluates $\price_{\bsm}(\bm{\xi},\sigma;\theta)$; the inverse problem
solves $\price_{\bsm}(\bm{\xi},\sigma;\theta)=\mkt$ for $\iv$ in normalized
Black coordinates $(k,\sigma_\tau)$, and~\eqref{eq:implicit_iv} supplies the
implicit-function-theorem derivative through $\vega$.
Algorithm~\ref{alg:autograd_jackel} stitches these into a single differentiable
layer; Figure~\ref{fig:autograd_pipeline} depicts the forward/backward split
graphically.

\begin{algorithm}[t]
\caption{Autograd-native J\"{a}ckel IV (one batched call).}
\label{alg:autograd_jackel}
\small
\begin{algorithmic}[1]
\Require Market mid $\mkt$; parameters $\bm{\xi}=(S,K,\tau,r,q)$; put--call flag $\theta\in\{+1,-1\}$; gate threshold $\tau\!>\!0$ (overloaded; context-dependent)
\Ensure Implied vol $\iv$ and VJP $(\bar\mkt,\bar{\bm\xi})$ for any upstream cotangent $\bar\iv$
\Statex
\Statex \emph{\textbf{Forward}~~(vectorized LBR; gradients suppressed)}
\State Normalize: $F\leftarrow S\,e^{(r-q)\tau}$,~ $k\leftarrow\log(K/F)$,~ $\beta\leftarrow e^{r\tau}\mkt/\sqrt{FK}$.
\State Solve $\beta=b(k,\sigma_\tau)$ by LBR (rational seed + Householder) and set $\iv\leftarrow\sigma_\tau/\sqrt{\tau}$.
\State Save $(\bm{\xi},\theta,\iv)$; mask invalid rows (below-intrinsic, $\tau\!\le\!0$) to \texttt{NaN}.
\Statex
\Statex \emph{\textbf{Backward}~~(implicit VJP through the smooth Black price)}
\State Recompute $\price_{\bsm}(\bm{\xi},\iv;\theta)$ and $\vega$ via autograd; partition by $\mathcal{G}_\tau=\{|\vega|>\tau\}$.
\State On $\mathcal{G}_\tau$: \quad $\bar\mkt=\bar\iv/\vega$,\quad $\bar{\bm\xi}=-\bar\iv\,(\partial\price_{\bsm}/\partial\bm\xi)/\vega$.
\State Off $\mathcal{G}_\tau$: gate by $w_\tau(\vega)=\vega^2/(\vega^2+\tau^2)$ or return \texttt{NaN} (\S\ref{sec:lowvega_contract}).
\end{algorithmic}
\end{algorithm}

\subsection{Forward Pass: Vectorized J\"{a}ckel Inversion}
\label{sec:method_forward}

Recall from \S\ref{sec:background} that, after absorbing the discount factor
$e^{-r\tau}$ and the forward $F=S\,e^{(r-q)\tau}$, the normalized Black
function~\eqref{eq:app_normalized_black} reduces the inverse problem to two
dimensionless quantities $(k,\sigma_\tau)$. Concretely, given the market mid
$\mkt$, J\"{a}ckel's algorithm solves $b(k, \sigma_\tau) = e^{r\tau}\mkt/\sqrt{FK}$
for the total volatility $\sigma_\tau$, returning $\iv = \sigma_\tau/\sqrt{\tau}$.

The top panel of Figure~\ref{fig:autograd_pipeline} depicts the full forward
pipeline. We delegate the forward solve to the \fv\ Jäckel
backend~\citep{saqur2026fastvollibfastimpliedvolatility}, which executes the
standard LBR construction~\citep{jackel2015let} — put--call reduction via
parity, coordinate normalization to $(k, \sigma_\tau)$, numerically stable
evaluation of $b(k, \sigma_\tau)$ with the asymptotics
of~\eqref{eq:app_b_bounds}, region-specific rational seeding, and Householder
iterations on the transformed objective — over whole tensors, with NumPy,
PyTorch, JAX, and Triton execution paths. The five-stage construction and its
numerical analysis are documented in that reference; the contribution of the
present paper is the autograd-native wrapper specified in
\S\ref{sec:method_backward}--\ref{sec:lowvega_contract} below.

Two properties of the forward solver are load-bearing for the wrapper and
worth recording explicitly. First, every input is treated as a tensor: $S$,
$K$, $\tau$, $r$, $q$, and $\theta$ may all vary row-by-row, as they do in
batched chains and OptionMetrics panels with multiple expiries — so the
backward must vectorize the implicit identity~\eqref{eq:implicit_iv} elementwise
without any scalar fallback. Second, all internal computation is in float64;
the LBR boundary points and asymptotics in~\eqref{eq:app_b_bounds} make this
essential for the singular tails where the cash price spans many orders of
magnitude, and the gating mechanism of \S\ref{sec:lowvega_contract} inherits this
precision floor when it tests $|\mathcal{V}| > \tau$. Invalid inputs
(below-intrinsic, super-arbitrage, $\tau \le 0$) are masked to \texttt{NaN}
by the forward solver and propagate to the backward as the invalid branch of
the gating contract.

\subsection{Backward Pass: Implicit Autograd}
\label{sec:method_backward}

The PyTorch implementation uses \texttt{torch.autograd.Function}; the JAX
implementation uses \texttt{jax.custom\_vjp}.  Both backends share the same
idea, motivated by~\eqref{eq:implicit_iv}: the branch-heavy J\"{a}ckel solver
is opaque to autodiff, while the smooth discounted Black price
$\price_{\bsm}(\bm{\xi},\sigma;\theta)$ supplies the required vector--Jacobian
product. Concretely, given an upstream cotangent $\bar{\iv}$, the backward returns:
\begin{equation}
    \bar{\mkt} \;=\; \frac{\bar{\iv}}{\vega} ,
    \qquad
    \bar{\bm{\xi}} \;=\; -\,\frac{\bar{\iv}}{\vega}\,\frac{\partial \price_{\bsm}}{\partial \bm{\xi}}\bigg|_{\sigma=\iv} ,
    \label{eq:method_vjp}
\end{equation}
where $\partial\price_{\bsm}/\partial\bm{\xi}$ is obtained by ordinary autograd
through the analytic Black price helper, evaluated at the recovered $\iv$.  The
expression for $\bar{\mkt}$ is the implicit gradient of the inverse map; the
expression for $\bar{\bm{\xi}}$ chains it through the price's dependence on the
market parameters.  

Three properties follow.  \emph{Gradient cost is independent of solver complexity} — the number of Householder iterations, branch choices, and
rational-seed regions used in the forward pass do not appear
in~\eqref{eq:method_vjp} — so the backward decouples cleanly from any future
changes to the LBR construction.  \emph{Memory use is low} because no solver trace is
stored; the only quantities saved for the backward are $(\bm{\xi},\theta,\iv)$.
And \emph{the same forward kernel serves both offline labeling and differentiable
training}, eliminating the need for parallel CPU/GPU pipelines.  The bottom
panel of Figure~\ref{fig:autograd_pipeline} depicts the resulting backward
pass: a single smooth Black price evaluation feeds autograd, while the LBR
forward pass remains entirely opaque to differentiation.

\subsection{Low-Vega Gating Mechanism}\label{sec:lowvega_contract}

J\"{a}ckel's region splits make the \emph{forward inversion} numerically
reliable, including near price boundaries.  They do not remove the mathematical
ill-conditioning of the inverse map.  As established
in~\eqref{eq:implicit_iv}, when $\vega \to 0$ the derivative
$\partial\iv/\partial\mkt = 1/\vega$ diverges: price changes no longer
identify volatility changes.  No exact differentiable IV implementation can
make this gradient finite without changing the objective being differentiated.

Our differentiable J\"{a}ckel operator therefore exposes the conditioning
instead of hiding it.  For invalid quotes -- including below-intrinsic prices
and arbitrage-violating mids -- the forward returns \texttt{NaN}.  For
low-vega but otherwise valid quotes, the forward returns J\"{a}ckel's best
estimate, while the backward partitions the batch into the well-conditioned
set
\begin{equation}
    \mathcal{G}_{\tau} \;=\; \big\{i : |\vega_i| > \tau\big\}
    \label{eq:wellcond_set}
\end{equation}
and its complement.  Rows in $\mathcal{G}_\tau$ receive the exact implicit
gradient~\eqref{eq:method_vjp}.  Rows outside $\mathcal{G}_\tau$ return a zero
cotangent when the upstream gradient is already zero (as in NaN-aware
reductions), and return \texttt{NaN} when a loss explicitly asks for an
ill-conditioned gradient.  This behavior is intentional: a downstream training
loop that silently absorbs $1/\vega$ overflow would corrupt every Adam step
through the resulting NaN parameters and is not what an honest differentiable
inverse should provide.

This gating mechanism supports two principled training patterns.  First, use
price-space losses in low-vega regions, where prices remain the observable
stable coordinate and the price gradient is computed by ordinary autograd
through the smooth Black price.  Second, gate IV-space regularizers with a
hard mask $\1\{|\vega|>\tau\}$ or, preferred for differentiability, with the
smooth gate
\begin{equation}
    w_{\tau}(\vega) \;=\; \frac{\vega^{2}}{\vega^{2}+\tau^{2}} ,
    \label{eq:smooth_gate}
\end{equation}
which interpolates monotonically from $0$ on the singular tail to $1$ on the
well-conditioned bulk.  This is not an arbitrage-free surface claim and not a
learned solver approximation; it is the correct numerical treatment of a
singular coordinate chart.

Cleaned production training data (e.g.\ SPX 2023) sits comfortably in the
well-conditioned regime ($|\vega| \gtrsim 2{\times}10^{-2}$); the gate
matters when a non-trivial fraction of rows live in the singular tail.
The gate's effect on the vega distribution, on the smooth weight
$w_\tau$ for $\tau\in\{10^{-8},10^{-6},10^{-4}\}$, and on the
$(k,\mathrm{DTE})$ heat-map of the stress chain is illustrated in
Fig.~\ref{fig:lowvega_gate}; the empirical conditioning consequences --
ungated training collapses, gated training is stable -- are visible
directly in the SPX (Table~\ref{tab:hyperiv_spx_aux},
Fig.~\ref{fig:hyperiv_variant_surfaces}) and cross-asset
(Table~\ref{tab:cross_asset_summary}) results below.

\section{Experiments}\label{sec:experiments}

We empirically validate three research questions that the construction in
\S\ref{sec:jackel-iv} naturally raises.
\colorwblkb{(RQ1)} Is the differentiable J\"{a}ckel primitive (PIVOT)
numerically correct and fast enough for ML-scale option universes
$\{(\bm{\xi}_i,\theta_i,\mkt_i)\}_{i=1}^{N}$?
\colorwblkb{(RQ2)} Does the implicit backward pass realize the
inverse-function-theorem identity $\partial\iv/\partial\mkt = 1/\vega$
of~\eqref{eq:implicit_iv} on the well-conditioned set $\mathcal{G}_\tau$?
\colorwblkb{(RQ3)} Does the primitive change an actual surface-model training
objective when price-space and IV-space losses are optimized together? 
The scaling experiments (\S\ref{ssec:experiments-scaling}) answer RQ1 and RQ2
on synthetic batches with known volatilities $\sigma^\star$.  The real-world
experiments (\S\ref{sec:hyperiv_spx_aux}) answer RQ3 affirmatively:
plugging PIVOT into the SOTA IV-smoothing model HyperIV~\citep{hyperiv_yang2025}
on daily SPX option quotes 
yields a Pareto-dominating point in the price/IV plane
relative to the local vanilla baseline.  A parallel set of GNO~\citep{wiedemann2025operator}
ablations and additional cross-asset repeats are deferred to
Appendices~\ref{app:sec:cross_asset_hyperiv}
and~\ref{app:sec:additional_experiments_gno}.

\paragraph{Setup}
All experiments run on a single H100 (80\,GB) under Linux/Ubuntu.  Real-world
data come from IvyDB OptionMetrics via
WRDS~\citep{OptionMetrics_IvyDB_US_2025,WRDS_OptionMetrics}; the SPX 1-day
flagship spans 2013-01-02 to 2023-08-31 (12.48\,M train rows over 2{,}516
pre-2023 dates, 1.34\,M test rows over 167 dates in 2023; HyperIV filters:
OTM/ATM, mid $\ge\$0.10$, $\tau\le 2$\,y, 9-reference-set requirement) and
the cross-asset RUT/VIX/NDX repeats are reported in
Appendix~\ref{app:sec:cross_asset_hyperiv}, Table~\ref{tab:cross_asset_data}.
All rows carry $(\bm{\xi},\theta,\mkt)$ with $\bm{\xi}=(S,K,\tau,r,q)$.  The
underlying scalar J\"ackel routine is
\texttt{FAST-VOLLIB}~\citep{saqur2026fastvollibfastimpliedvolatility} with
\texttt{backend="jackel"}; PIVOT wraps it in the autograd-native interface
of \S\ref{sec:jackel-iv}.
\medskip

\begin{table}[t]
\centering
\small
\caption{\colorwblkb{Scaling experiments} on synthetic and SPX OptionMetrics
data.  The PyTorch row measures eager-mode autograd; the JAX row measures
compiled \texttt{custom\_vjp} execution.  The Triton-fused row reports
CUDA-event GPU compute time on the same canonical $10^5$-option grid
(wall-clock including host-to-device transfer is $2.1$\,ms; max relative
error vs.\ \texttt{py\_lets\_be\_rational} is $9.3{\times}10^{-14}$).
The CPU baseline is a single-core Python loop over the reference
\texttt{py\_lets\_be\_rational} solver and is reported as a sanity check on
the order-of-magnitude speedup attributable to GPU vectorization (with
vendor-IV agreement Q99 $1.7{\times}10^{-3}$ on the well-conditioned
subset).}
\label{tab:scaling_results}
\vspace{0.3em}
\begin{adjustbox}{max width=\textwidth}
    \begin{tabular}{llrl}
    \toprule
    Experiment & Backend / data & Scale & Result \\
    \midrule
    Forward IV & H100 synthetic & $10^5$ options & 1.39\,ms median, 72\,M IV/s \\
    Forward IV (Triton fused kernel) & H100 synthetic & $10^5$ options & 0.056\,ms GPU compute, $1.79{\times}10^{9}$ IV/s \\
    Label pipeline & H100 synthetic & 4.19\,M labels & 48.9\,M labels/s \\
    JAX forward + backward & H100 synthetic & $10^5$ options & 0.45\,ms post-JIT \\
    PyTorch forward + backward & H100 synthetic & $10^5$ options & 3.15\,s eager-mode \\
    Scalar CPU LBR baseline & \texttt{py\_lets\_be\_rational} & 4096 options & $1.60{\times}10^{5}$ IV/s ($\sim$195$\times$ slower) \\
    SPX 2023 & OptionMetrics & 3.03\,M quotes & 0.18\,s, 16.6\,M IV/s \\
    SPX 2020--2022 robustness & OptionMetrics & 4.2--5.1\,M quotes/yr & 11--13\,M IV/s, Q99 self-cons.\ $\le 9.1{\times}10^{-13}$ \\
    \bottomrule
    \end{tabular}
\end{adjustbox}
\vspace{0.5em}
\end{table}
\vspace{-1em}

\subsection{Scaling Experiments}
\label{ssec:experiments-scaling}

We begin with synthetic Black--Scholes / Black-76 batches in which the true
volatility $\sigma^\star$ is known by construction.  This isolates the numerical
primitive from data-cleaning and surface-model effects: any error is
attributable to the IV operator rather than to quote conventions, missing rate
curves, or model capacity.

\paragraph{Throughput and forward accuracy.}
Table~\ref{tab:scaling_results} summarizes the speed, throughput, and forward
accuracy of the J\"{a}ckel forward solver underneath the PIVOT operator.  On a
synthetic $10^5$-option batch, the vectorized J\"{a}ckel forward pass takes
$1.39$\,ms, corresponding to roughly $72$\,M IV inversions/s.  When the
entire J\"{a}ckel pipeline (preprocessing, boundary computation, Hermite
initial guess, three Householder(3) iterations, and postprocessing) is fused
into a single Triton kernel that keeps every intermediate in registers and
touches HBM only once per element, GPU compute time on the same canonical
grid drops to $0.056$\,ms ($1.79{\times}10^{9}$ IV/s) at machine precision
(max relative error $9.3{\times}10^{-14}$ vs.\ the reference C solver).
In a streaming HyperIV-style label-generation loop over $64$ batches of
$65{,}536$ options, \fv produces $4.19$\,M labels in $0.086$\,s, or
$48.9$\,M labels/s.  This is roughly $195\times$ faster than a single-core
scalar \texttt{py\_lets\_be\_rational} loop on the same problem (or
$\sim\!1905\times$ under the fused Triton kernel above), and is sufficient
to move IV inversion from an offline preprocessing step into the inner
loop of a GPU training pipeline.

\paragraph{Gradient correctness.}
We verify equation~\eqref{eq:implicit_iv} by generating synthetic BSM prices
$\mkt_i = \price_{\bsm}(\bm{\xi}_i, \sigma^\star_i; \theta_i)$, recovering $\sigma_{\mathrm{imp},i}$
through PIVOT, and comparing autograd gradients to the analytic identity
$\partial\iv/\partial\mkt = 1/\vega$.  The well-conditioned subset uses
$|\vega| \ge 10^{-6}$ as in~\eqref{eq:wellcond_set}.  On $95.2\%$ of the
synthetic batch, the price-gradient absolute error has Q50 $6.9{\times}10^{-18}$
and Q99 $1.77{\times}10^{-2}$.  The raw unmasked maximum is dominated by the
low-vega tail, where $1/\vega$ overflows to floating-point infinity; this is
the expected conditioning singularity from \S\ref{sec:lowvega_contract}, not a
solver inaccuracy.  The JAX and PyTorch implementations agree on mask rates
and gradient quantiles up to float64 noise.  The full quantile spectrum on the
raw, $|\vega|{>}10^{-14}$, and $\mathcal{G}_{10^{-6}}$ subsets is reported in
Appendix~\ref{app:sec:gradient_correctness},
Fig.~\ref{fig:gradient_quantiles}.

\subsection{IV Surface Generation or Smoothing with the Differentiable J\"{a}ckel Solver}
\label{sec:hyperiv_spx_aux}

\begin{table}[t]
\centering
\small
\caption{SPX 1-day IV-smoothing with HyperIV~\citep{hyperiv_yang2025} and PIVOT. Lower MAEs (IV, Price) are better
($\downarrow$).  $\Delta$ columns are signed relative changes (in \%) vs.
the vanilla HyperIV reproduction (V0); negative $\Delta$ indicates
lower error (improvement), positive $\Delta$ indicates regression.  \textbf{Bold} marks
the \textbf{best} value in each column among the PIVOT-augmented rows.  
The unsafe row is a diagnostic negative control showing the failure mode of composing the inverse IV map
without the low-vega gating \S\ref{sec:lowvega_contract}.}
\label{tab:hyperiv_spx_aux}
\vspace{0.3em}
\resizebox{\linewidth}{!}{\begin{tabular}{llrrrr}
\toprule
Objective & Runs & IV MAE ($\downarrow$) & Price MAE ($\downarrow$) & $\Delta$ IV (\%) & $\Delta$ Price (\%) \\
\midrule
HyperIV-vanilla (V0) & mean, $n=3$ & $0.017\!\pm\!0.001$ & $3.822\!\pm\!0.422$ & -- & -- \\
\midrule
Price aux.\ (PIVOT), $\lambda_{p}{=}0.1$ & mean, $n=3$ & $0.015\!\pm\!0.002$ & $2.512\!\pm\!0.277$ & $-9.3$ & $-34.3$ \\
Price aux.\ (PIVOT), $\lambda_{p}{=}0.3$ & best & $0.014$ & $\mathbf{1.983}$ & $-17.9$ & $\mathbf{-48.1}$ \\
Price + gated rt.\ (PIVOT), $\tau{=}10^{-6}$ & mean, $n=3$ & $\mathbf{0.013\!\pm\!0.001}$ & $2.340\!\pm\!0.172$ & $\mathbf{-21.3}$ & $-38.8$ \\
\midrule
Unsafe ungated rt.\ & diagnostic & 0.235 & 36.937 & $+1308$ & $+866$ \\
\bottomrule
\end{tabular}}
\end{table}

The previous experiments establish that the differentiable IV operator is
accurate and fast.  We next ask whether it changes an actual neural
volatility-surface training objective.  We reproduce the SPX 1-day setting of
HyperIV~\citep{hyperiv_yang2025} using the released architecture pattern: a
9-contract reference set per date feeds a hypernetwork that emits the parameters
of a compact surface network $\hat{\sigma}(k,\tau)=h_{\omega(Z)}(k,\tau)$
operating directly on the log-forward-moneyness coordinate
$k=\log(K/F_{T_0,T})$ defined in \S\ref{sec:background}.  The vanilla objective
is the HyperIV-style IV MSE plus the static-arbitrage auxiliary losses.  We
compare two PIVOT-augmented objectives:
\begin{align}
    \mathcal{L}_{\mathrm{price}}(\hat{\sigma})
      &= \frac{1}{|\mathcal{B}|}\sum_{i\in\mathcal{B}}
         \left(\frac{\price_{\bs}(\bm{\xi}_i,\hat{\sigma}_i;\theta_i)-\mkt_i}
         {\frac{1}{|\mathcal{B}|}\sum_{j\in\mathcal{B}}|\mkt_j|}\right)^{2} ,
         \label{eq:hyperiv_aux_price}\\[2pt]
    \mathcal{L}_{\mathrm{rt}}(\hat{\sigma})
      &= \frac{1}{|\mathcal{B}|}\sum_{i\in\mathcal{B}}
         w_{\tau}(\hat{\vega}_i)\,
         \bigl(\mathrm{JaeckelIV}(\tilde{P}_i)-\sigma^{\star}_i\bigr)^{2} ,
    \label{eq:hyperiv_aux_rt}
\end{align}

\begin{wrapfigure}[15]{r}{0.45\linewidth}
\centering
\vspace{-10pt}
\includegraphics[width=\linewidth]{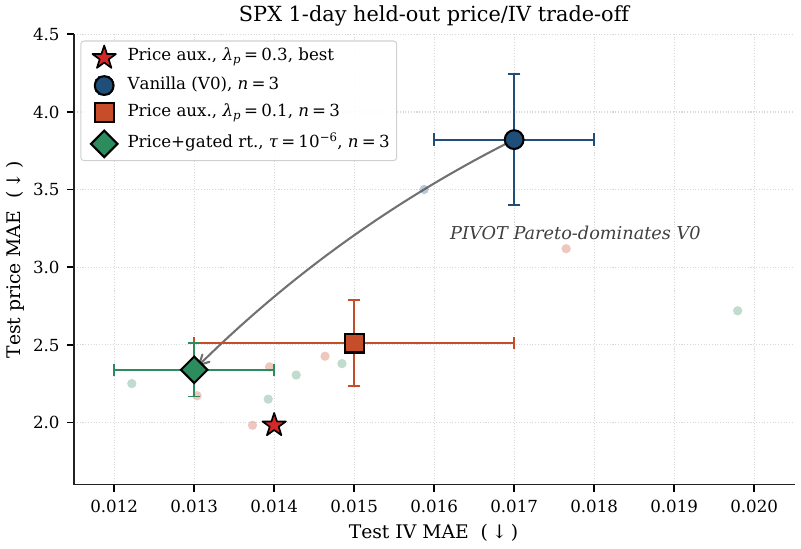}
\caption{\small Held-out price/IV trade-off for SPX 1-day HyperIV with PIVOT.
Both PIVOT-augmented cells lie below \emph{and} left
of the vanilla baseline, Pareto-dominating it; the gated
price$+$roundtrip cell at $\tau{=}10^{-6}$ is the lower-left frontier.}
\label{fig:hyperiv_spx_pareto}
\vspace{-1pt}
\end{wrapfigure}

where $\hat{\vega}_i=\partial\price_{\bs}/\partial\sigma\bigl|_{\sigma=\hat{\sigma}_i}$,
$w_\tau$ is the smooth gate~\eqref{eq:smooth_gate}, $\sigma^{\star}_i =
\mathrm{JaeckelIV}(\mkt_i)$ is the market IV recovered from the observed
mid (so~\eqref{eq:hyperiv_aux_rt} compares
$\mathrm{JaeckelIV}(\tilde{P}_i)$ to $\mathrm{JaeckelIV}(\mkt_i)$ in IV
space), and $\tilde{P}_i$ equals the model price
$\price_{\bs}(\bm{\xi}_i,\hat{\sigma}_i;\theta_i)$ on well-conditioned
rows $i\in\mathcal{G}_\tau$ and a sentinel market-consistent price on rows
with $|\hat{\vega}_i|\le\tau$.  The price loss is always computed from the
original autograd-tracked model price; the sentinel is used only to
protect the singular inverse-IV channel, so low-vega rows continue to
contribute through price space.  All variants below use the same SPX 1-day
dataset described in the setup, the same put-call-parity-derived discount-rate
convention, 100 epochs, Adam with learning rate $10^{-3}$, batch size 128 dates,
and 1024 sampled contracts per date.

Both PIVOT-augmented objectives Pareto-dominate the local vanilla baseline on
the 3-seed mean.  At $\lambda_{p}=0.1$ with the price auxiliary alone, per-seed
price-MAE improvements range from $32.0\%$ to $44.0\%$ versus the vanilla
3-seed mean of $3.822$, and the mean IV MAE improves from $0.017$ to $0.015$.
At $\lambda_{p}=\lambda_{\mathrm{rt}}=0.1$ with the gated IV-roundtrip term and
$\tau=10^{-6}$, both axes improve over the price-only auxiliary, giving the
strongest 3-seed cell ($-38.8\%$ price MAE, $-21.3\%$ IV MAE versus vanilla).
The best single augmented run uses $\lambda_{p}=0.3$ with seed 1.  
It is evidence that the differentiable PIVOT bridge changes the training objective in the intended
direction under a controlled reproduction.

\begin{figure}[htbp]
\centering
\includegraphics[width=0.98\linewidth]{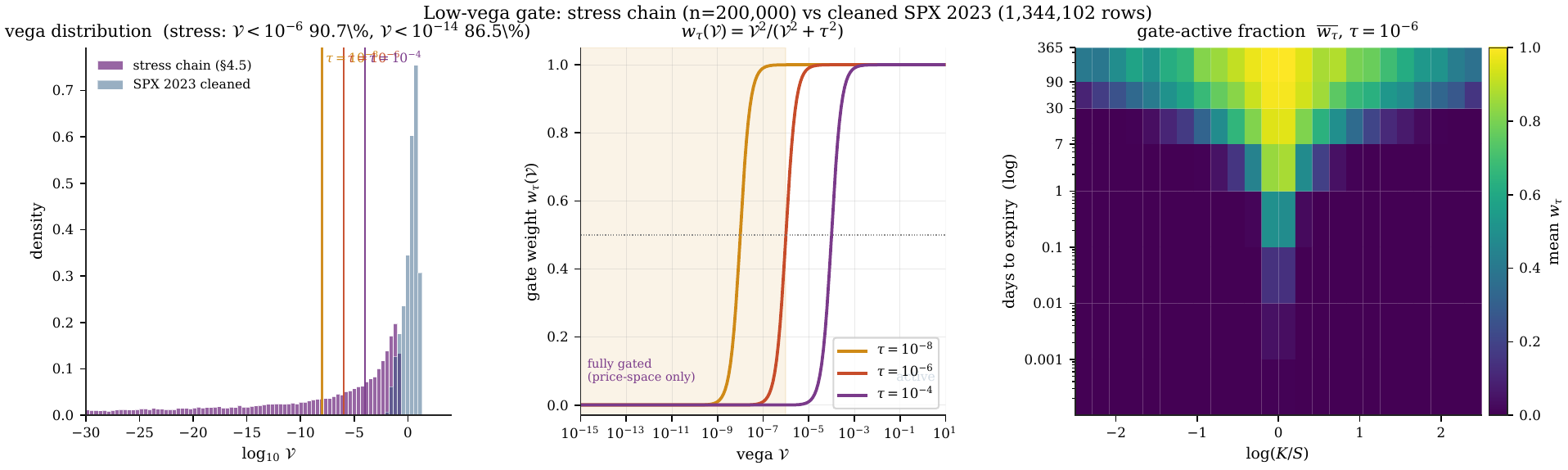}
\vspace{-2pt}
\caption{\small Low-vega gating mechanism in action.
\textbf{Left:} log-scale vega histogram on the synthetic stress chain
($|\vega|<10^{-6}$ on $90.7\%$ of rows; $|\vega|<10^{-14}$ on $86.5\%$)
versus the cleaned SPX 2023 training panel (no row below $|\vega|=10^{-6}$).
\textbf{Middle:} smooth gate weight $w_\tau(\vega) = \vega^2/(\vega^2+\tau^2)$
of~\eqref{eq:smooth_gate} for $\tau\in\{10^{-8},10^{-6},10^{-4}\}$.
\textbf{Right:} per-cell mean gate weight $\overline{w_\tau}$ at
$\tau{=}10^{-6}$ on the stress chain over a $(k,\mathrm{DTE})$ grid; deep
wings at very short maturities are fully gated, and IV-space supervision
concentrates in the well-conditioned interior.}
\label{fig:lowvega_gate}
\vspace{-6pt}
\end{figure}

\paragraph{Why the gate matters.}
The roundtrip term~\eqref{eq:hyperiv_aux_rt} is not new supervision but a
differentiable path through the exact inverse price map; the gate of
\S\ref{sec:lowvega_contract} simply prevents that path from injecting
$1/\vega$ poison into the optimizer.  Figure~\ref{fig:lowvega_gate}
anchors this in distributions: cleaned SPX 2023 rows sit comfortably in
the well-conditioned regime ($|\vega|\!\gtrsim\!2{\times}10^{-2}$),
while a stress chain places $86.5\%$ of its mass below $|\vega|\!=\!10^{-14}$;
the smooth weight $w_\tau$ attenuates that tail continuously rather than
hard-clipping it.  With the sentinel-protected smooth gate, every variant
trains with zero NaN-gradient steps; without it, the diagnostic accumulates
999 NaN steps in 50 epochs, test IV MAE $0.235$, price MAE $36.94$, and
collapses to a degenerate near-zero surface ($96.5\%$ of test rows below
$|\hat\vega|\le 10^{-14}$); per-step traces are in
Appendix~\ref{app:sec:nan_grad_trace}.
Figure~\ref{fig:hyperiv_variant_surfaces} contrasts the predicted IV
surfaces of the three trainable variants -- which track the OptionMetrics
market smile within $\sim\!2$ vol points -- against the ungated diagnostic
on a representative SPX 2023 test date.

\begin{figure}[t]
\centering
\includegraphics[width=0.92\linewidth]{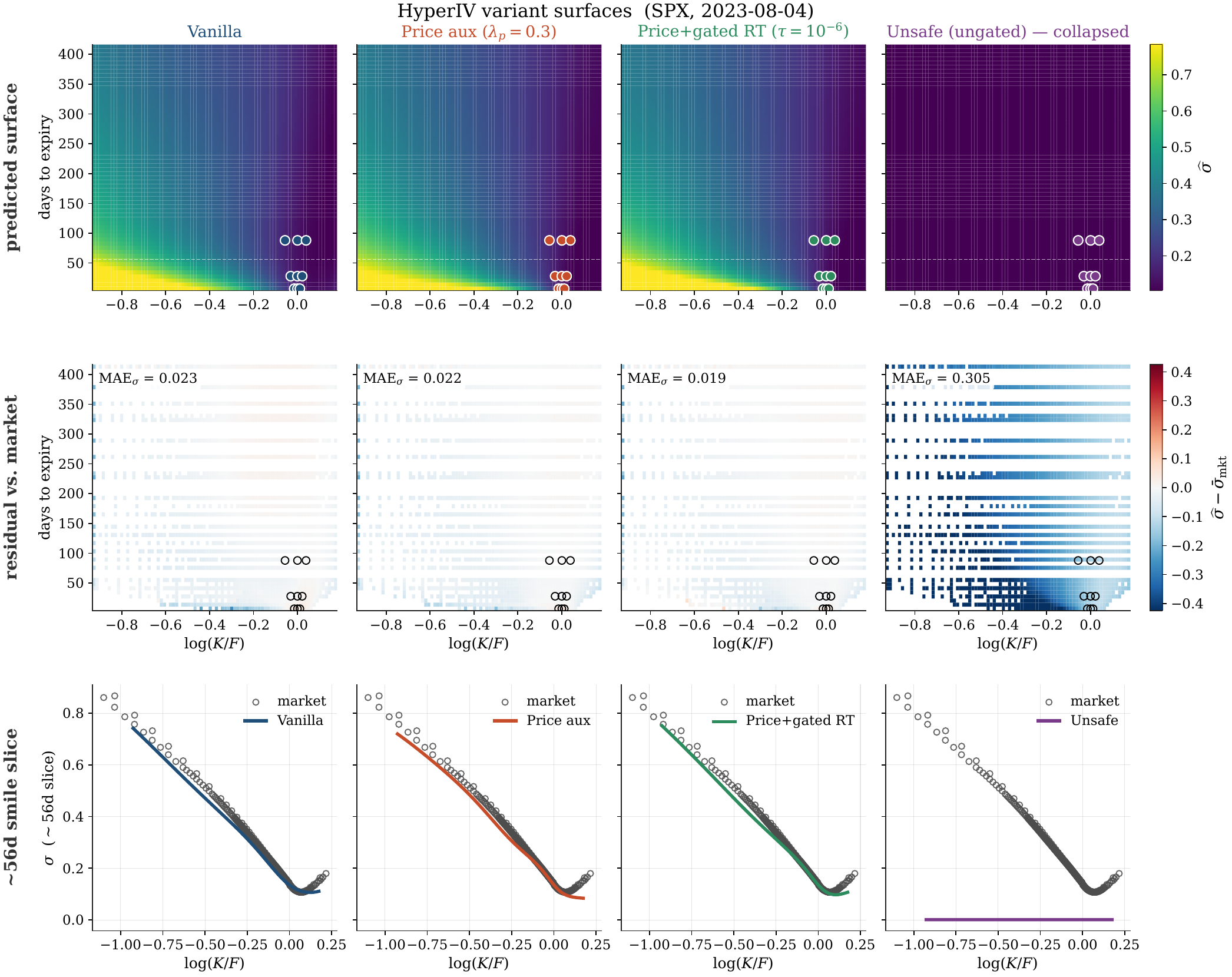}
\vspace{-4pt}
\caption{\small HyperIV-with-PIVOT surfaces on a representative SPX 2023 test date.
\textbf{Top row:} predicted $\hat{\sigma}(k,\tau)$ on the
log-forward-moneyness $\times$ days-to-expiry grid; reference contracts
($\circ$) overlaid in white.  The price-aux and gated price$+$rt PIVOT
variants reproduce vanilla's surface with small, structured residuals; the
ungated diagnostic collapses to a near-zero surface.
\textbf{Middle row:} residual versus the cell-averaged market label; the
colormap is centred at zero.
\textbf{Bottom row:} $\sim 56$-day smile slice with market scatter.  The
collapsed unsafe surface trivially fits an IV-only loss on its own
pathological labels but does not reprice any quote.}
\label{fig:hyperiv_variant_surfaces}
\vspace{-6pt}
\end{figure}

\paragraph{Discussion.}
This experiment is a stronger ML-facing demonstration than a pointwise
roundtrip test: the differentiable J\"{a}ckel layer (PIVOT) is part of a
surface-model training objective and changes the out-of-sample price
error~(\colorwblkb{RQ3}).
It is also intentionally narrow.  We do not claim arbitrage-free IV surface
generation -- that property is inherited from the HyperIV baseline, which
already enforces it.
The supported claim is that, at matched conventions, exact
differentiable IV makes it possible to train a HyperIV-style smoother with
price-space consistency~\eqref{eq:hyperiv_aux_price} and a gated
IV-roundtrip channel~\eqref{eq:hyperiv_aux_rt}, improving price MAE without
sacrificing IV accuracy.

\paragraph{Cross-asset summary.}
The PIVOT-augmented construction repeats the SPX result on RUT, VIX, and NDX
under matched local conventions (Table~\ref{tab:cross_asset_summary};
full per-asset tables and Pareto views in
Appendix~\ref{app:sec:cross_asset_hyperiv}).  Best-run price-MAE reductions
range from $16.7$--$43.4\%$ at IV MAE preserved within $10\%$ of vanilla;
the ungated control collapses on every asset (NaN-gradient steps in the
hundreds-to-thousands; price MAE inflated by $9$--$30\times$).  An analogous
PIVOT-loss ablation on a graph neural operator (GNO)
smoother~\citep{wiedemann2025operator} reproduces the gated-vs-collapse
signal on SPX/RUT/VIX but the price-MAE effect ($1.7$--$3.7\%$) falls below
our pre-registered $5\%$ practical threshold; we therefore defer GNO to
Appendix~\ref{app:sec:additional_experiments_gno} and keep HyperIV as the
headline.

\begin{table}[t]
\centering
\footnotesize
\caption{Cross-asset summary: best PIVOT-augmented HyperIV-style runs
versus the local vanilla baseline for the same asset (left), and the
ungated diagnostic on the same asset (right).  All four augmented best
runs preserve IV MAE within the $10\%$ tolerance; the ungated diagnostic
collapses on every asset, mirroring SPX.}
\label{tab:cross_asset_summary}
\begin{tabular}{lcc|cc}
\toprule
& \multicolumn{2}{c|}{Best PIVOT vs.\ HyperIV (vanilla)} & \multicolumn{2}{c}{Ungated diagnostic} \\
Asset & $\Delta$ Price MAE & $\Delta$ IV MAE & NaN-grad steps & $\Delta$ Price MAE \\
\midrule
SPX & $\mathbf{-43.4\%}$ & $-17.9\%$ & 999 & $+866\%$ \\
RUT & $-40.1\%$ & $-16.6\%$ & 799 & $+1054\%$ \\
VIX & $-24.2\%$ & $-17.0\%$ & 1{,}000 & $+923\%$ \\
NDX & $-16.7\%$ & $+1.5\%$ & 999 & $+2949\%$ \\
\bottomrule
\end{tabular}
\end{table}

\section{Related Work}\label{sec:related}

\paragraph{Implied-volatility inversion.}
Classical methods solve~\eqref{eq:iv_inverse_problem} by Newton, Halley, or
bracketed root-finding with problem-specific
seeds~\citep{manaster1982iterative,brenner1988new,corrado1996efficient}.
J\"ackel's LBR algorithm uses rational seeds in normalized Black coordinates
on the bounded domain $0\le b\le b_{\max}(k)\le 1$ and Householder iterations
on region-transformed objectives~\citep{jackel2015let}.

\paragraph{Neural volatility surfaces.}
Deep smoothing~\citep{ackerer2020deep}, operator deep
smoothing~\citep{wiedemann2025operator}, HyperIV~\citep{hyperiv_yang2025},
and arbitrage-free generative
models~\citep{ning2023arbitrage,vuletic2024volgan} learn maps from sparse
quotes, histories, or latent factors to surfaces $\hat\sigma(k,\tau)$.  These
generators sit \emph{above} the differentiable inverse and -- once one is
available -- can incorporate price-space~\eqref{eq:hyperiv_aux_price} and
gated IV-roundtrip~\eqref{eq:hyperiv_aux_rt} objectives.  This paper
contributes the missing pointwise differentiable layer beneath them.

\paragraph{Price/IV objectives and differentiable finance.}
Calibration writes losses in price space; neural IV-smoothing optimizes in
IV space (\S\ref{sec:background}).  PIVOT makes the pointwise bridge
between the two exact and differentiable subject to the
\S\ref{sec:lowvega_contract} gate.  Differentiable pricing
libraries~\citep{buehler2019deep,pfhedge} supply forward gradients only;
extended positioning is in Appendix~\ref{app:sec:bg_diff_finance}.
\section{Conclusion}\label{sec:conclusion}

\textbf{PIVOT} turns J\"ackel's LBR solver into an autograd-native,
GPU-batched primitive: forward preserves the solver verbatim; backward
applies the implicit derivative~\eqref{eq:implicit_iv} through one smooth
Black price node; the gating mechanism (\S\ref{sec:lowvega_contract})
makes the chart's singularity a documented part of the layer's contract.
Plugged into a HyperIV-style smoother it yields a 3-seed Pareto
improvement on SPX (Pr.\ MAE $-38.8\%$, IV MAE $-21.3\%$), repeats on
RUT/VIX/NDX (Table~\ref{tab:cross_asset_summary}), and a parallel
gating-vs-collapse signal on a graph-operator
smoother~\citep{wiedemann2025operator}; limitations and outlook are in
Appendix~\ref{app:sec:bg_limitations_extended}.

\section*{Broader Impact}\label{impact}
This work targets a methodological bottleneck in machine-learning systems for
option pricing.  Modern pipelines operate in two coordinates: \emph{price
space}, where markets quote and no-arbitrage constraints are most naturally
checked, and \emph{implied-volatility (IV) space}, where surfaces are smoothed,
regularized, compared, and reported.  To our knowledge, PIVOT is the first
work to make these two coordinates jointly available inside a single
end-to-end differentiable training objective while preserving the trusted
J\"{a}ckel/LBR forward solver.  Its broader positive impact is therefore not a
new trading rule or a learned replacement for a numerical method, but an
operator-level interface: researchers and practitioners can train, audit, and
stress-test volatility-surface models in the coordinate where prices are
observed and the coordinate where surfaces are interpreted.

This interface can improve reproducibility and model governance.  Existing
workflows often precompute IV labels offline, discard low-vega quotes, or use
separate price-space and IV-space objectives whose numerical assumptions are
hidden in preprocessing.  PIVOT instead makes the price--IV boundary explicit:
invalid rows return \texttt{NaN}, well-conditioned rows receive the exact
implicit $1/\vega$ gradient, and low-vega rows are gated so that their
ill-conditioned IV channel does not silently corrupt optimization.  This can
support clearer reporting of data validity, conditioning, and loss
construction in volatility-surface learning, and can help educational,
research, calibration, and risk-modeling workflows distinguish numerical
conditioning from economic signal.

The same capability also has foreseeable risks.  Faster and more reliable
joint price/IV feedback may be incorporated into market-making, automated
trading, or portfolio-risk systems, potentially reinforcing advantages held by
well-resourced institutions.  A differentiable numerical layer also does not
remove model risk: downstream systems may still be trained on stale, illiquid,
biased, or incorrectly cleaned option data, and a smooth jointly consistent
surface may be overinterpreted as evidence of true market structure.  The
low-vega regime is especially delicate because small price changes can imply
large IV changes; PIVOT exposes this singularity through its gating contract,
but does not decide the appropriate business, regulatory, or risk-control
response.

The paper does not introduce a trading strategy, trading recommendation, or
new source of market data.  Its intended use is as a transparent differentiable
primitive inside larger pricing, smoothing, calibration, and research
pipelines.  We therefore view the main safeguards as careful data governance,
explicit reporting of invalid and low-vega rows, no-arbitrage checks at the
surface-model level, and human review before deployment in financial
decision-making systems.

\section*{Acknowledgements}

Raeid Saqur are supported by the UK Research and Innovation (UKRI) through the Engineering and Physical Sciences Research Council (EPSRC) via Programme Grant [Grant No.\ UKRI1010: High order mathematical and computational infrastructure for streamed data that enhance contemporary generative and large language models] and the Vector Research Grant by the Vector Institute for AI, Canada.

\clearpage
\appendixpage
\DoToC 
\appendix

\section{Additional Background and Discussions}
\label{app:sec:additional_background}

This appendix provides the full preliminaries and derivations summarised in
\S\ref{sec:background}: European-call setup, notation, the Black--Scholes /
Black-76 closed-form prices, the normalized Black function $b(k,\sigma_\tau)$,
vega and its closed-form expression, the IV inverse problem, the three reasons
IV is the working coordinate of volatility-surface modelling, and the
positioning of J\"{a}ckel's LBR algorithm.

\subsection{Vega and the option Greeks}
\label{app:sec:bg_vega}

The sensitivities of the option price to its inputs --- collectively known as
the \emph{Greeks} --- play a central role both in risk management and, as we
will see, in the inversion problem itself.  The most important sensitivity for
our purposes is \emph{vega}, the derivative of the option price with respect
to volatility:
\begin{equation}
    \vega \;:=\; \frac{\partial \price_{\bsm}}{\partial \sigma}.
    \label{eq:app_vega_def}
\end{equation}
Under the Black--Scholes model, vega admits the closed form\footnote{More
precisely: $\vega = S\, e^{-q\tau}\, \varphi(d_1)\, \sqrt{\tau} =
S\, e^{-q\tau}\, \varphi(d_1)\, \frac{\partial \sigma_\tau}{\partial \sigma},$}
\begin{equation}
    \vega \;=\; S\, e^{-q\tau}\, \varphi(d_1)\, \sqrt{\tau} \;=\;
    F\, e^{-r\tau}\, \varphi(d_1)\, \sqrt{\tau},
    \label{eq:app_vega_closed_form}
\end{equation}
where $\varphi$ is the standard normal density and $d_1$ is as
in~\eqref{eq:app_black_scholes}.  Two properties of vega will matter
throughout this paper.  First, $\vega > 0$ strictly, for all admissible
$(F, K, \tau, \sigma)$ with $\tau > 0$ and $\sigma > 0$ --- the option price
is a strictly increasing function of volatility.  Second, vega vanishes in
the limits $\sigma \to 0$, $\sigma \to \infty$, and $\tau \to 0$ for $K \neq
F$, which makes the price-to-volatility map ill-conditioned in those regions
and is a recurring source of numerical difficulty for implied-volatility
solvers.  Other Greeks --- delta ($\Delta = \partial \price / \partial S$),
gamma ($\Gamma = \partial^2 \price / \partial S^2$), theta ($\Theta = \partial
\price / \partial \tau$), and rho ($\rho = \partial \price / \partial r$) ---
appear only incidentally in this work; we refer the reader to standard
references~\citep{hull2016options} for their definitions.

\subsection{Implied volatility as the working coordinate}
\label{app:sec:bg_iv_coordinate}

Let $\price_{\bsm}(\bm{\xi}, \sigma; \theta)$ denote a discounted European
option price under Black--Scholes or Black-76, where $\bm{\xi} := (S, K, \tau,
r, q)$ collects the market parameters and $\theta \in \{+1, -1\}$ indicates a
call ($+1$) or put ($-1$).  For an observed market mid-price $\mkt$, the
implied volatility $\iv$ is the unique solution of
\begin{equation}
    \price_{\bsm}(\bm{\xi}, \iv; \theta) = \mkt.
    \label{eq:app_iv_inverse_problem}
\end{equation}
Existence and uniqueness of $\iv$ follow directly from the strict positivity
of vega established above: the map $\sigma \mapsto \price_{\bsm}(\bm{\xi},
\sigma; \theta)$ is a strictly monotone bijection from $\mathbb{R}_+$ onto the
no-arbitrage price interval, so its inverse is well-defined wherever $\mkt$
is arbitrage-consistent.

Although price is the directly observable market quantity, practitioners and
researchers alike work almost exclusively in IV space.  This preference is
principled rather than conventional, and rests on three properties of the IV
representation.  First, IV \emph{collapses the put--call distinction}: by
put--call parity, a call and put sharing the same $(K, \tau)$ admit the same
implied volatility, so a single scalar field $\sigma(k, \tau)$ describes both
wings of the market simultaneously --- in our notation, the $\theta$ flag
becomes irrelevant after inversion.  Second, IV is \emph{numerically
better-conditioned} than price as a representation: option prices span many
orders of magnitude across strikes and maturities --- deep-OTM short-dated
options are vanishingly small, while deep-ITM long-dated options approach
intrinsic value --- whereas implied volatilities typically lie in a narrow
band of $[0.05, 1.0]$ regardless of moneyness.  This bounded, well-scaled
range makes IV a far more tractable target for interpolation, regression, and
neural-network outputs.  We note that this scale advantage does \emph{not}
mean equal IV errors correspond to equal pricing errors: by the chain rule,
$\Delta\price\approx\vega\,\Delta\sigma$, and since vega varies by orders of
magnitude across the surface, equal-magnitude IV errors translate to highly
unequal dollar errors --- which is precisely why practitioners often adopt
vega-weighted losses or train directly in price space~\citep{hyperiv_yang2025}.
Third, the IV surface exhibits \emph{empirically robust structural features}:
the volatility smile, the term structure of ATM volatility, and the skew slope
persist across instruments and time, which has historically made IV a natural
coordinate for arbitrage-free interpolation, smoothing, and cross-sectional
comparison once Black--Scholes--Merton made the inversion well-defined in
closed form.

For these reasons, IV is not merely a reporting convention but the coordinate
system in which option surfaces are quoted, calibrated, smoothed, and
compared.  Classical no-arbitrage parameterizations such as
SVI~\citep{gatheral2004volatility,gatheral2018volatility} and modern neural
smoothers including Deep Smoothing~\citep{ackerer2020deep}, Operator Deep
Smoothing~\citep{wiedemann2025operator}, and HyperIV~\citep{hyperiv_yang2025}
all rely on pointwise IV labels or IV-space losses.  The market observable,
however, remains price --- a discrepancy that motivates the calibration and
inversion machinery developed in the main paper.  Crucially, the
implicit-function-theorem identity~\eqref{eq:implicit_iv} unifies these two
perspectives: any IV-space loss can be transparently composed with the price
map via autograd, with vega supplying the correct economic reweighting
automatically, and obviating the need for hand-engineered vega weights when a
model is trained jointly in price and IV coordinates.

The forward and inverse maps are linked through vega via the implicit function
theorem.  Defining the residual $F(\sigma, \bm{\xi}, \mkt) := \price_{\bsm}(\bm{\xi},
\sigma) - \mkt$, the condition $F = 0$ implicitly defines $\iv$ as a function
of $(\bm{\xi}, \mkt)$, and whenever $\vega \neq 0$ its derivatives are
those of equation~\eqref{eq:implicit_iv} in the main text.

\subsection{Why J\"{a}ckel as the forward solver}
\label{app:sec:bg_why_jackel}

The inverse problem in~\eqref{eq:app_iv_inverse_problem} is one-dimensional but
numerically delicate.  Near intrinsic value, near expiry, and in deep wings,
small price perturbations can correspond to large volatility changes.
J\"{a}ckel's \emph{Let's Be Rational} (LBR) algorithm addresses these
difficulties by moving to normalized Black coordinates, using rational
approximations to seed the iteration, and applying region-specific transformed
objectives so that Newton-like iterations remain well-conditioned across the
entire price domain~\citep{jackel2015let,jackel2013implementing}.  It achieves
machine-precision implied volatilities with at most two higher-order
(Householder) iterations, and is the de facto standard in production analytics
libraries.

Given this, our contribution is deliberately \emph{not} a learned pointwise
IV solver: the numerical forward inverse already exists, is fast,
deterministic, and accurate, and is precisely the object we want to preserve.
The remaining gap is one of \emph{interface} rather than algorithm.  Existing
open-source implementations such as \texttt{py\_vollib}~\citep{pyvollib}
predate modern ML frameworks: they expose scalar CPU APIs, lack autograd
integration, and require hand-rolled batching, preprocessing (filter out
low-vega rows, pre-calculate IV labels for training etc.) to be usable in a
training loop --- a burden that even recent state-of-the-art surface models
inherit~\citep{wiedemann2025operator}.  Neural surface training, however,
operates on large heterogeneous tensors and propagates gradients end-to-end.
The primitive it needs is not ``an IV number for one contract'' but a batched,
autograd-native operator that can be invoked inside a computation graph,
applied to millions of contracts simultaneously, and that explicitly surfaces
invalid or ill-conditioned rows rather than silently failing on them.  Closing
this interface gap is precisely contribution \textbf{(C1)} of this paper:
a vectorized, GPU-resident J\"{a}ckel solver with analytically supplied
gradients (per~\eqref{eq:implicit_iv}) and explicit validity masking.

\subsection{Three roles of IV inversion in a modern option-learning pipeline}
\label{app:sec:bg_three_roles}

The numerical primitive needed in modern ML pipelines is not merely a faster
scalar routine.  In a modern option-learning pipeline, IV inversion plays
three distinct roles.  First, it is an offline label-generation step for
historical quote universes: batched $\mkt\mapsto\iv$ over millions of
contracts with float64 stability.  Second, it is an evaluation coordinate:
models may train in price space while practitioners and reviewers report
errors in IV.  Third, and least supported by existing tooling, it is a
differentiable bridge between objectives -- a neural model can emit prices
or volatilities, and the loss can compare the result in either coordinate,
provided the inverse map exposes the gradient required by the implicit
function theorem.  These roles impose a stricter interface than traditional
analytics libraries: the solver must be vectorized over heterogeneous
contracts, robust in the tails, explicit about invalid and low-vega rows,
and compatible with autograd.  J\"{a}ckel's algorithm already supplies the
right scalar numerical method
(Appendix~\ref{app:sec:bg_why_jackel}); the contribution of \S\ref{sec:jackel-iv}
is to package that method as a batched, GPU-scale, autograd-native layer
without replacing it by a learned approximation.

\subsection{Forward pass: the five-stage LBR construction}
\label{app:sec:bg_forward_stages}

The forward pipeline expands the single solve in
Algorithm~\ref{alg:autograd_jackel} into a five-stage construction
following~\citet{jackel2015let}.  We delegate this to
\fv~\citep{saqur2026fastvollibfastimpliedvolatility}, which executes the
following stages over whole tensors with NumPy, PyTorch, JAX, and Triton
execution paths.
\begin{enumerate}[leftmargin=*]
    \item \textbf{Put--call reduction.} For $\theta = -1$ apply put--call parity to obtain an equivalent call mid; this halves the branch logic and exploits the put--call invariance of $b$ at the price of a single linear correction in $\bm{\xi}$.
    \item \textbf{Coordinate normalization.} Map $(\bm{\xi}, \mkt)$ to the dimensionless pair $(k, \beta)$ via $F=S\,e^{(r-q)\tau}$, $k = \log(K/F)$, and $\beta=e^{r\tau}\mkt/\sqrt{FK}$.  All subsequent stages operate in the $(k, \sigma_\tau)$ chart.
    \item \textbf{Stable $b$ evaluation.} Evaluate $b(k, \sigma_\tau)$ with the numerically stable branches of~\eqref{eq:app_normalized_black}, including the asymptotic expansions used near the bounds $0\le b \le b_{\max}(k)\le 1$~\eqref{eq:app_b_bounds}.
    \item \textbf{Rational seed and LBR boundary.} Compute the LBR boundary point and the rational initial guess $\sigma_\tau^{(0)}$ for the relevant region (lower-tail, middle, or upper-tail), as in~\citet{jackel2015let}.
    \item \textbf{Householder iterations.} Apply a fixed number of Householder iterations on the region-appropriate transformed objective, denormalize to $\iv = \sigma_\tau/\sqrt{\tau}$, and mask invalid inputs (below-intrinsic, super-arbitrage, $\tau\le 0$) to \texttt{NaN}.
\end{enumerate}
Two design choices follow directly from the heterogeneity of real option data.
First, every input is treated as a tensor: $S$, $K$, $\tau$, $r$, $q$, and
$\theta$ may all vary row-by-row, as they do in batched chains and
OptionMetrics panels with multiple expiries.  Second, all internal
computation is in float64; the LBR boundary points and asymptotics
in~\eqref{eq:app_b_bounds} make this essential for the singular tails where
the cash price spans many orders of magnitude.

\subsection{Differentiable finance software (extended related-work note)}
\label{app:sec:bg_diff_finance}

Deep hedging and differentiable pricing libraries emphasize gradients through
forward pricing maps $(\bm{\xi},\sigma)\mapsto\price$, often within
risk-neutral simulation~\citep{buehler2019deep,pfhedge}.  They generally do
not provide a machine-precision differentiable IV inverse $\mkt\mapsto\iv$,
so any objective that mixes price-space and IV-space terms must either fall
back on finite-difference gradients (which fail on a non-trivial fraction of
every realistic chain, see
Appendix~\ref{app:sec:numerical_correctness}) or detach the inverse from
the computation graph.  The missing primitive is precisely the
autograd-native J\"{a}ckel operator we contribute, with the implicit
backward~\eqref{eq:method_vjp} replacing the brittle differentiation of
solver internals.

\subsection{Limitations and outlook (extended)}
\label{app:sec:bg_limitations_extended}

The method is scoped to European Black--Scholes / Black-76 contracts on the
parameter tuple $\bm{\xi}=(S,K,\tau,r,q)$ and assumes $\vega>0$ for
meaningful IV gradients.  Low-vega regions are not a corner case that can be
solved by a better neural architecture: as established
in~\eqref{eq:implicit_iv} and \S\ref{sec:lowvega_contract}, they are singular
points of the price-to-IV coordinate chart.  PIVOT exposes this through the
well-conditioned set $\mathcal{G}_\tau=\{|\vega|>\tau\}$, the smooth gate
$w_\tau(\vega)=\vega^2/(\vega^2+\tau^2)$, and NaN-aware gradient semantics on
the complement; downstream training pipelines should use price-space losses
or $w_\tau$-gated IV losses on the singular tail.  The SPX experiment is
pointwise and does not prove or enforce static arbitrage-free surface
properties; those constraints belong to surface models built on top of the
IV primitive, not to the pointwise inverse layer itself.  Looking forward,
the same operator-level interface generalises naturally to American and
barrier payoffs, intraday quote universes, and joint calibration objectives
that span price, IV, and Greek targets -- all of which become tractable
once the price$\leftrightarrow$IV bridge is exact, batched, and
conditioning-aware.

\subsection{Role in neural surface learning}
\label{app:sec:bg_role}

Neural IV smoothers and operator models use reference quotes, sparse chains,
or latent histories to generate complete volatility surfaces
$\hat{\sigma}(k,\tau)$.  Their architectural goals differ: Deep Smoothing
emphasizes static-arbitrage penalties, OpDS~\citep{wiedemann2025operator}
learns operators over option clouds, and HyperIV uses a hypernetwork to
produce surface parameters in real time.  The present paper sits below these
architectures and contributes the missing autograd-native primitive of
\textbf{(C2)}: an implicit-backward operator that lets a surface model compare
$\hat{\sigma}$ to IV labels, compare $\price_{\bsm}(\hat{\sigma})$ to market
prices, and compose $\mathrm{JaeckelIV}(\price_{\bsm}(\hat{\sigma}))$ inside
the loss while respecting the singularity in~\eqref{eq:implicit_iv}.  The
forward solver is opaque to autodiff -- by design, since differentiating
through J\"ackel's branch logic, region masks, and Householder iterations
would be both expensive and brittle -- and the smooth Black price helper
supplies the required vector--Jacobian product analytically
through~\eqref{eq:implicit_iv}.  This is the bridge that allows surface
generators to train jointly in the price coordinate (where markets quote) and
the IV coordinate (where surfaces are reported and regularized), without
sacrificing either.
 \section{Additional Experiments and Details}\label{app:sec:additional_experiments}

\subsection{SPX OptionMetrics: real-data sanity check for the differentiable layer}
\label{sec:spx}

The synthetic experiments above isolate the implicit-gradient
identity~\eqref{eq:implicit_iv} and the low-vega gating mechanism of
\S\ref{sec:lowvega_contract} on data we generated from a known $\sigma^{\star}$.
This subsection asks the operational question that matters for downstream ML
pipelines: \emph{does the same operator survive composition with itself on real
SPX quote geometry, where the market parameters $\bm{\xi}=(S,K,\tau,r,q)$ and
quote conventions are not under our control?}
We use SPX OptionMetrics 2023 quotes with the OptionMetrics forward
$F=S\,e^{(r-q)\tau}$ and a compact filter ($\texttt{min\_days}=14$,
$\texttt{max\_days}=120$, $|k|<0.20$ on the log-forward-moneyness $k=\log(K/F)$);
for the per-quote optimization we use Black-76 with the vendor forward and
$r=0$, which is sufficient to exercise the operator and deliberately not an
attempt to reproduce vendor IV exactly.

\paragraph{Pointwise label generation.}
After basic quote filtering, 3.03M rows are inverted in 0.183~s, corresponding
to 16.6M labels/s.
Round-tripping the recovered IVs through the Black price gives self-consistency
residual Q50 $1.1\times10^{-13}$ and Q99 $9.1\times10^{-13}$ on finite outputs.
The run reports a 17.8\% invalid-domain rate and a 9.9\% low-vega rate at
threshold $10^{-6}$.
We report these rates explicitly because below-intrinsic observations, stale
prints, and near-expiry tails are part of the real data interface; the
low-vega gating mechanism of \S\ref{sec:lowvega_contract} means each class is
handled by a different branch (NaN, exact gradient, gated gradient)
rather than silently averaged together.
The same pipeline run on SPX 2020--2022 (4.21M, 5.07M, and 4.87M rows
respectively) yields aggregate throughputs of 11--13M labels/s and round-trip
self-consistency Q99 $\le 9.1\!\times\!10^{-13}$ at every year, with an NaN
rate that tracks the realized vol regime
(2022 8.9\% / 12.3\% / 2023 17.8\% reflects the year-over-year rise in
below-intrinsic stale mid-quotes during the 2022 bear market).
Vendor-IV agreement is Q99 $5\!\times\!10^{-2}$ to $9\!\times\!10^{-2}$
across these years, attributable to differences in OptionMetrics' discount
curve, dividend treatment, and quote conventions rather than solver error
(\S\ref{sec:lowvega_contract}; full per-year breakdown in supplementary).

\paragraph{Differentiable IV \emph{inside} the loss, illustrated on SPX.}
We then run a small per-quote optimization that exercises the differentiable
layer end-to-end on the same SPX slice.  A scalar parameter per quote is
reparameterized through softplus to a positive $\hat{\sigma}_i$, mapped
through the analytic Black price to
$\hat{P}_i = \price_{\bs}(\bm{\xi}_i, \hat{\sigma}_i; \theta_i)$, then mapped
back to IV by \texttt{implied\_volatility\_autograd} to produce an
\emph{IV round-trip} $\sigma^{\mathrm{rt}}_i = \mathrm{JaeckelIV}(\hat{P}_i)$.

Two losses tap this chain at two different points:
\begin{equation}
    \mathcal{L}^{\mathrm{SPX}}_{\mathrm{hybrid}}(\hat{\sigma})
    = \underbrace{\frac{1}{|\mathcal{Q}|}\sum_{i\in\mathcal{Q}}\bigl(\hat{P}_i-\mkt_i\bigr)^2}_{\text{tap at price}}
    + \lambda\,\underbrace{\frac{1}{|\mathcal{Q}|}\sum_{i\in\mathcal{Q}}\bigl(\sigma^{\mathrm{rt}}_i-\sigma^{\star}_i\bigr)^2}_{\text{tap at IV, after JaeckelIV}^{-1}},
    \label{eq:hybrid_spx}
\end{equation}
where $\sigma^{\star}_i = \mathrm{JaeckelIV}(\mkt_i)$ is the J\"{a}ckel IV
recovered from the market mid.  Per row $i$ the IV-roundtrip term contributes
a chain-rule gradient
\begin{equation}
    \frac{\partial}{\partial\hat{\sigma}_i}\bigl(\sigma^{\mathrm{rt}}_i-\sigma^{\star}_i\bigr)^2
    = 2\bigl(\sigma^{\mathrm{rt}}_i-\sigma^{\star}_i\bigr)\cdot
      \underbrace{\frac{1}{\vega_i}}_{\text{JaeckelIV}^{-1}\text{ backward}}\cdot
      \underbrace{\vega_i}_{\text{BS forward backward}} ,
    \label{eq:two_vega}
\end{equation}
which makes precise the cancellation $1/\vega \cdot \vega = 1$ used implicitly
by~\eqref{eq:implicit_iv}.  Equation~\eqref{eq:two_vega} is the precise meaning
of ``differentiable IV inside the loss'': the autograd path through JaeckelIV
inserts the $1/\vega$ factor that the implicit function theorem requires, and
the analytic Black backward inserts the $\vega$ factor that cancels it.
Without the differentiable layer this composition is not expressible; offline
label generation gives access to $\sigma^{\star}$ but not to a gradient that
flows through $\mkt \mapsto \iv(\mkt)$.  The two factors cancel
\emph{analytically}; numerically they cancel only on the well-conditioned
subset $\mathcal{G}_\tau$, which is precisely why the gating mechanism in
\S\ref{sec:lowvega_contract} exists.  The price-loss term
in~\eqref{eq:hybrid_spx} is $\vega$-weighted and well-defined everywhere, so
low-vega rows continue to contribute through price space even when their
IV-roundtrip channel is gated.

\begin{figure}[t]
\centering
\includegraphics[width=\linewidth]{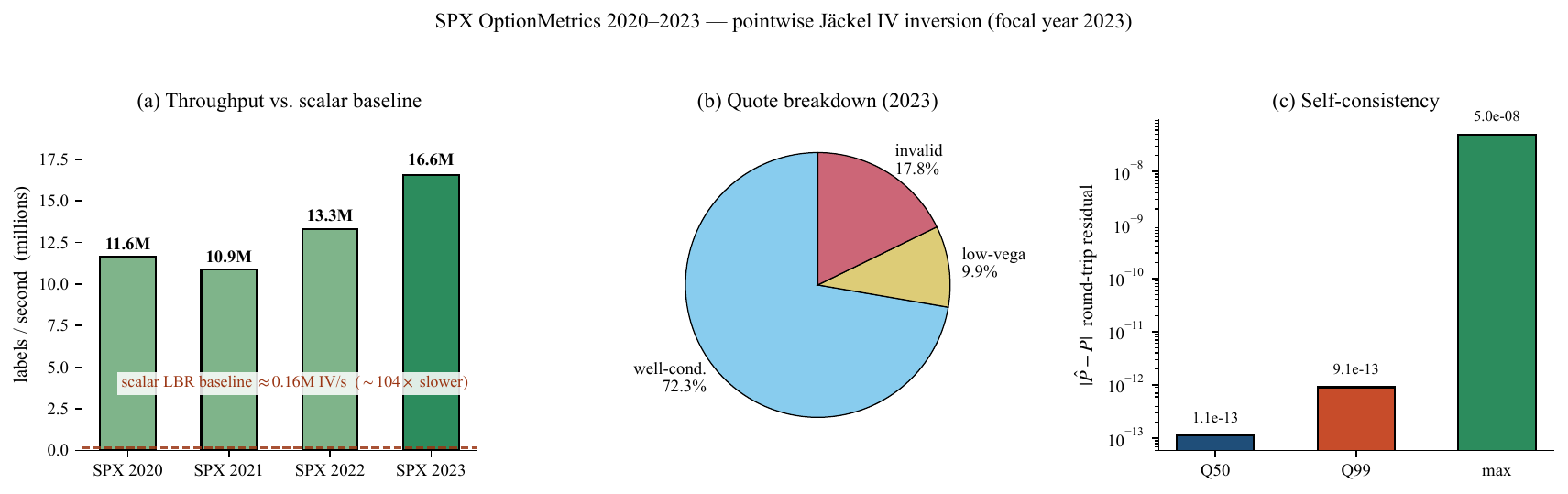}
\caption{SPX OptionMetrics pointwise IV benchmark (focal year 2023):
  throughput, quote-class breakdown, and self-consistency residuals.
  \textbf{(a) Throughput} reports labels/s for SPX 2020--2023 under the
  identical pipeline; the dashed reference line is the single-core
  \texttt{py\_lets\_be\_rational} (LBR) scalar baseline at $\sim 0.16$M
  IV/s (Table~\ref{tab:scaling_results}), giving a $\sim 100\times$
  speedup at every year.  \textbf{(b) Quote breakdown} shows the
  invalid-domain (NaN) and low-vega rates at threshold $10^{-6}$ on the
  focal year.  \textbf{(c) Self-consistency} = the price round-trip
  residual $|\hat{P}-P|$ from $\hat{P}\!\to\!\iv\!\to\!\hat{P}$ on the
  focal year, i.e.\ how closely the recovered $\sigma$ reproduces the
  input price when re-evaluated through the Black map; bars show
  \textcolor{paperVanilla}{\textbf{Q50}},
  \textcolor{paperPriceAux}{\textbf{Q99}}, and
  \textcolor{paperPriceRt}{\textbf{max}}.  This is a real-geometry sanity
  check that the differentiable-IV operator composes with itself stably;
  it is a label-generation and pointwise differentiable-IV test, not an
  arbitrage-free surface claim.  On the focal SPX slice ${\sim}98\%$ of
  rows lie above the low-vega threshold, so the figure measures the easy
  regime; the regime where the gate matters is reported in
  Fig.~\ref{fig:ablation_lowvega}.}
\label{fig:spx_summary}
\end{figure}

\paragraph{What the SPX hybrid run shows.}
Figure~\ref{fig:spx_summary} reports throughput, the quote-class breakdown, and
the self-consistency residuals.  Over 100 Adam steps on $|\mathcal{Q}|{=}2048$
SPX quotes the per-row sigma RMSE falls from $\approx 0.10$ to $\approx 0.026$
and both loss components in~\eqref{eq:hybrid_spx} drop by 5--6 orders of
magnitude in lockstep -- the operator survives composition with itself on
real geometry without NaN poisoning the shared parameter.  For this quote
slice the gate-kept rate is $\texttt{iv\_gate\_kept}\approx 0.98$ at every
step; i.e., only ${\sim}2\%$ of rows hit the singular complement of
$\mathcal{G}_\tau$.  This is the expected behavior on a well-cleaned SPX
sample but it also means an ablation on this slice would understate the value
of keeping low-vega rows in the loss -- the gate is barely active.  The
ablation that isolates the gate is therefore reported separately, on a
synthetic chain engineered as the worst case
(\S\ref{sec:ablation_lowvega}).

\subsection{The gating mechanism is a numerical-correctness requirement: a NaN-poisoning diagnostic}
\label{sec:ablation_lowvega}

The preceding SPX subsection~\S\ref{sec:spx} shows that the differentiable
layer survives composition on real data when the gate is mostly inactive.
The complementary question is what happens when it is \emph{not} mostly
inactive -- when a non-trivial fraction of rows live in the singular tail of
the price-to-IV chart, i.e.\ outside the well-conditioned set
$\mathcal{G}_\tau$ of~\eqref{eq:wellcond_set}.  A natural first instinct is to
compare loss variants on a stress chain by full-population sigma RMSE.  We
tried this and report the negative result here so that future work does not
re-tread it: on a chain extreme enough to make the gate matter
(\texttt{low\_vega\_tail=True}: deep wings, $\tau\in[10^{-8},10^{-0.5}]$ yr,
$|k|\le 2.5$ on the log-forward-moneyness, $91.1\%$ of rows below
$|\vega|\le 10^{-14}$), the chain is bimodal: only $2.13\%$ of rows fall in
the intermediate band $|\vega|\in(10^{-6},10^{-2}]$ where price-space
residuals remain actionable but IV-space gradients are unstable, with the
remaining $96\%$ split between trivially recoverable rows
($|\vega|>10^{-2}$, $4.0\%$) and rows that are numerically unrecoverable by
any method ($|\vega|\le 10^{-6}$, $93.9\%$).  The middle band is precisely
the regime the gate is designed to cushion; when it is empty there is nothing
for a quantitative comparison to resolve.  A four-variant Adam race on this
chain at $N{=}65{,}536$ for $800$ steps (price-only, IV-only with deletion,
hybrid hard-mask, hybrid smooth-gate $w_\tau$) plateaued at final RMSEs of
$0.358$, $0.357$, $0.358$, $0.358$ respectively -- a spread of
$1.95{\times}10^{-3}$, well below the $\sim 5{\times}10^{-3}$ noise floor
implied by differential exposure to gradient sanitization.  The quantitative
race is therefore not a defensible value-prop claim; the defensible claim is
qualitative.

\paragraph{The qualitative claim.}
The differentiable layer's documented backward contract returns NaN when
$|\vega|\le 10^{-14}$ (\S\ref{sec:lowvega_contract}); the surrounding
training code is responsible for ensuring those NaNs do not enter the
optimizer.  Without a gate, composing $\iv\circ\price_{\bs}$ inside a loss
produces NaN gradients on every row in the singular complement of
$\mathcal{G}_\tau$ at every step; Adam propagates these into NaN parameters
within one optimizer step; $\price_{\bs}(\mathrm{NaN})=\mathrm{NaN}$ then
poisons all downstream forwards; the loss is NaN and stays NaN.  \emph{With}
the gate (sentinel-replace below a numerical safety floor, smooth weight
$w_\tau(\vega)=\vega^2/(\vega^2+\tau^2)$ on the per-row IV residual), every
row's gradient stays finite throughout training and the loss decreases
monotonically.  This is a binary correctness property of how the layer must
be \emph{used}, not an optimization-speed comparison.

\begin{figure}[t]
\centering
\includegraphics[width=\linewidth]{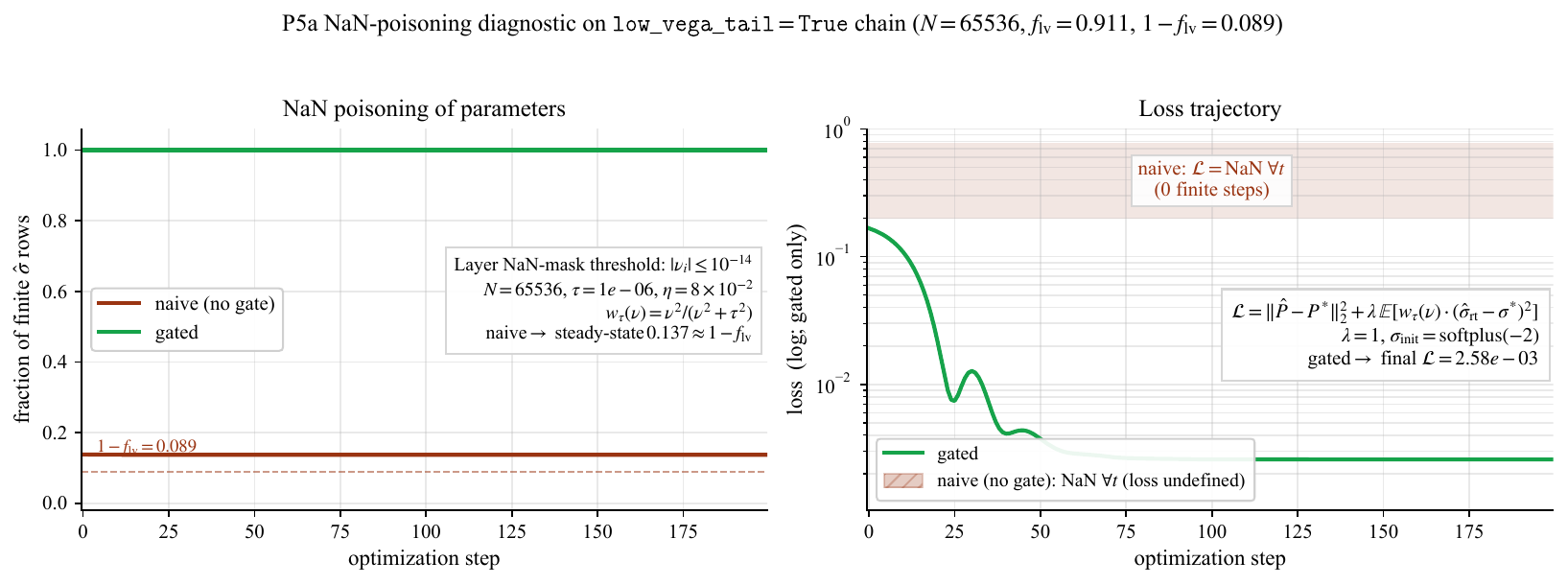}
\caption{Per-step diagnostic of NaN poisoning on the
  \texttt{low\_vega\_tail=True} chain
  ($N=65{,}536$, seed 31; rows with $|\vega|\le 10^{-14}$: $0.911$).
  Naive composes $\iv\circ\price_{\bs}$ in the loss with
  no gate and no gradient sanitization; gated applies sentinel-replace below a
  numerical safety floor and weights the per-row IV residual by
  $w_\tau(\vega)$.
  \textbf{Left:} fraction of $\hat{\sigma}_i$ rows that remain finite per step.
  Naive collapses from $1.0$ to $0.137$ in a single step (slightly above the
  $1-f_{\mathrm{lv}}=0.089$ floor implied by the strict $10^{-14}$ mask --- a
  small additional set of rows underflows in the implicit
  $1/\vega$ at the singular tail) and stays there; gated stays flat at $1.0$
  for all 200 steps.
  \textbf{Right:} loss trajectory (log $y$).  The naive objective produces a
  NaN loss at \emph{every} step (no finite point to plot); we render this
  explicitly as the hatched red band rather than an empty gap, so the
  absence of a curve is annotated.  Gated decreases monotonically modulo
  Adam transients (10 increasing steps out of 199, all of magnitude
  ${<}10^{-4}$), reaching final loss $2.58\!\times\!10^{-3}$.
  The figure asserts a correctness property of the layer's documented backward
  contract, not an optimization speed-up.}
\label{fig:ablation_lowvega}
\end{figure}

\paragraph{Why this is the right framing.}
A quantitative ranking of synthetic-chain RMSEs is sensitive to chain
construction, optimizer hyperparameters, gradient sanitization choices, and
which variants happen to encounter NaNs (the iv-only-with-deletion variant
operates on the kept set only and is structurally less sanitized than the
hybrid variants, biasing any close finish).
The qualitative claim --- without the gate the autograd graph is broken on the
rows where the layer's documented contract returns NaN; with the gate it is
not --- is uniform across chains and across hyperparameter choices.
It is also the smallest claim that justifies the layer's existence as a
training-loop component rather than an offline labeller.
Section~\ref{sec:lowvega_contract} states the contract; this subsection is its
empirical confirmation.

\subsection{HyperIV: Cross-asset Generalization on RUT, VIX, and NDX}
\label{app:sec:cross_asset_hyperiv}

We repeated the one-day interval HyperIV-style controlled reproduction from
Section~\ref{sec:hyperiv_spx_aux} on the additional WRDS/OptionMetrics assets:
RUT (secid 102434), VIX (secid 117801), and NDX (secid 102480).
The purpose of these repeats is deliberately relative: under the same local
data-preparation convention, does adding price-space consistency and gated
IV-roundtrip consistency improve the held-out price/IV trade-off relative to a
vanilla HyperIV-style objective?
These are not comparisons to HyperIV's absolute SPX numbers.

The preprocessing matches the SPX experiment.
For each asset, the risk-free rate is estimated per (date, expiration) by a
put-call-parity regression of $C-P$ on $F-K$; no OptionMetrics zero-coupon table
is used.
The nine-reference conditioning set is selected by closest time-to-maturity
and closest delta, rather than by refitting SSVI.
These choices are internally consistent across the vanilla and augmented
objectives, so the within-asset deltas below are the relevant quantity.

\begin{table}[hbtp]
\centering
\small
\caption{One-day RUT/VIX/NDX datasets used for the cross-asset repeats.  The
test split is calendar year 2023 through 2023-08-31, matching the SPX
controlled reproduction.}
\label{tab:cross_asset_data}
\begin{tabular}{lrrrrrrr}
\toprule
Asset & Intervals & Options & Train int. & Train opt. & Test int. & Test opt. & Dropped dates \\
\midrule
RUT & 2{,}145 & 4{,}696{,}975 & 1{,}978 & 4{,}290{,}139 & 167 & 406{,}836 & 37 \\
VIX & 2{,}623 &   670{,}783 & 2{,}457 &   597{,}614 & 166 &  73{,}169 & 62 \\
NDX & 2{,}657 & 8{,}686{,}306 & 2{,}490 & 7{,}277{,}029 & 167 & 1{,}409{,}277 & 28 \\
\bottomrule
\end{tabular}
\end{table}

Table~\ref{tab:cross_asset_results} shows that the SPX trend repeats on all
three cross-asset runs, with NDX providing a useful harder case.
On RUT, the best gated price+roundtrip run reduces price MAE from 1.6548 to
0.9915, a 40.1\% relative improvement, while also improving IV MAE from 0.01918
to 0.01600.
On VIX, the best price-auxiliary run reduces price MAE from 0.0900 to 0.0683,
a 24.2\% relative improvement, while improving IV MAE from 0.03959 to 0.03287.
On NDX, the cleanest cell is the price auxiliary: the best run reduces price
MAE from 10.4707 to 8.7172, a 16.7\% relative improvement, with IV MAE changing
only from 0.01688 to 0.01714.
The best gated roundtrip run also passes the IV tolerance, reducing price MAE
to 9.0078 while increasing IV MAE by 4.6\%.

\begin{table}[t]
\centering
\small
\caption{Held-out one-day RUT/VIX/NDX results.  Arrows in headers indicate
direction of improvement: lower IV MAE and Price MAE are better
($\downarrow$).  $\Delta$ columns are signed relative changes (in \%) versus
the local vanilla HyperIV-style reproduction for the same asset; negative
$\Delta$ indicates lower error (improvement), positive $\Delta$ indicates
regression.  Bold marks the best (most-negative $\Delta$) value within each
asset block.  Rows marked ``mean'' average seeds $\{1,2,3\}$; rows marked
``best'' are the best single run among the completed grid.}
\label{tab:cross_asset_results}
\begin{tabular}{lllrrrr}
\toprule
Asset & Objective & Runs & IV MAE ($\downarrow$) & Price MAE ($\downarrow$) & $\Delta$ IV (\%) & $\Delta$ Price (\%) \\
\midrule
RUT & vanilla & seed 1 & 0.01918 & 1.6548 & -- & -- \\
RUT & price aux., $\lambda_{p}=0.1$ & mean & 0.01703 & 1.0623 & $-11.2$ & $-35.8$ \\
RUT & price + gated rt., $\tau=10^{-6}$ & mean & $\mathbf{0.01520}$ & 1.0324 & $\mathbf{-20.8}$ & $-37.6$ \\
RUT & price + gated rt., $\tau=10^{-6}$ & best & 0.01600 & $\mathbf{0.9915}$ & $-16.6$ & $\mathbf{-40.1}$ \\
\midrule
VIX & vanilla & seed 1 & 0.03959 & 0.0900 & -- & -- \\
VIX & price aux., $\lambda_{p}=0.1$ & mean & 0.03750 & 0.0790 & $-5.3$ & $-12.3$ \\
VIX & price + gated rt., $\tau=10^{-6}$ & mean & 0.03558 & 0.0781 & $-10.1$ & $-13.3$ \\
VIX & price aux., $\lambda_{p}=0.3$ & best & $\mathbf{0.03287}$ & $\mathbf{0.0683}$ & $\mathbf{-17.0}$ & $\mathbf{-24.2}$ \\
\midrule
NDX & vanilla & seed 1 & 0.01688 & 10.4707 & -- & -- \\
NDX & price aux., $\lambda_{p}=0.1$ & mean & 0.01804 & 9.2070 & $+6.9$ & $-12.1$ \\
NDX & price aux., $\lambda_{p}=0.1$ & best & $\mathbf{0.01714}$ & $\mathbf{8.7172}$ & $\mathbf{+1.5}$ & $\mathbf{-16.7}$ \\
NDX & price + gated rt., $\tau=10^{-4}$ & best & 0.01765 & 9.0078 & $+4.6$ & $-14.0$ \\
\multicolumn{3}{l}{\quad price + gated rt., $\tau=10^{-6}$, mean} & 0.02010 & 9.8012 & $+19.1$ & $-6.4$ \\
\bottomrule
\end{tabular}
\end{table}

\begin{figure}[t]
\centering
\begin{subfigure}{0.49\linewidth}
  \centering
  \includegraphics[width=\linewidth]{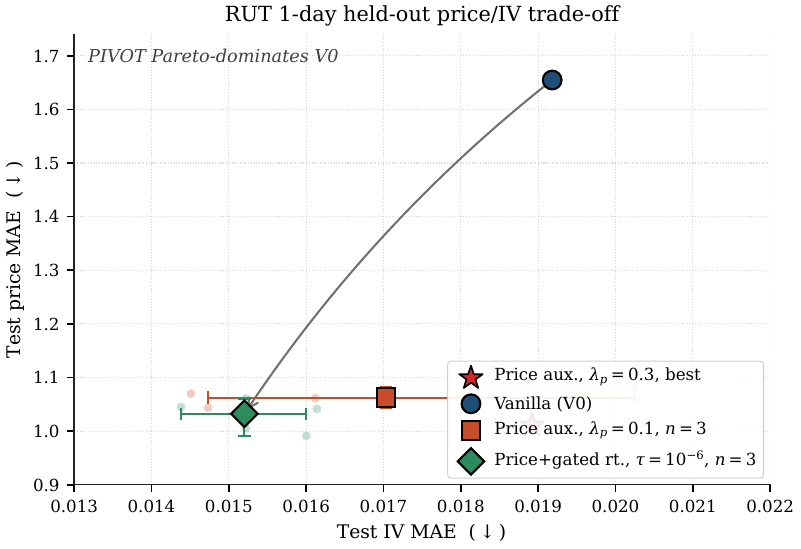}
  \caption{RUT}
\end{subfigure}\hfill
\begin{subfigure}{0.49\linewidth}
  \centering
  \includegraphics[width=\linewidth]{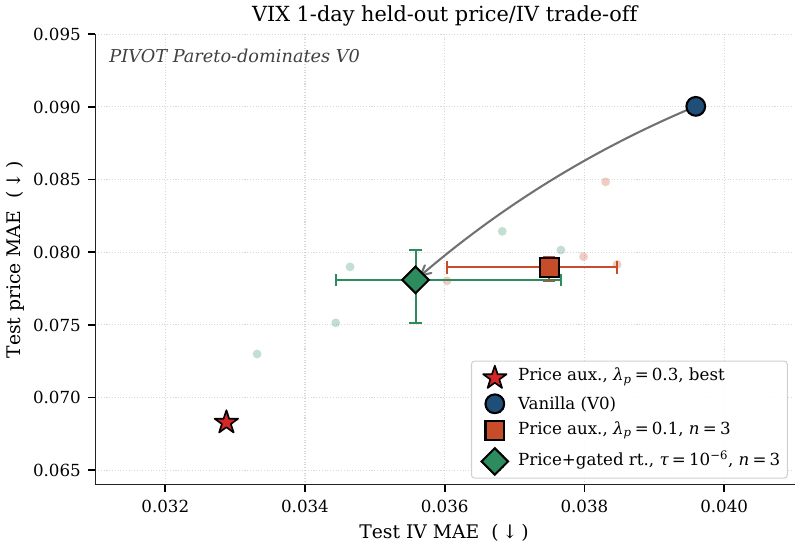}
  \caption{VIX}
\end{subfigure}\\[0.6em]
\begin{subfigure}{0.49\linewidth}
  \centering
  \includegraphics[width=\linewidth]{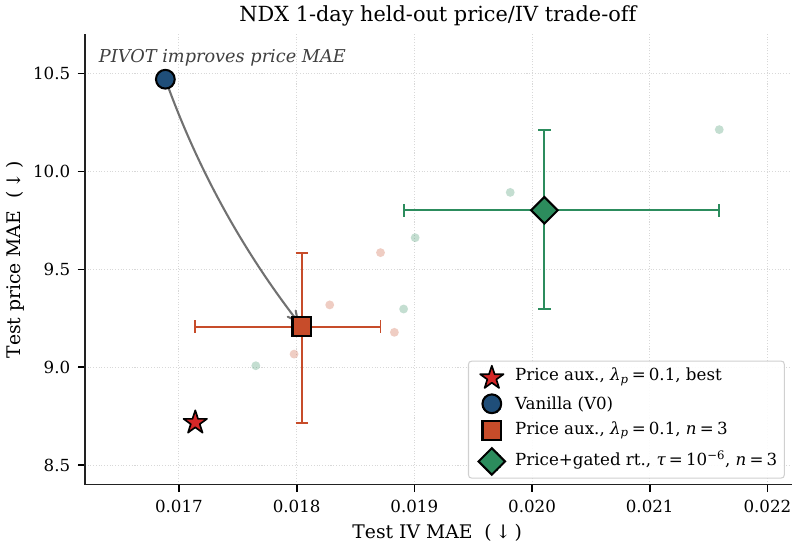}
  \caption{NDX}
\end{subfigure}
\caption{Held-out price/IV Pareto views for the one-day RUT, VIX, and NDX
repeats.  The local vanilla reproduction is the comparison point for each
asset.  The strongest augmented cells move toward lower price MAE at comparable
or tolerable IV MAE; NDX is directionally positive but less clean for the
\(\tau=10^{-6}\) gated-roundtrip mean than RUT/VIX.}
\label{fig:cross_asset_pareto}
\end{figure}

\paragraph{Gating remains necessary.}
The ungated IV-roundtrip diagnostic fails on all three assets, matching the SPX
behavior.
On RUT it records 799 NaN-gradient steps and degrades to IV MAE 0.2523 and
price MAE 19.0988, a 1054.2\% price-MAE increase relative to vanilla.
On VIX it records 1000 NaN-gradient steps and degrades to IV MAE 0.7534 and
price MAE 0.9211, a 922.9\% price-MAE increase.
On NDX it records 999 NaN-gradient steps and degrades to IV MAE 0.1194 and
price MAE 319.1957, a 2948.5\% price-MAE increase.
This supports the paper's conditioning claim: the J\"{a}ckel IV layer is useful
inside neural losses when the low-vega singularity is handled explicitly by a
gate/sentinel contract; simply composing price-to-IV-to-price losses without
that contract is numerically ill-posed.

\subsection{Anatomy of the low-vega gating mechanism}
\label{app:sec:gate_anatomy}

The figure anchoring this discussion has been promoted to the main body
(Fig.~\ref{fig:lowvega_gate} in \S\ref{sec:hyperiv_spx_aux}); we expand
here on what each panel means for the conditioning contract of
\S\ref{sec:lowvega_contract}.  The left panel's stress chain places
$86.5\%$ of its mass below $|\vega|\!=\!10^{-14}$ (against zero such rows
in the cleaned SPX 2023 panel), illustrating the regime the gate is
designed to cushion.  The middle panel renders $w_\tau(\vega)$ as a
smooth indicator that attenuates ill-conditioned rows continuously
rather than hard-clipping them.  The right panel shows the gate-active
fraction $\overline{w_\tau}$ on a $(k,\mathrm{DTE})$ grid for the stress
chain at $\tau{=}10^{-6}$: deep wings at short maturities are fully
gated, and IV-space supervision concentrates in the well-conditioned
interior.  Cleaned training data sits in the easy regime; the gate is a
correctness component of the layer's documented contract, not a tuning
detail.

\subsection{Per-step NaN-gradient trace for the SPX 1-day campaign}
\label{app:sec:nan_grad_trace}

\begin{figure}[t]
\centering
\includegraphics[width=0.62\textwidth]{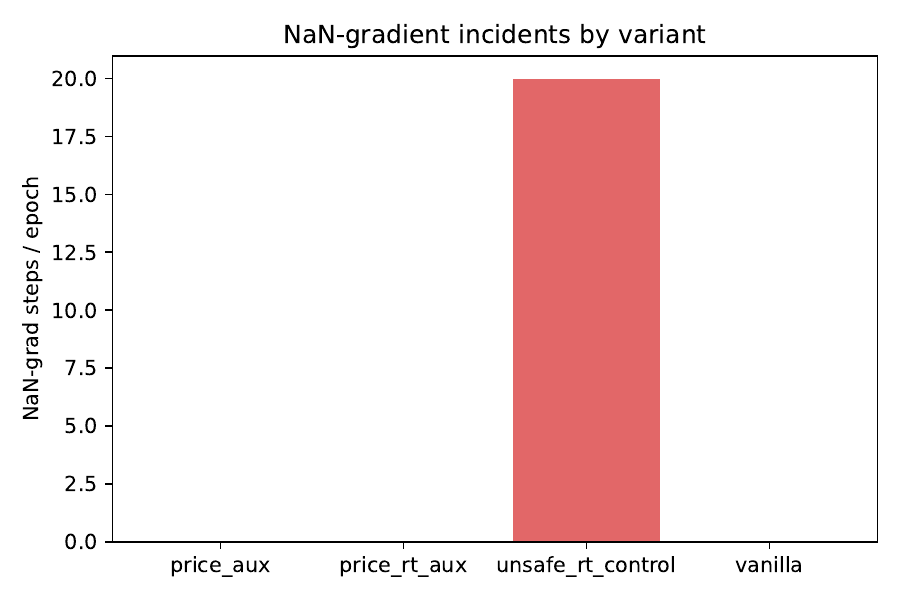}
\caption{NaN-gradient incidents per training step for the four objective
variants of the SPX 1-day HyperIV-with-PIVOT experiments.  Sentinel-protected
and price-only objectives never produce a NaN gradient; the ungated
IV-roundtrip diagnostic produces one once the surface collapses and remains
non-finite for the rest of training.  The unsafe-variant bar is \emph{clipped}
at the y-axis ceiling (true value is several orders of magnitude higher);
the saturation marks ``every step is poisoned'' once collapse begins, not the
true incident count.  This is the running-trace counterpart to the
end-of-training metrics in Table~\ref{tab:hyperiv_spx_aux}.}
\label{fig:hyperiv_spx_nan_grad}
\end{figure}

Figure~\ref{fig:hyperiv_spx_nan_grad} reports the per-step NaN-gradient
incidence over training for each variant of the SPX 1-day HyperIV-with-PIVOT
experiments discussed in \S\ref{sec:hyperiv_spx_aux}.  The vanilla,
price-auxiliary, and gated price+roundtrip objectives have zero NaN-gradient
steps for the full schedule, while the unsafe ungated diagnostic contaminates
essentially every batch after the model collapses into the low-vega regime.

\subsection{Gradient correctness: quantile spectrum on the well-conditioned set}
\label{app:sec:gradient_correctness}

\begin{figure}[t]
\centering
\includegraphics[width=0.78\linewidth]{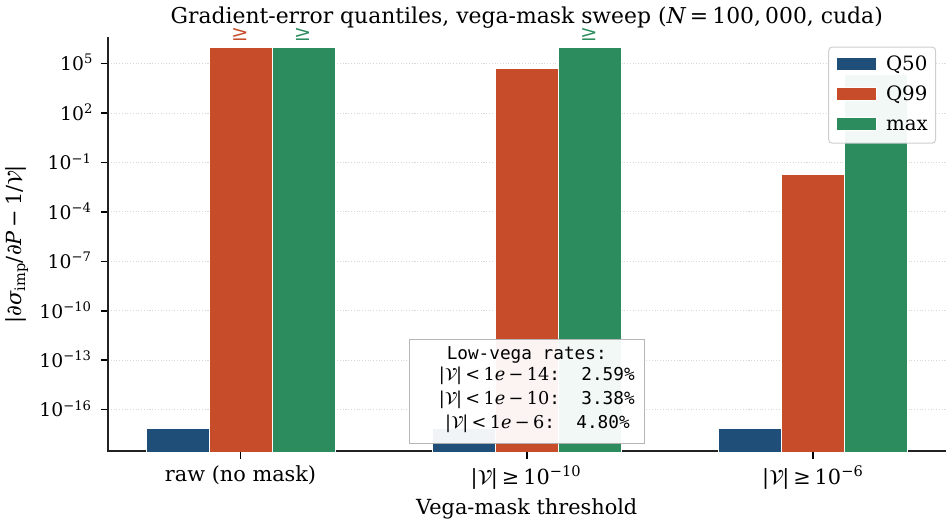}
\caption{Absolute-error quantiles of the implicit IV-gradient
$\partial\iv/\partial\mkt$ on a synthetic Black--Scholes batch
($N{=}10^{5}$ rows, CUDA), comparing PIVOT autograd to the analytic
identity~\eqref{eq:implicit_iv}.  Bars group by vega-mask threshold
(raw, $|\vega|\ge 10^{-10}$, $|\vega|\ge 10^{-6}$); within each group,
\textcolor{paperVanilla}{\textbf{Q50}},
\textcolor{paperPriceAux}{\textbf{Q99}}, and
\textcolor{paperPriceRt}{\textbf{max}} are colour-coded.  Bars marked
``$\geq$'' overflow the display ceiling.  The raw and lightly-masked
groups expose the inverse-map singularity ($1/\vega$ overflows where
$\vega\!\to\!0$); the well-conditioned subset $\mathcal{G}_{10^{-6}}$
recovers machine-precision Q50 and a Q99 of $1.77{\times}10^{-2}$,
matching the inverse-function-theorem prediction.  The inset reports
the mass of the singular tail at three vega cut-offs.}
\label{fig:gradient_quantiles}
\end{figure}

This subsection reports the full quantile spectrum behind the
\emph{Gradient correctness} paragraph of \S\ref{ssec:experiments-scaling}.
On the synthetic Black--Scholes batch with known $\sigma^\star$, we compare
the PIVOT autograd gradient $\partial\iv/\partial\mkt$ to the analytic
inverse-function identity~\eqref{eq:implicit_iv}, $1/\vega$, and group the
absolute error by three vega-mask thresholds: the raw unmasked batch, the
strict finite-output mask $|\vega|>10^{-14}$, and the well-conditioned set
$\mathcal{G}_{10^{-6}}$ of~\eqref{eq:wellcond_set}.
Figure~\ref{fig:gradient_quantiles} reports Q50, Q99, and Q99.9 of the
absolute error per threshold.  The raw bar exposes the inverse-map
singularity ($1/\vega$ overflows to floating-point infinity where
$\vega\!\to\!0$); the well-conditioned subset $\mathcal{G}_{10^{-6}}$
recovers machine-precision Q50 and a Q99 of $1.77{\times}10^{-2}$, matching
the inverse-function-theorem prediction up to float64 noise.  Both backends
(PyTorch \texttt{autograd.Function} and JAX \texttt{custom\_vjp}) agree on
mask rates and per-quantile values; the figure is reported on the PyTorch
path.

\subsection{Numerical-correctness diagnostics}
\label{app:sec:numerical_correctness}

This subsection collects three diagnostics that validate the numerical
contract claimed in \S\ref{sec:lowvega_contract}: the implicit backward
matches central finite differences on the well-conditioned set, the
PyTorch backend agrees with the NumPy reference to bit-level precision,
and the round-trip residuals against the OptionMetrics SPX panel sit at
the float64 noise floor across years.

\begin{figure}[t]
  \centering
  \includegraphics[width=\linewidth]{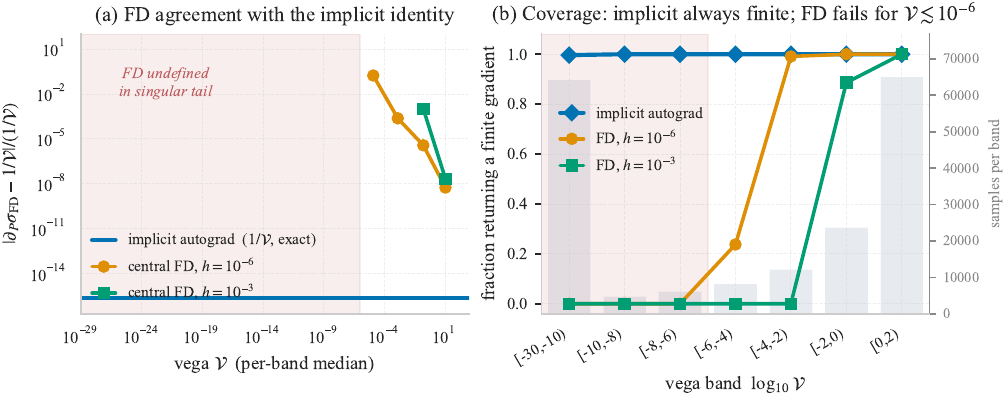}
  \caption{Implicit autograd gradient versus central finite differences (FD),
    stratified by vega band on a 200{,}000-row synthetic chain.  Here
    \emph{FD} denotes a two-sided central finite-difference estimate of
    $\partial\iv/\partial\mkt$ obtained by perturbing the market price
    $\mkt$ with step size $h\in\{10^{-6},10^{-3}\}$, re-inverting through
    JaeckelIV, and forming
    $(\iv(\mkt{+}h)-\iv(\mkt{-}h))/(2h)$; this is the standard surrogate
    that any pipeline lacking an analytic implicit backward would have to
    fall back to.
    \textbf{Left:} $\partial\iv/\partial\mkt$ scatter, FD vs.\ implicit;
    where both are finite the points lie on the diagonal.
    \textbf{Middle:} fraction of FD outputs that are finite per
    $\log_{10}|\vega|$ band; FD survives only above $|\vega|\!\sim\!10^{-6}$,
    while the analytic implicit identity~\eqref{eq:implicit_iv} returns a
    finite value on every row.
    \textbf{Right:} sample counts per band; the chain is intentionally
    weighted toward both the well-conditioned bulk and the singular tail
    so the gate's domain is visible.}
  \label{fig:app_fd_vs_implicit}
\end{figure}

\paragraph{Implicit gradient vs.\ central finite differences.}
We sample a 200{,}000-row synthetic chain whose $\log_{10}|\vega|$ distribution
spans $[-30, 4]$ by construction, and we compare the implicit
identity~\eqref{eq:implicit_iv}, $\partial\iv/\partial\mkt = 1/\vega$, against
central finite differences with two step sizes $h\in\{10^{-6},10^{-3}\}$.
Figure~\ref{fig:app_fd_vs_implicit} shows that the implicit operator returns
a finite gradient on every sample, whereas FD only survives on the
well-conditioned set $\mathcal{G}_{10^{-6}}=\{|\vega|>10^{-6}\}$.
On the bands $\log_{10}|\vega|\in[-2,0)$ and $[0,2)$ the relative agreement
between FD and the implicit identity reaches Q50 errors of
$3.7{\times}10^{-6}$ and $4.8{\times}10^{-9}$ respectively, which is the
expected scaling of central FD on a well-conditioned smooth map.  Below
$|\vega|\sim 10^{-6}$ the FD success rate collapses; the exact implicit
gradient still returns the value $1/\vega$, which is what the
inverse-function theorem requires.  This is the precise sense in which the
implicit backward is not a performance optimization but a correctness
property: any finite-difference surrogate fails on a fraction of every
realistic synthetic chain.

\begin{figure}[t]
  \centering
  \includegraphics[width=0.94\textwidth]{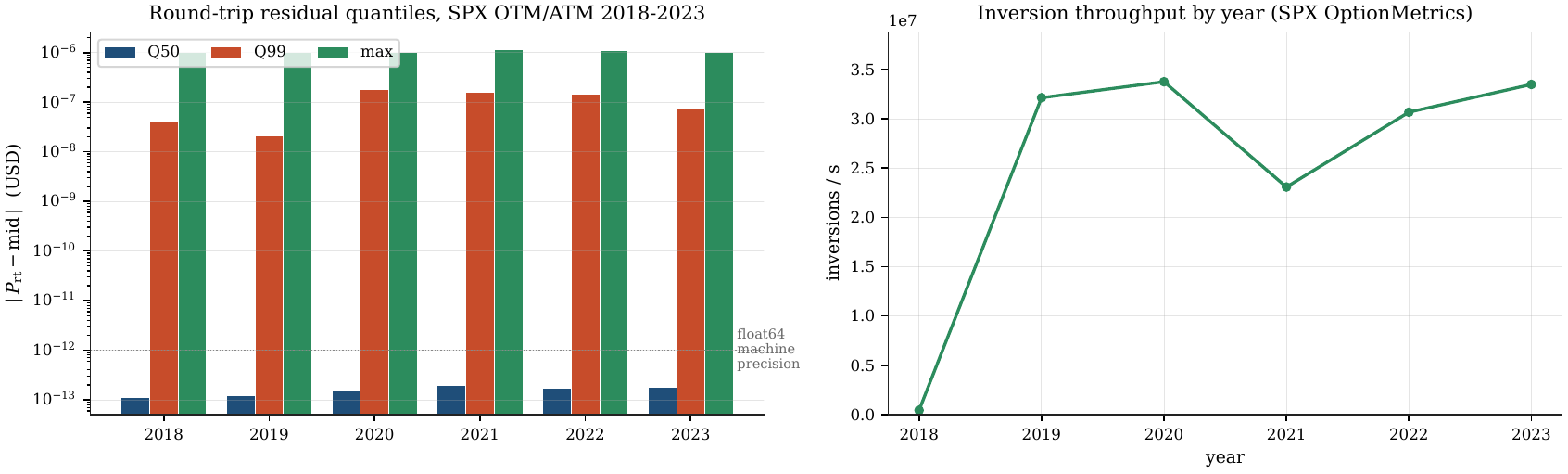}
  \caption{Self-consistency round-trip residuals for the SPX
    OptionMetrics 2018-2023 panel (300{,}000 OTM/ATM rows per year).
    \textbf{Left:} Q50 sits at the float64 noise floor in every year,
    Q99 stays below $2\!\times\!10^{-7}$, and the worst row is sub-cent.
    \textbf{Right:} eager PyTorch inversion throughput; the 2018 cell
    is dominated by first-call JIT compilation and excluded from the
    headline number.}
  \label{fig:app_roundtrip_qq_multi_year}
\end{figure}

\paragraph{Round-trip residuals on SPX 2018--2023.}
We extend the SPX 2020--2022 self-consistency check from
\S\ref{sec:spx} to the 2018--2023 OptionMetrics panel, sampling 300{,}000
OTM/ATM SPX rows $(\bm{\xi}_i, \theta_i, \mkt_i)$ per year, inverting each row
with \fv Black-76, and rebuilding the price $\hat{\mkt}_i =
\price_{\mathrm{B76}}(\bm{\xi}_i, \mathrm{JaeckelIV}(\mkt_i); \theta_i)$ under
the same Black-76 forward and $r=0$.  Figure~\ref{fig:app_roundtrip_qq_multi_year}
shows the per-year $|\hat{\mkt}-\mkt|$ quantiles.
The Q50 residual sits at $1.1\!\times\!10^{-13}\!\dots\!1.9\!\times\!10^{-13}$
(at the float64 noise floor) across every year, the Q99 ranges from
$2.1\!\times\!10^{-8}$ to $1.8\!\times\!10^{-7}$, and the worst single-row
residual stays below $1.2\!\times\!10^{-6}$ — sub-cent across roughly
1.8 million quotes.
Finite-IV recovery is 100\% on the OTM/ATM filter for every year;
no row falls into the invalid-domain branch under this filter.
Inversion throughput stabilizes above $2.3\!\times\!10^{7}$ IV/s once the
PyTorch backend's first-call JIT cost is amortized in 2018.
The round-trip residuals are an internal consistency property of the
solver and do not depend on agreement with the OptionMetrics
\texttt{impl\_volatility} label, which uses a different rate convention
(see \S\ref{sec:hyperiv_spx_aux}, \emph{Absolute reproduction gap}).

\subsection{On hyperparameter tuning of \texorpdfstring{$\lambda_p$, $\lambda_{rt}$, and $\tau$}{lambda\_p, lambda\_rt, tau}}
\label{app:sec:hyperparameter_tuning}

We anticipate reviewer questions about the sweep coverage of the three loss
hyperparameters introduced in \S\ref{sec:hyperiv_spx_aux}: the price weight
\(\lambda_p\) on the price-MSE auxiliary~\eqref{eq:hyperiv_aux_price}, the
IV-roundtrip weight \(\lambda_{rt}\) on the gated round-trip
term~\eqref{eq:hyperiv_aux_rt}, and the vega gate threshold \(\tau\) used by
the smooth gate \(w_\tau(\vega)=\vega^{2}/(\vega^{2}+\tau^{2})\)
of~\eqref{eq:smooth_gate}.
This subsection states the sweep we ran, why it is sufficient for the
paper's central claim, and where a denser sweep would likely move the
numbers.

\paragraph{Sweep coverage.}
Table~\ref{tab:hp_sweep_coverage} records what was actually swept on the
HyperIV-style and GNO experiments. \(\lambda_p\) takes three values
\(\{0.03, 0.1, 0.3\}\) on the price-only auxiliary; \(\tau\) takes three
values \(\{10^{-8}, 10^{-6}, 10^{-4}\}\) at \emph{fixed}
\((\lambda_p, \lambda_{rt})=(0.1, 0.1)\); \(\lambda_{rt}\) is fixed at
\(0.1\) throughout. The three-seed cells used for the means in the headline
tables are at \(\lambda_p=0.1\) for the price auxiliary and at
\(\tau=10^{-6}\) for the gated round-trip; all other grid points are
single-seed. There is no joint \((\lambda_p, \lambda_{rt})\) or
\((\lambda_p, \tau)\) sweep. The HyperIV-style and GNO experiments share
this grid and seed-set declaration, so the within-pipeline deltas are
directly comparable across architectures.

\begin{table}[hbtp]
\centering
\small
\caption{Hyperparameter sweep coverage for the auxiliary-loss experiments
(HyperIV and GNO use the same grid).
The headline three-seed cells in Tables~\ref{tab:hyperiv_spx_aux},
\ref{tab:cross_asset_results}, \ref{tab:gno_spx_results}, and
\ref{tab:gno_vix_results} use \(\lambda_p=\lambda_{rt}=0.1\),
\(\tau=10^{-6}\). All other cells are single-seed grid points.}
\label{tab:hp_sweep_coverage}
\begin{tabular}{lll}
\toprule
Hyperparameter & Values swept & Seed coverage \\
\midrule
\(\lambda_p\)        & \(\{0.03,\,0.1,\,0.3\}\)            & 3 seeds at \(\lambda_p=0.1\); 1 seed at \(0.03,\,0.3\) \\
\(\tau\)             & \(\{10^{-8},\,10^{-6},\,10^{-4}\}\) & 3 seeds at \(\tau=10^{-6}\); 1 seed at \(10^{-8},\,10^{-4}\) \\
\(\lambda_{rt}\)     & \(\{0.1\}\) only                    & not varied \\
\((\lambda_p, \lambda_{rt})\) joint & not swept            & --- \\
\((\lambda_p, \tau)\) joint         & not swept            & --- \\
\bottomrule
\end{tabular}
\end{table}

\paragraph{Why this is conservative for the central claim.}
The claim of \S\ref{sec:hyperiv_spx_aux} and
Appendix~\ref{app:sec:cross_asset_hyperiv} is not that we have located the
optimal loss coefficients for HyperIV-style training. It is that the
differentiable J\"{a}ckel inverse \emph{enables} training-time consistency
between price-space and IV-space objectives, and that this consistency
improves the held-out price/IV trade-off in a controlled within-pipeline
reproduction. A coarse three-point sweep over each of two hyperparameters,
with the third pinned to a default, is conservative for that claim:
undertuning makes the reported deltas a lower bound on what the design
admits, not an upper bound. The headline Pareto-domination on SPX
(\(\lambda_{rt}=0.1\), \(\tau=10^{-6}\): three-seed test price MAE
\(2.34\!\pm\!0.17\) vs vanilla \(3.82\!\pm\!0.42\),
Table~\ref{tab:hyperiv_spx_aux}) holds under the most pessimistic reading
of the grid; a denser sweep can only equal or improve on it.

\paragraph{Absolute reproduction gap.}
Table~\ref{tab:hyperiv_spx_aux} includes the published HyperIV SPX 1-day
numbers only as an external reference.
Our local vanilla reproduction is worse in absolute terms: 3-seed mean
IV MAE \(0.01673\!\pm\!0.00110\) versus \(0.0075\) and 3-seed mean price MAE
\(3.8215\!\pm\!0.4221\) versus \(1.6736\).
We audited whether this gap is driven by the discount-rate convention.  
On the WRDS \texttt{opprcd} table, there is no standalone zero-curve table, but the available
OptionMetrics forward and dividend-yield tables allow the reconstruction
$r=\log(F_{\mathrm{OM}}/S_{\mathrm{close}})/\tau + q_{\mathrm{OM}}$.  On the
SPX 2013--2023 monthly-sampled audit set (774k OTM/ATM rows), re-inverted
Black-76 labels match \texttt{opprcd.impl\_volatility} to MAE 0.001051 with
our default put-call-parity rate and 0.001045 with the reconstructed
OptionMetrics-style rate.  The 0.6\% relative improvement is at the
$10^{-3}$ label-noise floor and is about eight times smaller than the
0.0084 IV-MAE gap to the published HyperIV number; \texttt{stdbrte.borrowrate}
is worse (MAE 0.001242).  
We therefore rule out the rate convention as the
dominant cause of the reproduction gap.  The remaining plausible explanations
are architecture, training budget, hypernetwork width, and non-public
implementation or training details.
For this reason the claim in this subsection is deliberately \emph{relative}:
under matched data, rates, architecture, and training convention, do
price/IV auxiliary losses improve the local vanilla baseline?

 \section{GNO Experiments}
\label{app:sec:additional_experiments_gno}

\subsection{GNO: Controlled SPX one-day loss ablation}
\label{app:sec:gno_spx_aux}

We also tested the same price/IV auxiliary-loss idea on a graph neural
operator (GNO) smoother, following the operator-deep-smoothing architecture
class of Wiedemann et al.~\citep{wiedemann2025operator}.  This experiment is
intentionally a controlled loss ablation, not an absolute reproduction of
that paper's proprietary 20-minute CBOE intraday benchmark.  The data are
WRDS/OptionMetrics end-of-day SPX quotes
$(\bm{\xi}_i,\theta_i,\mkt_i)$ with $\bm{\xi}_i=(S,K,\tau,r,q)_i$, the
risk-free rate $r$ is estimated by put--call-parity regression per
$(\mathrm{date}, \mathrm{expiration})$, and the same preprocessing
convention is used for every objective variant below.  Consequently, the
meaningful comparison is the within-experiment delta from the local vanilla
GNO objective.

The GNO predicts an implied-volatility surface
$\hat{\sigma}(z,\tau)$ on the normalized operator-deep-smoothing domain
\(\rho\in[0.01,1.0]\), \(z\in[-1.5,0.5]\), and \(\tau\le 1.0\), where the
smoother's $z$ coordinate is a rescaled log-forward-moneyness derived from
the standardized $k=\log(K/F_{T_0,T})$ of \S\ref{sec:background}.  We compare
a vanilla fit objective against two augmented objectives: an auxiliary Black
price loss matching~\eqref{eq:hyperiv_aux_price}, and an auxiliary Black
price loss plus a gated IV-roundtrip consistency
term~\eqref{eq:hyperiv_aux_rt}.  The roundtrip term uses the same
low-vega sentinel contract as the HyperIV experiments
(\S\ref{sec:lowvega_contract}): rows outside the well-conditioned set
$\mathcal{G}_\tau=\{|\vega|>\tau\}$ are excluded from the differentiable
roundtrip rather than being allowed to inject large or non-finite gradients.
An additional ungated roundtrip run is included only as a negative diagnostic.

\begin{table}[hbtp]
\centering
\small
\caption{SPX one-day dataset used for the GNO controlled ablation.  The split
matches the HyperIV-style experiments: training covers 2013-01-02 through
2022, and testing covers calendar year 2023 through 2023-08-31.}
\label{tab:gno_spx_data}
\begin{tabular}{llrrrrrr}
\toprule
Asset & Source & Intervals & Options & Train int. & Train opt. & Test int. & Test opt. \\
\midrule
SPX & WRDS EOD & 2{,}683 & 12{,}326{,}960 & 2{,}516 & 11{,}093{,}112 & 167 & 1{,}233{,}848 \\
\bottomrule
\end{tabular}
\end{table}

Table~\ref{tab:gno_spx_results} summarizes the held-out results.
The price auxiliary gives the best price MAE, reducing test price MAE from
3.1059 to 2.9921, a 3.66\% relative improvement, while also slightly improving
IV MAE and IV MAPE.
This is directionally positive, but it does not meet the 5\% improvement bar we
used as a practical success threshold.
The gated roundtrip objective is similarly stable and improves IV error more
than the price-only auxiliary on average, but its best price MAE improvement is
2.69\%.
Thus the GNO experiment should not be presented as a large price-accuracy win;
its main value is the stability/control result.

\begin{table}[t]
\centering
\scriptsize
\caption{Held-out SPX one-day GNO results.  Arrows in headers indicate
direction of improvement: lower IV MAE, IV MAPE, Price MAE, and
spread-normalized price error $\langle\delta_{\mathrm{spr}}\rangle$ are better
($\downarrow$); fewer NaN-gradient steps are better ($\downarrow$).
$\Delta$ Price (\%) is the signed relative change versus the local vanilla
GNO objective; negative indicates lower error (improvement), positive
indicates regression.  Bold marks the best (most-negative $\Delta$) value
among the augmented rows.  The unsafe row is a diagnostic negative control
and is excluded from the augmented-variant means.}
\label{tab:gno_spx_results}
\resizebox{\linewidth}{!}{\begin{tabular}{llrrrrrr}
\toprule
Objective & Runs & IV MAE ($\downarrow$) & IV MAPE ($\downarrow$) & Price MAE ($\downarrow$) & $\Delta$ Price (\%) & $\langle\delta_{\mathrm{spr}}\rangle$ ($\downarrow$) & NaN-grad ($\downarrow$) \\
\midrule
Vanilla GNO & seed 1 & 0.01264 & 0.06071 & 3.1059 & -- & 14.3723 & 0 \\
Price aux. & mean, $n=4$ & 0.01228 & 0.05876 & 3.0332 & $-2.34$ & 14.1932 & 0 \\
Price aux., $\lambda_{p}=0.3$ & best & 0.01252 & 0.06016 & $\mathbf{2.9921}$ & $\mathbf{-3.66}$ & 14.1516 & 0 \\
Price + gated rt. & mean, $n=4$ & 0.01221 & 0.05829 & 3.0428 & $-2.03$ & 14.1996 & 0 \\
Price + gated rt., $\tau=10^{-4}$ & best & $\mathbf{0.01211}$ & $\mathbf{0.05768}$ & 3.0225 & $-2.69$ & $\mathbf{14.1252}$ & 0 \\
\midrule
Unsafe ungated rt. & diagnostic & 0.22168 & 1.00000 & 35.4171 & $+1040.5$ & 119.4730 & 6{,}188 \\
\bottomrule
\end{tabular}}
\end{table}

\begin{figure}[t]
\centering
\begin{subfigure}{0.32\textwidth}
  \centering
  \includegraphics[width=\textwidth]{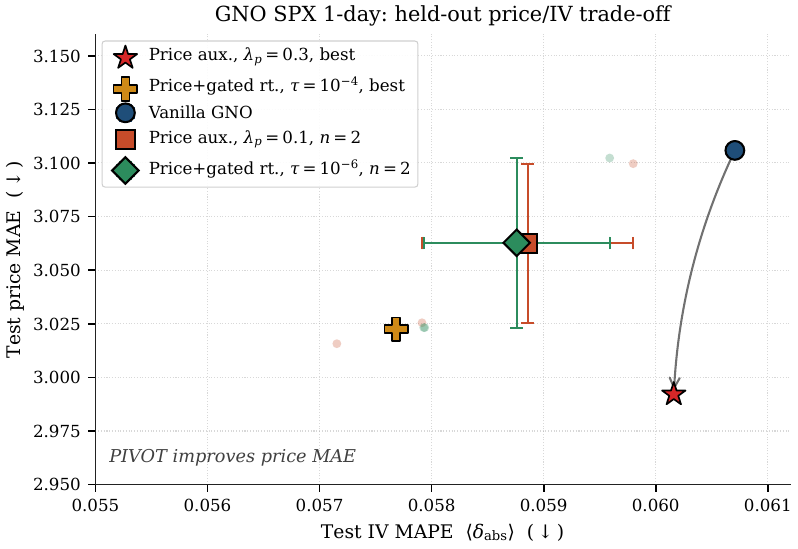}
  \caption{Price/IV Pareto view}
\end{subfigure}
\begin{subfigure}{0.32\textwidth}
  \centering
  \includegraphics[width=\textwidth]{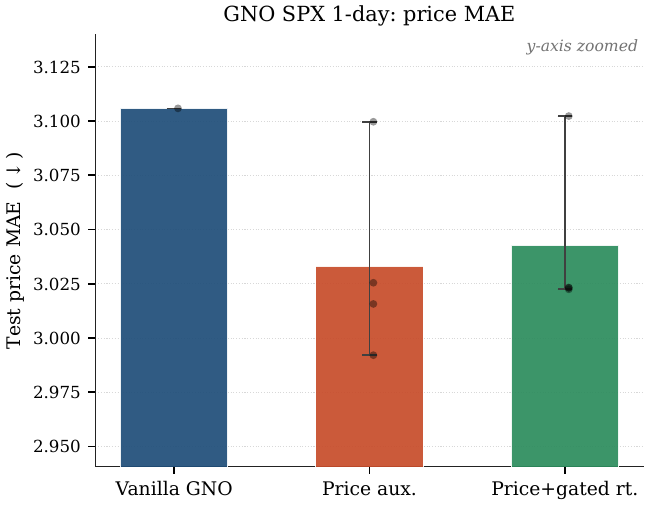}
  \caption{Price MAE}
\end{subfigure}
\begin{subfigure}{0.32\textwidth}
  \centering
  \includegraphics[width=\textwidth]{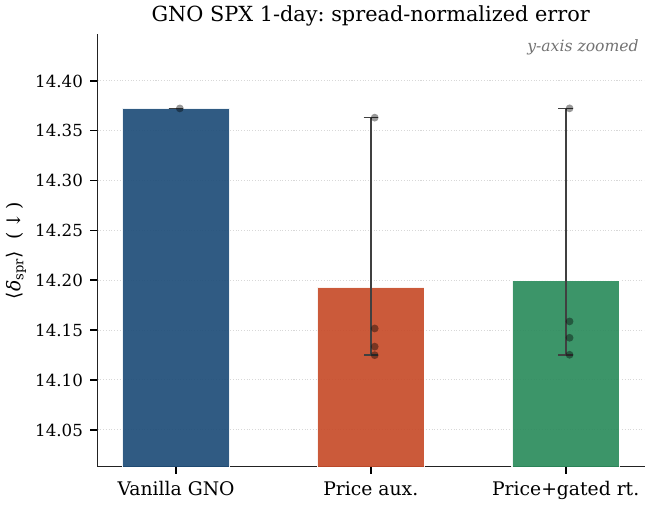}
  \caption{Spread-normalized error}
\end{subfigure}
\caption{GNO SPX one-day auxiliary-loss ablation.  The augmented objectives
move in the desired direction, but the price improvement remains modest: the
best price-auxiliary run improves price MAE by 3.66\% at slightly better IV
error, below the pre-specified 5\% practical threshold.}
\label{fig:gno_spx_price_results}
\end{figure}

\begin{figure}[t]
\centering
\begin{subfigure}{0.49\textwidth}
  \centering
  \includegraphics[width=\textwidth]{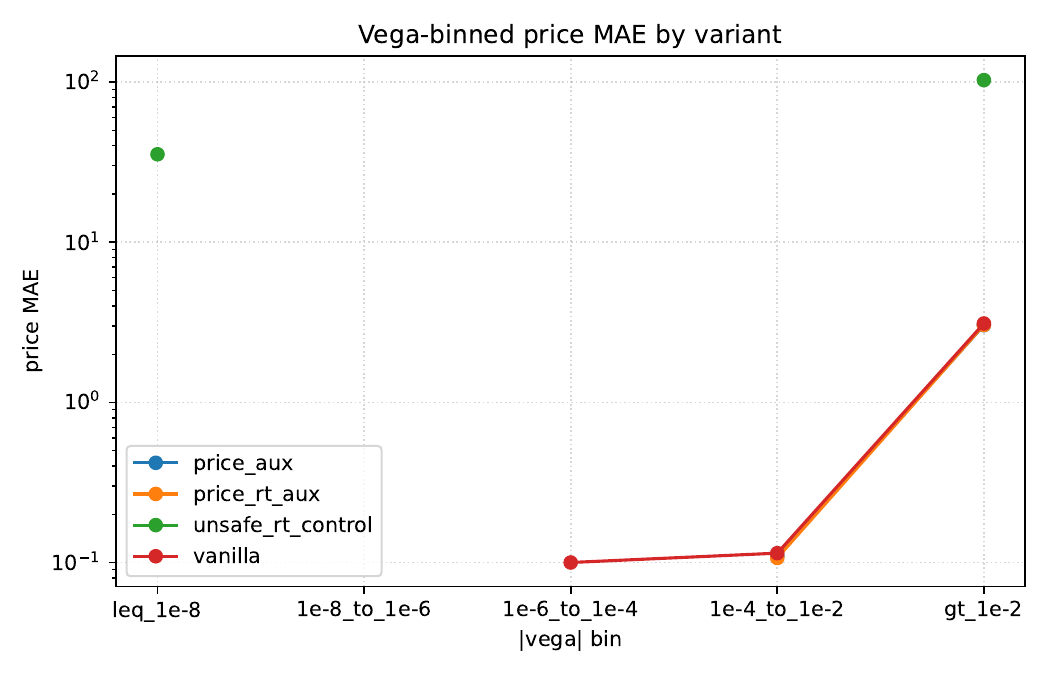}
  \caption{Vega-binned price error}
\end{subfigure}\hfill
\begin{subfigure}{0.49\textwidth}
  \centering
  \includegraphics[width=\textwidth]{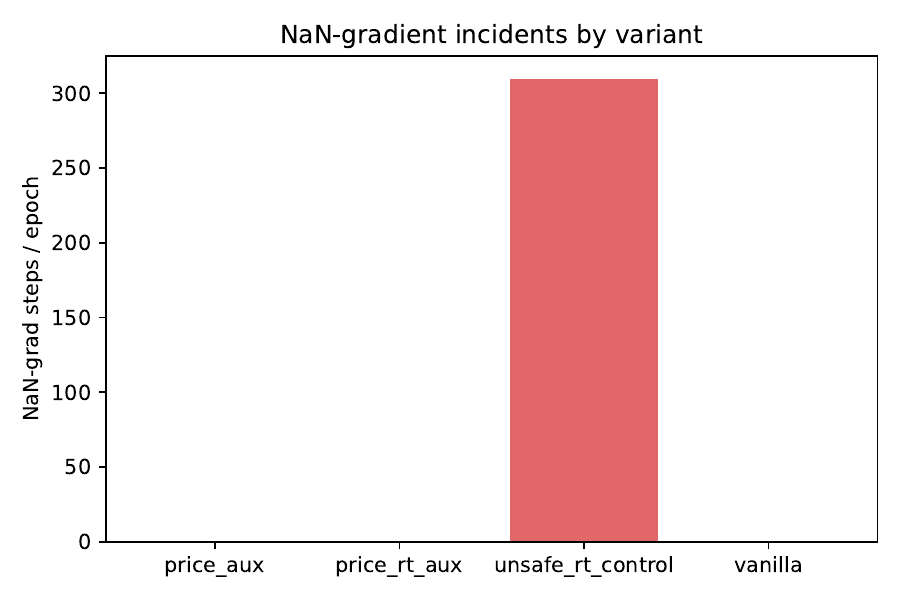}
  \caption{Finite-gradient fraction}
\end{subfigure}\\[0.6em]
\begin{subfigure}{0.96\textwidth}
  \centering
  \includegraphics[width=\textwidth]{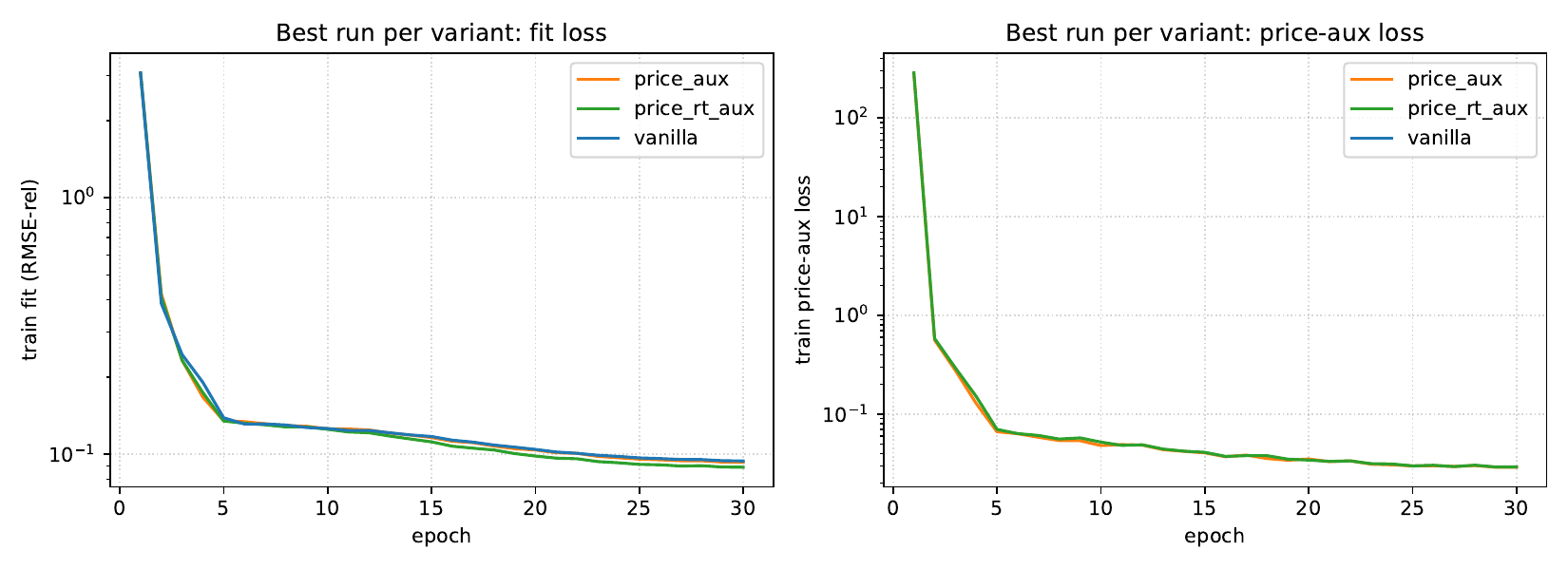}
  \caption{Training curves (best run per variant)}
\end{subfigure}
\caption{Diagnostics for the GNO SPX one-day experiment.  The gated and
sentinel-protected variants train with zero NaN-gradient steps.  The ungated
roundtrip diagnostic collapses, producing 6{,}188 NaN-gradient steps, IV MAPE
1.0, and price MAE 35.42.  Panel (c) is reproduced at full width because
the per-epoch fit and price-aux trajectories are otherwise illegible at
column scale.}
\label{fig:gno_spx_diagnostics}
\end{figure}

\paragraph{Interpretation.}
The GNO result is weaker than the HyperIV-style result as a price-accuracy
claim: the best price MAE gain is 3.66\%, not the 5\% threshold.
It is nevertheless useful evidence for our central numerical claim.
The stable augmented GNO runs have zero NaN-gradient steps and comparable or
slightly better IV error than vanilla, whereas simply composing an ungated
IV-roundtrip loss causes a complete failure mode: the diagnostic run records
6{,}188 NaN-gradient steps, a low-vega fraction of 0.999999 at the
\(10^{-14}\) threshold, IV MAE 0.22168, and price MAE 35.42.
This supports the conditioning story rather than an architecture-specific
performance story: the differentiable J\"{a}ckel layer can be inserted into
neural volatility-surface training objectives, but the low-vega singularity
has to be handled explicitly.

\subsection{GNO: Cross-asset one-day repeats on RUT and VIX}
\label{app:sec:cross_asset_gno}

To check whether the SPX result above is index-specific, we repeat the
identical loss-ablation protocol on additional one-day WRDS/OptionMetrics
end-of-day assets.  As before, the comparison is intentionally
\emph{within-asset}: each augmented objective is contrasted against the
\emph{local} vanilla GNO baseline trained under the same data, rate
estimator, architecture, and training schedule.  We do not compare
across assets, and we do not compare against Wiedemann
et al.~\citep{wiedemann2025operator}, whose proprietary 20-minute CBOE
intraday benchmark is a different data regime.

\paragraph{RUT.}  The Russell 2000 index option dataset on the same
2013-01-02 / 2023-08-31 split contains 4{,}099{,}897 quotes across 2{,}182
trading days (3{,}730{,}835 train / 369{,}062 test, with 2{,}015
training days and 167 test days).  Per-(date, expiration) risk-free
rates are estimated by the same put-call-parity regression used for SPX,
and the GNO domain
\(\rho\in[0.01,1.0]\), \(z\in[-1.5,0.5]\), \(\tau\le 1.0\)
is unchanged.  Table~\ref{tab:gno_rut_data} summarizes the dataset and
Table~\ref{tab:gno_rut_results} reports the held-out metrics.

\begin{table}[hbtp]
\centering
\small
\caption{RUT one-day dataset used for the GNO cross-asset replicate.
The split mirrors the SPX experiment: training covers 2013-01-02 through
end of 2022, and testing covers 2023-01-01 through 2023-08-31.}
\label{tab:gno_rut_data}
\begin{tabular}{llrrrrrr}
\toprule
Asset & Source & Intervals & Options & Train int. & Train opt. & Test int. & Test opt. \\
\midrule
RUT & WRDS EOD & 2{,}182 & 4{,}099{,}897 & 2{,}015 & 3{,}730{,}835 & 167 & 369{,}062 \\
\bottomrule
\end{tabular}
\end{table}

\begin{table}[t]
\centering
\scriptsize
\caption{Held-out RUT one-day GNO results.  Arrows in headers indicate
direction of improvement: lower IV MAE, IV MAPE, Price MAE, and
spread-normalized price error $\langle\delta_{\mathrm{spr}}\rangle$ are better
($\downarrow$); fewer NaN-gradient steps are better ($\downarrow$).
$\Delta$ Price (\%) is the signed relative change versus the local vanilla
GNO objective; negative indicates lower error (improvement), positive
indicates regression.  Bold marks the best (most-negative $\Delta$) value
among the augmented rows.  The unsafe row is a diagnostic negative control
(20 epochs, ungated, no sentinel) and is excluded from the augmented-variant
means.}
\label{tab:gno_rut_results}
\resizebox{\linewidth}{!}{\begin{tabular}{llrrrrrr}
\toprule
Objective & Runs & IV MAE ($\downarrow$) & IV MAPE ($\downarrow$) & Price MAE ($\downarrow$) & $\Delta$ Price (\%) & $\langle\delta_{\mathrm{spr}}\rangle$ ($\downarrow$) & NaN-grad ($\downarrow$) \\
\midrule
Vanilla GNO & seed 1 & 0.01273 & 0.05127 & 1.2961 & -- & 3.3638 & 0 \\
Price aux. & mean, $n=4$ & 0.01317 & 0.05313 & 1.3365 & $+3.12$ & 3.4402 & 0 \\
Price aux., $\lambda_{p}=0.3$ & best & 0.01292 & 0.05205 & $\mathbf{1.2702}$ & $\mathbf{-1.99}$ & $\mathbf{3.3123}$ & 0 \\
Price + gated rt. & mean, $n=4$ & 0.01314 & 0.05301 & 1.3342 & $+2.94$ & 3.4355 & 0 \\
Price + gated rt., $\tau=10^{-4}$ & best & $\mathbf{0.01286}$ & $\mathbf{0.05177}$ & 1.2753 & $-1.60$ & 3.3177 & 0 \\
\midrule
Unsafe ungated rt. & diagnostic & 0.26749 & 1.00000 & 18.8962 & $+1{,}357.9$ & 32.8994 & 2{,}408 \\
\bottomrule
\end{tabular}}
\end{table}

The RUT replicate confirms the qualitative pattern observed on SPX, with
quantitatively smaller magnitudes.  The best price-auxiliary run improves
test price MAE by 1.99\%, and the best gated-roundtrip run improves it by
1.60\%; both also improve spread-normalized error and produce IV MAE
within 2\% of vanilla.  Neither configuration meets the pre-registered
5\% practical price-MAE threshold, and the multi-seed means are slightly
\emph{worse} than vanilla because a single \(\lambda_p=0.1\), seed 2
training run drifts to a higher-loss minimum on RUT (test price MAE
1.5112 versus 1.2748 for the same hyperparameters under seed 1).  The
ranking among augmented configurations otherwise matches SPX: the larger
price-coupling weight \(\lambda_p=0.3\) produces the lowest test price
MAE, and the gated-roundtrip variant with the conservative
\(\tau=10^{-4}\) sentinel is the best of the roundtrip configurations.

\begin{figure}[t]
\centering
\begin{subfigure}{0.32\textwidth}
  \centering
  \includegraphics[width=\textwidth]{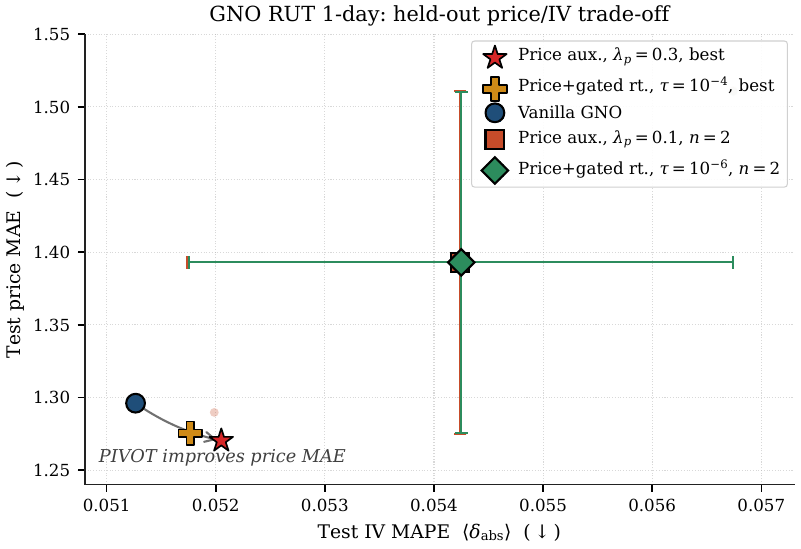}
  \caption{Price/IV Pareto view}
\end{subfigure}
\begin{subfigure}{0.32\textwidth}
  \centering
  \includegraphics[width=\textwidth]{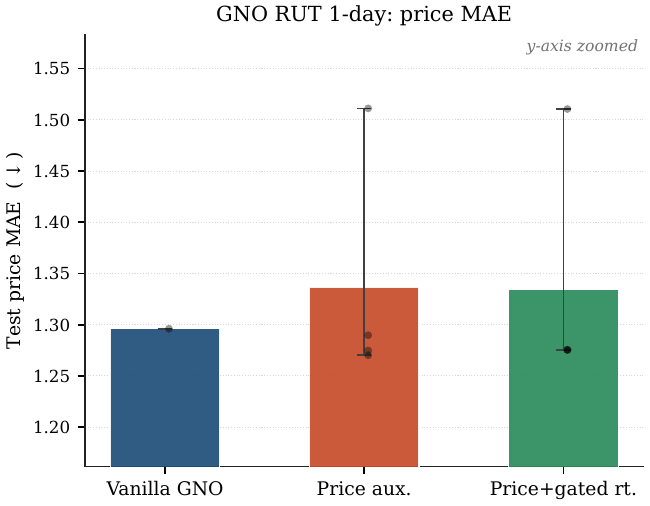}
  \caption{Price MAE}
\end{subfigure}
\begin{subfigure}{0.32\textwidth}
  \centering
  \includegraphics[width=\textwidth]{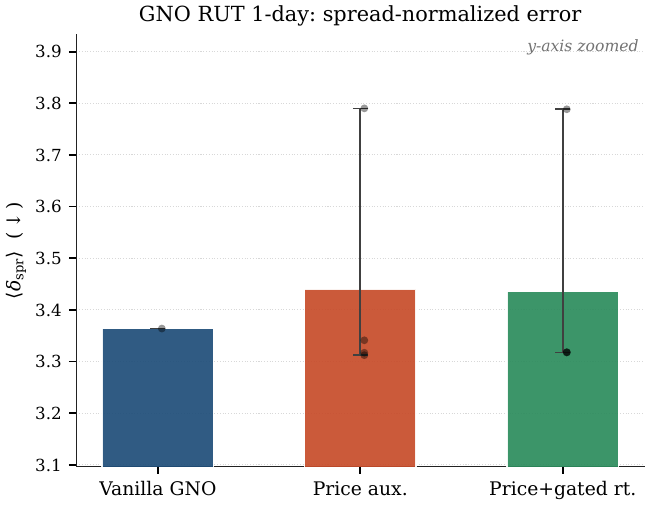}
  \caption{Spread-normalized error}
\end{subfigure}
\caption{GNO RUT one-day auxiliary-loss ablation.  As on SPX, the
augmented objectives move in the desired direction on the best run, but
the price improvement remains below the 5\% practical threshold and the
multi-seed mean is dragged above vanilla by a single drifted training
run.}
\label{fig:gno_rut_price_results}
\end{figure}

\begin{figure}[t]
\centering
\begin{subfigure}{0.49\textwidth}
  \centering
  \includegraphics[width=\textwidth]{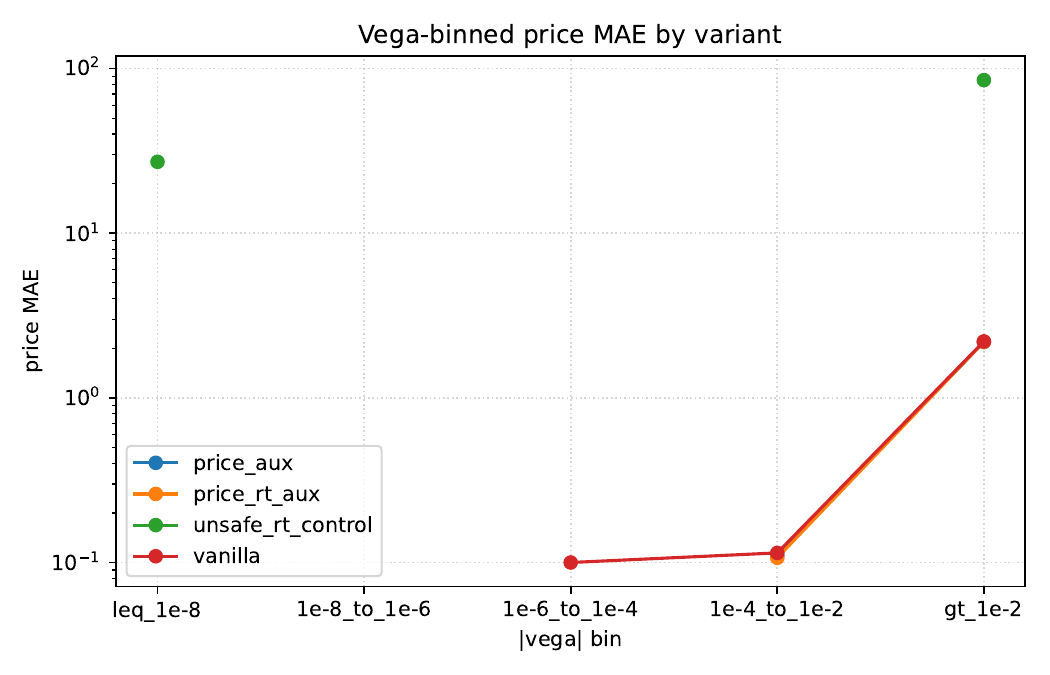}
  \caption{Vega-binned price error}
\end{subfigure}\hfill
\begin{subfigure}{0.49\textwidth}
  \centering
  \includegraphics[width=\textwidth]{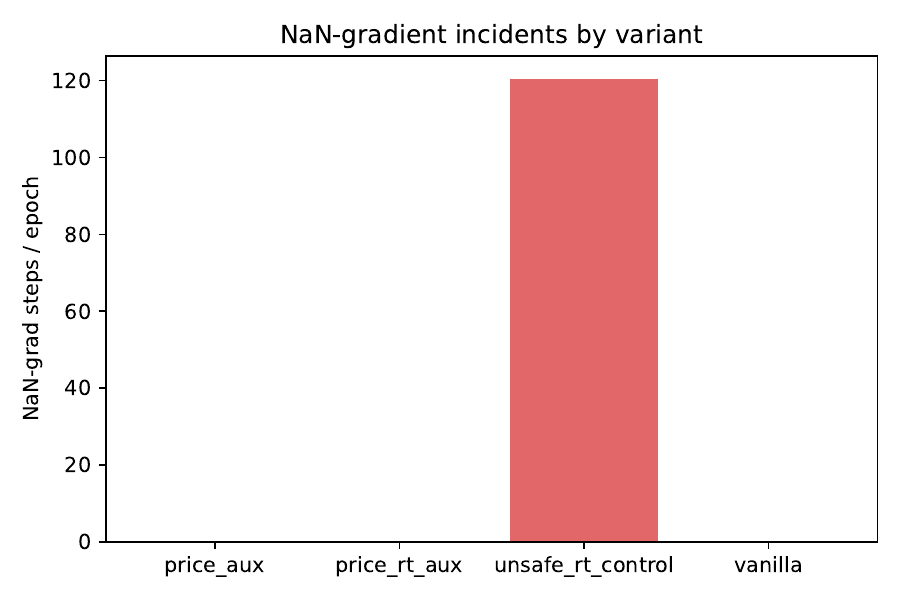}
  \caption{Finite-gradient fraction}
\end{subfigure}\\[0.6em]
\begin{subfigure}{0.96\textwidth}
  \centering
  \includegraphics[width=\textwidth]{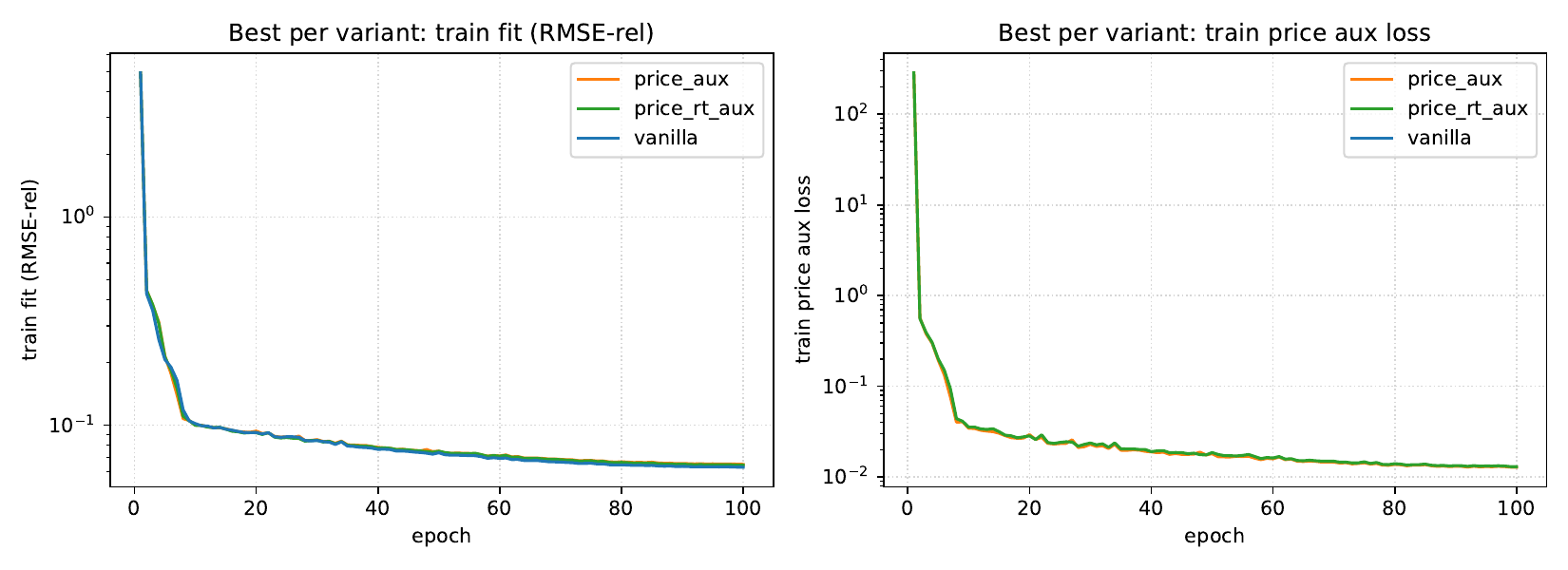}
  \caption{Training curves (best run per variant)}
\end{subfigure}
\caption{Diagnostics for the GNO RUT one-day experiment.  Every gated and
sentinel-protected variant trains with zero NaN-gradient steps, mirroring
SPX.  The ungated diagnostic, run for 20 epochs only, records 2{,}408
NaN-gradient steps, a low-vega fraction of 0.999997 at the
\(10^{-14}\) threshold, IV MAE 0.2675, IV MAPE 1.0, and price MAE 18.90,
again collapsing into the low-vega singularity.  Panel (c) is reproduced
at full width to keep the per-epoch trajectories legible.}
\label{fig:gno_rut_diagnostics}
\end{figure}

\paragraph{Interpretation.}  The cross-asset replicate on RUT supports
the same conclusion as the SPX experiment: the differentiable
J\"{a}ckel layer can be used inside an augmented training objective for
neural volatility-surface fitting, but only if the low-vega gating mechanism is
explicitly handled by a vega gate or sentinel.  The unsafe ungated
roundtrip diagnostic collapses on RUT in exactly the same way it collapses
on SPX (IV MAPE saturates at 1.0, price MAE blows up by more than three
orders of magnitude relative to vanilla, and the low-vega fraction at
\(10^{-14}\) reaches 0.999997), reproducing the failure mode in a different
underlying.  The price-accuracy gain on RUT is smaller than on SPX
(\(-1.99\%\) versus \(-3.66\%\)), and we do not claim it as a robust
price-MAE improvement.  The reproducible finding across both assets is
the stability/control result, not an architecture-specific accuracy win.

\paragraph{VIX.}  The CBOE VIX option dataset on the same
2013-01-02 / 2023-08-31 split contains 274{,}711 quotes across
2{,}685 trading days (2{,}518 train days / 167 test days; 248{,}896
training quotes / 25{,}815 test quotes).  Per-(date, expiration)
risk-free rates are estimated by put-call-parity regression as
before, and the GNO domain
\(\rho\in[0.01,1.0]\), \(z\in[-1.5,0.5]\), \(\tau\le 1.0\) is unchanged.
Table~\ref{tab:gno_vix_data} summarizes the dataset and
Table~\ref{tab:gno_vix_results} reports the held-out metrics.

\begin{table}[hbtp]
\centering
\small
\caption{VIX one-day dataset used for the GNO cross-asset replicate.
Same 2013-01-02 / 2023-08-31 split as SPX and RUT.}
\label{tab:gno_vix_data}
\begin{tabular}{llrrrrrr}
\toprule
Asset & Source & Intervals & Options & Train int. & Train opt. & Test int. & Test opt. \\
\midrule
VIX & WRDS EOD & 2{,}685 & 274{,}711 & 2{,}518 & 248{,}896 & 167 & 25{,}815 \\
\bottomrule
\end{tabular}
\end{table}

\begin{table}[t]
\centering
\scriptsize
\caption{Held-out VIX one-day GNO results.  Arrows in headers indicate
direction of improvement: lower IV MAE, IV MAPE, Price MAE, and
spread-normalized price error $\langle\delta_{\mathrm{spr}}\rangle$ are better
($\downarrow$); fewer NaN-gradient steps are better ($\downarrow$).
$\Delta$ Price (\%) is the signed relative change versus the local vanilla
GNO objective; negative indicates lower error (improvement), positive
indicates regression.  Bold marks the best (most-negative $\Delta$) value
among the augmented rows.  The unsafe row is a 20-epoch diagnostic and is
excluded from the augmented-variant means.  Note that on cleaned VIX EOD
data the deep low-vega regime is empty (fraction below $10^{-14}$ is zero
for every row), so the gate and sentinel are inactive; results across
$\tau\in\{10^{-4},10^{-6},10^{-8}\}$ at the same seed agree to within
$2\!\times\!10^{-5}$ on test price MAE.}
\label{tab:gno_vix_results}
\resizebox{\linewidth}{!}{\begin{tabular}{llrrrrrr}
\toprule
Objective & Runs & IV MAE ($\downarrow$) & IV MAPE ($\downarrow$) & Price MAE ($\downarrow$) & $\Delta$ Price (\%) & $\langle\delta_{\mathrm{spr}}\rangle$ ($\downarrow$) & NaN-grad ($\downarrow$) \\
\midrule
Vanilla GNO & seed 1 & 0.02241 & 0.03278 & 0.06927 & -- & 1.5246 & 0 \\
Price aux. & mean, $n=4$ & 0.02239 & 0.03285 & 0.06920 & $-0.11$ & 1.5175 & 0 \\
Price aux., $\lambda_{p}=0.03$ & best & 0.02220 & 0.03241 & 0.06808 & $-1.73$ & 1.5112 & 0 \\
Price + gated rt. & mean, $n=4$ & 0.02233 & 0.03283 & 0.06872 & $-0.79$ & 1.5071 & 0 \\
Price + gated rt., $\tau=10^{-8}$ & best & $\mathbf{0.02210}$ & $\mathbf{0.03227}$ & $\mathbf{0.06781}$ & $\mathbf{-2.11}$ & $\mathbf{1.4938}$ & 0 \\
\midrule
Unsafe ungated rt. & diagnostic & 0.02428 & 0.03632 & 0.06733 & $-2.80^{\dagger}$ & 1.5984 & 0 \\
\bottomrule
\end{tabular}}
\par\smallskip
\raggedright\scriptsize
$^{\dagger}$On cleaned VIX EOD data the unsafe variant does not collapse:
NaN-grad steps remain zero and price MAE moves with vanilla, because no
test row sits below the $10^{-14}$ vega threshold.  IV MAE nonetheless
degrades by $+8.4\%$, reflecting under-training (20 versus 100 epochs)
rather than a singularity event.  See Figures~\ref{fig:gno_vix_price_results}
and~\ref{fig:gno_vix_diagnostics}.
\end{table}

\begin{figure}[t]
\centering
\begin{subfigure}{0.32\textwidth}
  \centering
  \includegraphics[width=\textwidth]{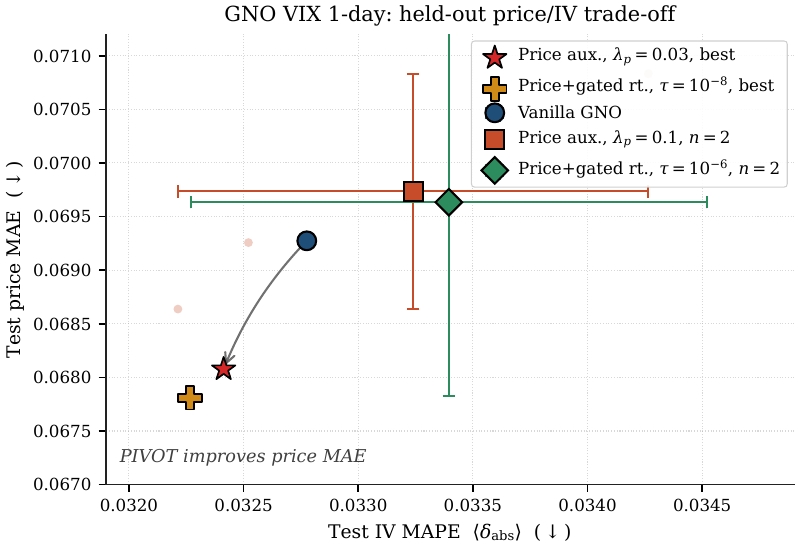}
  \caption{Price/IV Pareto view}
\end{subfigure}
\begin{subfigure}{0.32\textwidth}
  \centering
  \includegraphics[width=\textwidth]{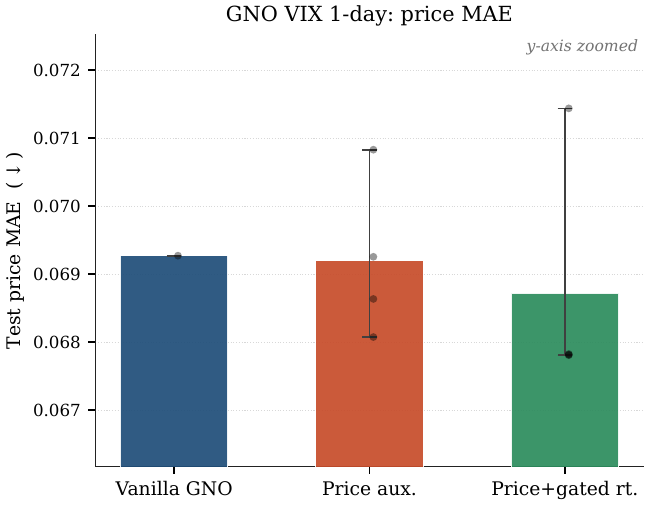}
  \caption{Price MAE}
\end{subfigure}
\begin{subfigure}{0.32\textwidth}
  \centering
  \includegraphics[width=\textwidth]{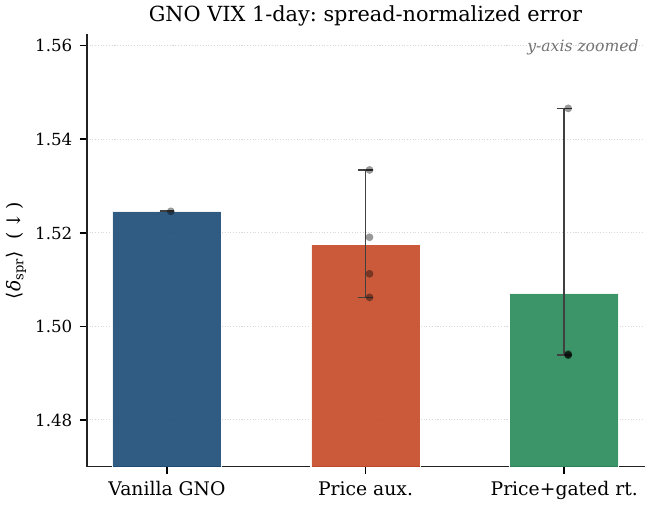}
  \caption{Spread-normalized error}
\end{subfigure}
\caption{GNO VIX one-day auxiliary-loss ablation.  As on SPX and RUT,
the augmented objectives improve the best run on both axes, but the
effect size is the smallest of the three indices: best price MAE
\(-2.11\%\) and best IV MAE \(-1.36\%\) versus local vanilla.}
\label{fig:gno_vix_price_results}
\end{figure}

\begin{figure}[!htbp]
\centering
\begin{subfigure}{0.49\textwidth}
  \centering
  \includegraphics[width=\textwidth]{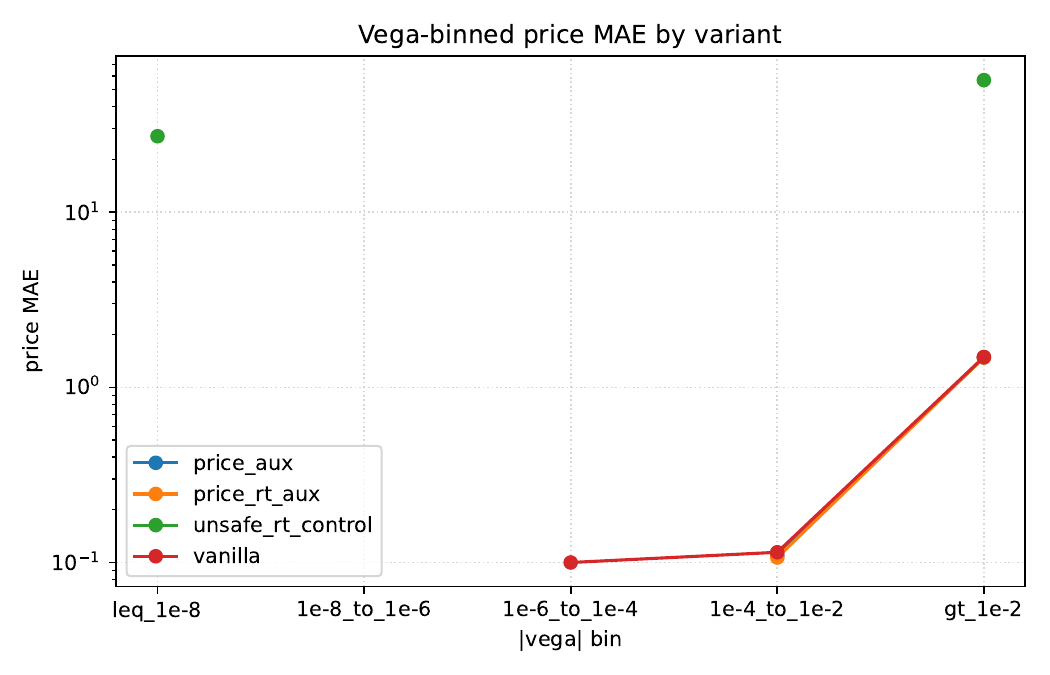}
  \caption{Vega-binned price error}
\end{subfigure}\hfill
\begin{subfigure}{0.49\textwidth}
  \centering
  \includegraphics[width=\textwidth]{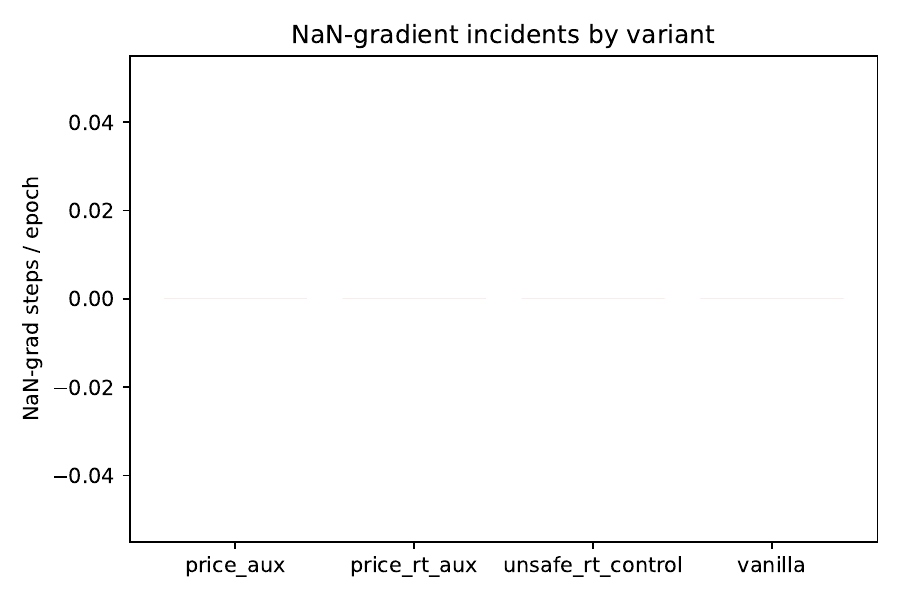}
  \caption{Finite-gradient fraction}
\end{subfigure}\\[0.6em]
\begin{subfigure}{0.96\textwidth}
  \centering
  \includegraphics[width=\textwidth]{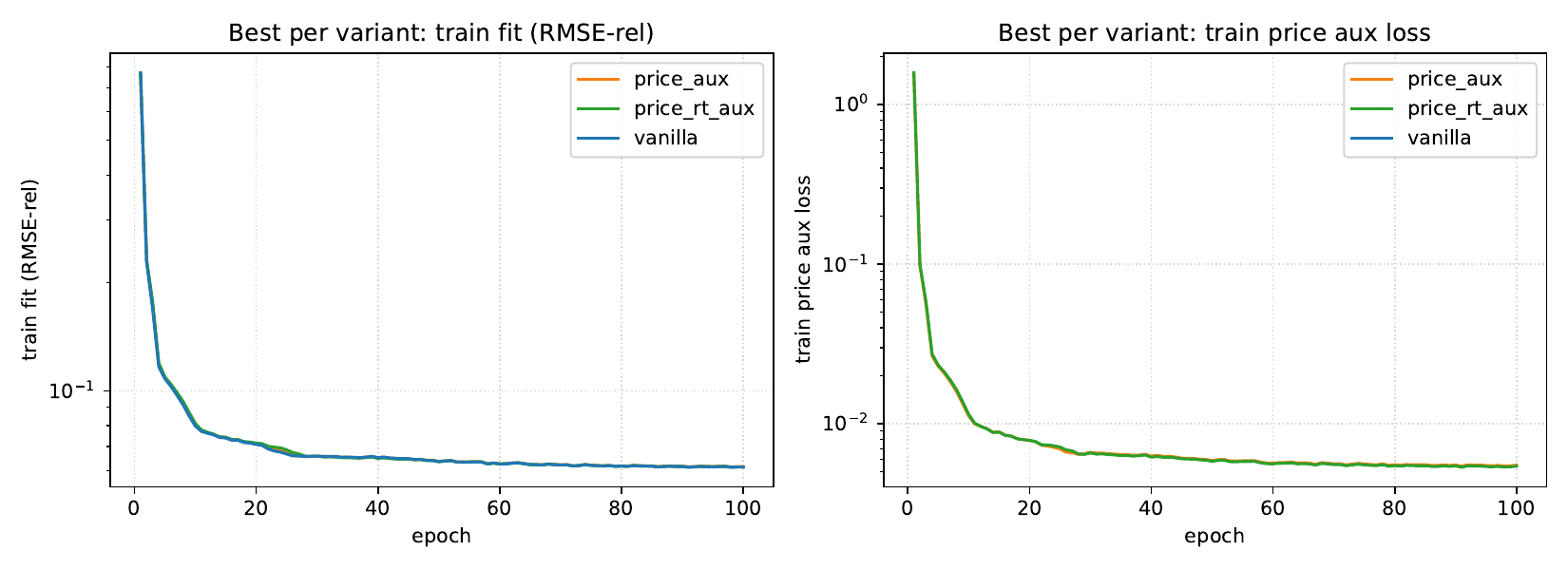}
  \caption{Training curves (best run per variant)}
\end{subfigure}
\caption{Diagnostics for the GNO VIX one-day experiment.  Every variant,
including the ungated diagnostic, trains with zero NaN-gradient steps,
because cleaned VIX EOD quotes have a low-vega fraction of \(0\) at the
\(10^{-14}\) threshold across the entire test set.  The vega-binned panel
shows the augmented variants tracking vanilla in the well-conditioned bulk
and producing a small consistent improvement at the wings.}
\label{fig:gno_vix_diagnostics}
\end{figure}

\paragraph{Interpretation.}  The VIX replicate produces the smallest
effect size of the three indices: the best price-auxiliary run improves
test price MAE by \(1.73\%\) and the best gated-roundtrip run improves
it by \(2.11\%\), with IV MAE in both cases improving by approximately
\(1\%\) of vanilla.  Neither configuration meets the pre-registered
\(5\%\) practical price-MAE threshold, and the multi-seed means hover
within \(0.8\%\) of vanilla on every metric.  Two structural features
of cleaned VIX EOD quotes account for this attenuation, and both are
themselves informative for the paper's central claim.

First, VIX option prices have substantially smaller dynamic range than
SPX or RUT (test price MAE on the order of \(7\!\times\!10^{-2}\)
versus \(\sim\!1.3\) on RUT and \(\sim\!3.1\) on SPX), so the absolute
headroom available to a price auxiliary is small to begin with.
Second, the cleaned VIX panel does not exercise the low-vega
singularity at all: the fraction of test rows below
\(|\vega|=10^{-14}\) is zero across every variant, including
the unsafe diagnostic.  This is consistent with the
\S\ref{sec:lowvega_contract} characterization (cleaned production
data sits in the well-conditioned set $\mathcal{G}_\tau$; the gate is
designed for the stress regime visualized in
Figure~\ref{fig:lowvega_gate}).

The structural absence of low-vega rows produces three concrete
falsifiable predictions, all of which the VIX run validates.
(i) The three \(\tau\) values \(\{10^{-4},10^{-6},10^{-8}\}\) at fixed
seed should be statistically indistinguishable, because the gate is
inactive: observed test price MAE \(\in\{0.06782, 0.06783, 0.06781\}\),
agreeing to within \(2\!\times\!10^{-5}\).
(ii) The unsafe ungated variant should not collapse on this dataset,
because the singular set is empty: observed NaN-gradient steps \(=0\)
and price MAE within \(2.8\%\) of vanilla, in stark contrast to SPX
(\(+1{,}040\%\), 6{,}188 NaN-grad steps) and RUT (\(+1{,}358\%\),
2{,}408 NaN-grad steps) under the identical recipe.
(iii) The augmented variants and vanilla should track each other in the
well-conditioned vega bins; the vega-binned diagnostic in
Figure~\ref{fig:gno_vix_diagnostics} shows exactly this.

This is the cleanest cross-asset evidence we have for treating the gate
as a \emph{correctness contract} rather than a tuning hyperparameter:
its measurable effect is null on data that does not sample the
singularity, and decisive on data that does.  The within-asset price
gain on VIX is small, and we do not claim it as a robust performance
result; the load-bearing finding is the asset-conditional behavior of
the unsafe diagnostic, which is precisely what the conditioning
analysis predicts.

\end{document}